\documentclass{phd}  % comment out if a MS thesis
\usepackage{buthm}
\usepackage{amsmath,amssymb,amsxtra,subfigure,lscape}
\usepackage{graphicx}
\usepackage{citesort}
\newcommand{\perk}{\raise-1mm\hbox{\large$\Box$}}

\def\beq{\begin{equation}}
\def\eeq{\end{equation}}
\def\beqn{\begin{eqnarray}}
\def\eeqn{\end{eqnarray}}

\begin{tiny}\label{} \end{tiny}%math definitions

\def\slash#1{#1\hskip-6pt/\hskip6pt}

\def\p4{\Phi_4}
\def\pp4{\Phi^{'}_4}
\def\pb4{\bar{\Phi}_4}
\def\ppb4{\bar{\Phi}^{'}_4}
\def\p#1{{\Phi_{#1}}}

\def\pp#1{{\Phi^{'}_{#1}}}
\def\pb#1{{{\overline{\Phi}}_{#1}}}

\def\ppb#1{{{\overline{\Phi}}^{'}_{#1}}}

\def\FD2pv{FD2$^{'}$V }
\def\FD2p{FD2$^{'}$ }

\def\inbar{\,\vrule height1.5ex width.4pt depth0pt}

\def\IC{\relax\hbox{$\inbar\kern-.3em{\rm C}$}}
\def\IQ{\relax\hbox{$\inbar\kern-.3em{\rm Q}$}}
\def\IR{\relax{\rm I\kern-.18em R}}
 \font\cmss=cmss10 \font\cmsss=cmss10 at 7pt

\def\IZ{\relax\ifmmode\mathchoice
 {\hbox{\cmss Z\kern-.4em Z}}{\hbox{\cmss Z\kern-.4em Z}}
 {\lower.9pt\hbox{\cmsss Z\kern-.4em Z}}
 {\lower1.2pt\hbox{\cmsss Z\kern-.4em Z}}\else{\cmss Z\kern-.4em Z}\fi}

\def\Io{\relax\ifmmode\mathchoice
 {\hbox{\cmss 1\kern-.4em 1}}{\hbox{\cmss 1\kern-.4em 1}}
 {\lower.9pt\hbox{\cmsss 1\kern-.4em 1}}
 {\lower1.2pt\hbox{\cmsss 1\kern-.4em 1}}\else{\cmss 1\kern-.4em 1}\fi}

\title{Strange Quark Contribution to the Nucleon}
\author{Dean F. Darnell}
\mentor{Walter M. Wilcox, Ph.D.}
\reader{Ronald B Morgan, Ph.D.}
\readerThree{Gerald Cleaver, Ph.D.}
\readerFour{Greg Benesh, Ph.D.} 
\readerFive{Jay Dittmann, Ph.D.}

\confDate{August 2006}
\makeCopyrightPage

\graduateDean{J. Larry Lyon, Ph.D.} \deptChair{Greg Benesh, Ph.D.}

\abstract{The strangeness contribution to the electric and magnetic properties of the nucleon has been under investigation experimentally for many years. Lattice Quantum Chromodynamics (LQCD) gives theoretical predictions of these measurements by implementing the continuum gauge theory on a discrete, mathematical Euclidean space-time lattice which provides a cutoff removing the ultra-violet divergences. In this dissertation we will discuss effective methods using LQCD that will lead to a better determination of the strangeness contribution to the nucleon properties. 
\enlargethispage*{4\baselineskip}
Strangeness calculations are demanding technically and computationally. Sophisticated techniques are required to carry them to completion. In this thesis, new theoretical and computational methods for this calculation such as twisted mass fermions, perturbative subtraction, and General Minimal Residual (GMRES) techniques which have proven useful in the determination of these form factors will be investigated. Numerical results of the scalar form factor using these techniques are presented. These results give validation to these methods in future calculations of the strange quark contribution to the electric and magnetic form factors.}
          
\acknowledgements{``As iron sharpens iron, so as man sharpens another." - Proverbs 27:17

First and foremost, I would like to thank my favorite $roomate$. To my beautiful wife, Amanda Nicole Darnell for being patient, supportive, and loving during the trying ``graduate school years". Without her love I would have been lost. I would also like to thank our families, Ray and Joyce Darnell as well as Tom and Pat Shirreffs for insight and support. I would like to especially thank my dad, Ray, for being my inspiration and hero for all of these years. Without him I would have never had the courage to pursue this task.

I would like to thank Dr. Walter M. Wilcox and Dr. Ronald B. Morgan for their advisement and mentoring in physics and numerical mathematics. These men not only helped me become a better scientist and mathematician, but through their daily lives, friendship, and mentoring taught me to be a better person. I will be forever grateful for their gifts to me.

To the roommates! I would like to honor my brothers; James Bach, Dan Hernandez, and Matt Vial for encouragement, adventure, and unconditional friendship. I look forward to many more years of the same kind of interactions that made the last decade so special. Also, I would like to honor John Perkins, Dan Dries, and Darren Gross for insightful conversations and similar mischief. Friends make life special! I would also like to acknowledge Mike Hutcheson, Tim Logan, and Carl ``the debug monkey" Bell for their technical support and friendship. Carl, would you like fries with that?

Finally, I would like to thank the physics department at Baylor University for teaching me the necessary tools and skills to allow me to conduct research. The warmth and encouragement here at Baylor is unparalleled anywhere else. Specifically, I would like to acknowledge Dr. Gerald and Lisa Cleaver and Dr. Jay and Jeannie Dittmann. Thanks. I would also like to acknowledge the National Science Foundation for funding our research and my graduate experience.
}
\dedication{\begin{center}To ``The Roomates"\end{center}}
\begin{document}

\DeclareGraphicsExtensions{.pdf,.png,.gif,.jpg}

\pagenumbering{arabic}

\chapter{Introduction}\label{intro}
In physics today there exist four fundamental forces: the strong force, the weak force, electromagnetism and gravity. The focus of this thesis is the strong force and related particles. Quantum Chromodynamics (QCD) is the study of the strong interaction.

A hadron is a particle constructed of quarks and gluons, which are the fundamental strong force particles. There exist six different flavors of quarks: up (u), down (d), strange (s), charmed (c), bottom (b), and top (t). Of these six quarks the up, down, and strange are known as the light quarks. The up and down quarks have masses of a few MeV while the strange quark has a mass of approximately 120 MeV. The light quarks are present in low energy nuclear physics which is a topic of investigation in future chapters of this thesis.  

QCD is a gauge theory based on the non-abelian SU(3) gauge group. The eight independent generators of SU(3) give rise to eight massless gluons carrying a color charge. Gluons are the strong ``force carriers" in QCD.   

The Lagrangian density of QCD is
\begin{eqnarray}\label{QCDL}
L_{QCD} &=& 1/4 F^{a}_{\mu \nu} F^{a \mu \nu} +\bar q (\slash{D} -  m_{q})q 
\end{eqnarray}
where the field tensor $F^{a}_{\mu \nu}$ is
\begin{eqnarray}
F^{a}_{\mu \nu} &=& \partial_{\mu}G^{a}_{\nu}(x) - \partial_{\nu}G^{a}_{\mu}(x) + ig_{o}[G^{a}_{\mu}(x), G^{a}_{\nu}(x)],
\end{eqnarray}
and $G^{a}_{\mu}$ are the gluon fields. The index $a$ is a color index. The free parameters in the QCD Lagrangian density are the gauge coupling constant, $g_{o}$, and the quark masses, $m_{q}$.

QCD has been well investigated with perturbation theory in the high energy regime. In the low energy limit, QCD should describe nuclear physics and the hadron mass spectrum. Hadron masses depend on the gauge coupling constant like $M_{hadron} \sim e^{-1/g_{o}^{2}}$. When the QCD coupling constant is large, perturbation theory is not valid and a new recipe is needed. The only solution in present day physics is Lattice Quantum Chromodynamics (LQCD). Lattice QCD was first introduced by Kenneth Wilson in 1974 ~\cite{Wilson}.

Lattice QCD implements field quantization through path integrals and the discretization of space-time onto a four-dimensional Euclidean lattice. The path integrals on this space-time lattice allow the lattice gauge theory to be studied numerically with Monte Carlo simulations. These simulations share similarities with statistical models in Solid State physics. These similarities allow the particle physicist to use similar analysis techniques as used in the Solid State models to extract meaningful results from the lattice. 

The strangeness contribution to the electric and magnetic properties of the nucleon has been under investigation experimentally for many years. Lattice calculations of the strange quark in the presence of a nucleon are both computationally expensive making meaningful results difficult to extract. New computational and numerical techniques are needed to determine the nucleon properties. In this dissertation we will discuss effective methods using LQCD that will lead to a better understanding of the strangeness contribution to the nucleon.

\section{Experimental Motivation}

More accurate theoretical predictions of the disconnected strangeness matrix elements are needed to compare with experiment. The current experimental measurement of the low-momentum transfer of the strange nucleon form factors are being conducted by groups at HAPPEX ~\cite{collaboration-2006-635}, A4 ~\cite{mass}, and SAMPLE ~\cite{spayde}. The most recent experimental results published by these groups and the group at Thomas Jefferson National Accelerator Facility (JLab) are summarized in Fig \ref{fig:happex}.

{\setlength{\abovecaptionskip}{0mm}
\setlength{\belowcaptionskip}{5mm}
\begin{figure}[t!]
\centering
\includegraphics[scale=1, height=10cm, width=13cm]{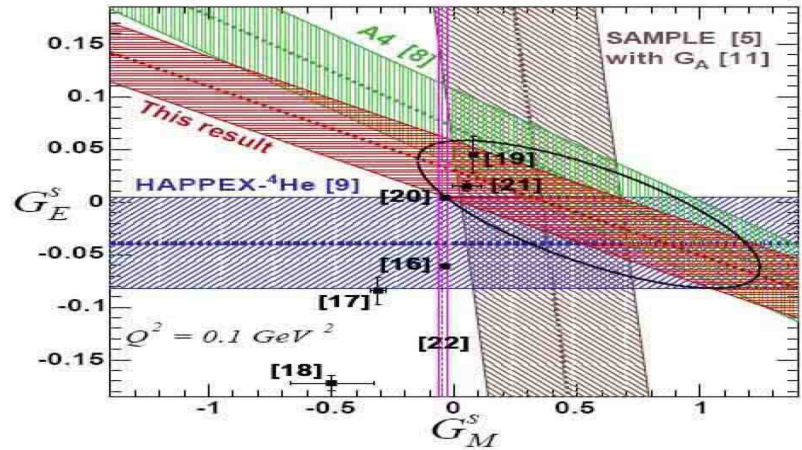}
\caption{Experimental results for the simultaneous strange electric and magnetic form factors at small four-momentum transfer.}
\label{fig:happex}
\end{figure}}

\enlargethispage{\baselineskip}
Figure \ref{fig:happex} is a plot of linear combinations of the electric $G_{E}(q^{2})$ and magnetic $G_{M}(q^{2})$ form factors using  a parity violating electron-proton scattering process. The ellipsed region is the experimental $95\%$ confidence region. The leading lattice result marked as $[21]$ is well within this $95\%$ confidence region indicating small positive values for the electric and magnetic form factors. Result $[21]$ from the authors of reference ~\cite{lewis-2003-67} are from a quenched lattice calculation employing chiral perturbation models to extend to the continuum theory. This result can be improved by introducing smaller quark masses to the simulation so that a stronger connection can be made with chiral models. The agreement with experimental results is strong motivation to look deeper into the strange disconnected form factor. 

To make better connection with these experimental results smaller quark masses must be used in the lattice calculation. The Wilson QCD action can suffer from gauge configurations which produce unphysical results that prohibit the calculation of small quark masses. Therefore, theorists must turn to other methods that can avoid these types of damaging configurations. One such method that removes the unphysical results and produces more reliable physics is twisted mass QCD (tmQCD). The tmQCD action is used in this thesis to improve the strangeness calculation.

In this thesis, we will discuss the basic lattice techniques that are used in this hadron calculation. In chapter two, the basics of lattice gauge theory are reviewed. Next, there is a review of twisted mass LQCD and the symmetries that are preserved in this formalism. We consider the lattice techniques necessary to extract meaningful results in chapter four. 

New work is presented in chapter five. This work discusses new mathematical algorithms to efficiently solve linear systems of equations giving quark propagators for both the Wilson and twisted mass formalism. In addition to these new methods, a perturbative method to calculate the strange quark vacuum expectation values is discussed in chapter six. Here, an extension to the existing method in reference ~\cite{Wnoise} is employed and an introduction to a twisted mass disconnected perturbative technique is given.  Finally, the simulation details and numerical results are presented in chapter seven. Conclusions of the strangeness calculation and plans for future work are summarized in chapter eight. 

\chapter{Lattice Gauge Theory}
Lattice gauge theory is the discretization of the QCD action onto a four-dimensional hyper-cubic lattice with a finite lattice spacing. There are, of course, an infinite number of ways to define a discrete gluonic and fermionic action on the lattice but the simplest method is the Wilson gauge action using the Wilson Dirac operator. These methods retain the necessary symmetries that continuum QCD requires. In this chapter the fundamental concepts of lattice gauge theory are discussed. A more complete discussion of lattice gauge theory can be found in many texts and journals ~\cite{Creutz,Huang,Close,Green,gupta-1998,Peskin}. 
 
\section{Lattice Gauge Fields}
The continuum gauge fields are represented by $A_{\mu}$, which belong to the gauge algebra. The corresponding lattice gauge fields, $U_{\mu}(x)$, belong to the the gauge group $G$. The role of the lattice gauge fields is to move color locally between nearest neighbor lattice sites. On any plane of the lattice we define two unit vectors $\hat\mu$ and $\hat\nu$ that define the directional orientation of the gauge links (See Figure \ref{fig:lattice}). 

{\setlength{\abovecaptionskip}{0mm}
\setlength{\belowcaptionskip}{0mm}
\begin{figure}[h!]
\centering
\includegraphics[scale=1, height=6cm, width=6cm]{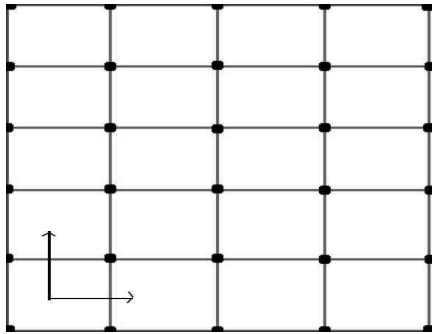}
\caption{A plane in the lattice showing the gauge link structure.}
\label{fig:lattice}
\end{figure}}

Let $a$ be the lattice spacing. If $U_{\mu}(x+a)$ is the gauge link between space-time points $x_{i}$ and $x_{i} + a$ in the $\mu$ direction, then the gauge field that moves in the opposite direction from $x_{i} + a$ to $x_{i}$ is the Hermitian conjugate of $U_{\mu}(x+a)$ due to the unitarity of gauge fields. 

The continuum and lattice gauge fields are related by
\begin{eqnarray}
 U_{\mu}(x) &=& e^{-iag_{o}A_{\mu}(x)} 
\end{eqnarray}
where $a$ is the lattice spacing, $g_{o}$ is the coupling constant and $A_{\mu}(x)$ are the continuum gauge fields. Gauge fields on the lattice must obey local gauge transformations as they do in the continuum theory. To apply a local gauge transformation to a link we must specify a gauge transformation at the beginning and end-point of that gauge link. Let the local gauge transformation be $G(x)$. The gauge link and fermion fields under a local gauge transformation $G(x)$are  
\begin{eqnarray}
 U_{\mu}(x)  &\rightarrow& G(x)U_{\mu}G^{\dagger}(x+a_{\mu}),
\end{eqnarray} 
\begin{equation}
\psi(x) \rightarrow G(x)\psi(x) .
\end{equation}
With these definitions we are now able to construct gauge invariant operators on the lattice. For example, in the pure gauge theory it is now possible to construct a closed Wilson loop. A Wilson loop is constructed by taking the trace of four links around a closed loop in the $\mu$ - $\nu$ plane. This operator is independent of starting position and is invariant under gauge transformations. The simplest non-trivial Wilson loop is the average plaquette. A plaquette is a closed loop, gauge invriant object constructed of gaugelinks on the lattice. The average plaquette is an order parameter of the Wilson theory.  

According to Wilson, the discrete gauge field action is given by
\begin{equation}
S_{G}[U] = \frac{1}{g_{o}^{2}} \sum_{p} Re\left\{ tr\{1 - U(p)\}\right\} ,
\end{equation} 
where the sum is over all elementary plaquettes, $U(p)$. Wilson showed that this action is equivalent to the continuum action to leading order in the lattice spacing $a$.

\section{Lattice Fermions}
The Euclidean continuum fermion action for QCD is
\begin{equation} \label{SF}
S_{F}^{cont.} = - \int d^{4}x\bar\psi^{cont.}(x) (\slash{D}_{\mu} + m)\psi^{cont.}(x).
\end{equation}

The four components of $\slash{D}$ are the usual $\slash{D} = D_{\mu}\gamma_{\mu}$. The $\gamma_{\mu}$ matrices are a set of four matrices that satisfy the algebra
\begin{eqnarray}
[\gamma_{\mu},\gamma_{\nu}]_{+} &=& 2 \delta_{\mu\nu} \\
\gamma^{\dagger}_{\mu} &=& \gamma_{\mu}. \nonumber
\end{eqnarray}
We also define the quantities
\begin{eqnarray}
\bar\psi &=& \psi^{\dagger}\gamma_{4}, \\
\gamma_{5} &=& \gamma_{1}\gamma_{2}\gamma_{3}\gamma_{4}, \nonumber \\
\gamma_{5} &=& \gamma^{\dagger}_{5}.
\end{eqnarray}
The representation for the 4$\times$4 $\gamma$ matrices we use is 
\\
\begin{displaymath}
\gamma_{i}=
\left(
\begin{array}{rr}
 0 & \sigma_{i}\\
 \sigma_{i} & 0\\
\end{array}
\right),
%\end{displaymath}
%\\
%\begin{displaymath}
\gamma_{4}=
\left(
\begin{array}{rr}
 1 & 0\\
 0 & -1\\
\end{array}
\right) ,
\end{displaymath}
\\
\begin{displaymath}
\gamma_{5}=
\left(
\begin{array}{rr}
 0 & -i\\
 i & 0\\
\end{array}
\right) .
\end{displaymath}
\\
where the index $i = 1,2,3$ and the $\sigma_{i}$ are the $2 \times 2$ Pauli matrices.
 
A discrete representation of equation (\ref{SF}) is needed for lattice calculations. We require that the fermion fields and operators only exist on the lattice sites themselves. This is in contrast to the links that only exist between lattice points. The lattice fermion fields are Grassmann-valued fields that carry flavor, color, and Dirac indices.

\subsection{N$\ddot{a}$ive Fermion Action}
Lattice fermions in Euclidean space are represented by anticommuting Dirac spinors, $\psi(x)$, that satisfy the relations
\begin{equation}
[\psi,\psi]_{+} = [\psi^{\dagger},\psi]_{+} = [\psi^{\dagger},\psi^{\dagger}]_{+} = 0.
\end{equation}
To find a discrete fermion action for these fields, Wilson replaced the covariant derivative in the continuum action with a symmetrized difference equation. By using the correct choice for gauge links as well, the discrete fermion action remains gauge invariant. To leading order in $a$, the n$\ddot{a}$ive action for the fermion fields is 
\begin{eqnarray}
S^{N\ddot{a}ive}_{F} &=& m_{q}\sum_{x}\bar\psi(x)\psi(x)\\
         &+& \frac{1}{2a}\sum_{x}\bar\psi(x)\gamma_{\mu}[U_{\mu}(x)\psi(x+\mu) - U^{\dagger}_{\mu}(x-\mu)\psi(x-\mu)] \nonumber \\
         & \equiv & \sum_{x}\bar\psi(x) M^{N\ddot{a}ive}_{xy}[U]\psi(y)
\end{eqnarray}
where the n$\ddot{a}$ive interaction matrix is
\begin{equation}\label{Mnaive}
M^{N\ddot{a}ive}_{xy}[U] = m_{q}\delta_{xy} + \frac{1}{2a}\sum_{\mu}\gamma_{\mu}[U_{x,\mu}\delta_{x,y-\mu} - U^{\dagger}_{x-\mu,y}\delta_{x,y+\mu}].
\end{equation}
In equation \ref{Mnaive}, $m_{q}$ is the quark mass and the sum is over Dirac indices. The n$\ddot{a}$ive fermion action creates huge problems on the lattice. Consider the inverse of the free field propagator in momentum space: 
\begin{eqnarray}
S^{-1}(p) &=& \sum_{x,y}M^{N\ddot{a}ive}_{x,y}[U=1]e^{i p \cdot (x-y)}. \\
          &=& m_{q} + \frac{i}{a}\sum_{\mu}\gamma_{\mu}sin(p_{\mu}a).
\end{eqnarray}
In the limit as $m_{q} \rightarrow 0$, the inverse propagator creates $2^{4}$ zeros in the momentum space unit cell. Each of these zeros corresponds to a species of fermion on the lattice. This is obviously an unacceptable result. This phenomena is known as fermion doubling because there are two species in each direction of the lattice. 

\subsection{Corrected Fermion Actions}

There are many possible corrections to the n$\ddot{a}$ive fermion action that will remove the doubling problem and still remain a ``good action" in the continuum limit. Three good choices for actions are the Wilson, Kogut-Susskind, and twisted mass fermion actions. The advantages and disadvantages of each of these actions will be presented.

\subsection{Wilson Fermions}

One approximation to the n$\ddot{a}$ive action is the Wilson fermion action. Wilson added a second derivative term to the n$\ddot{a}$ive fermion action that results in a rescaled factor that is related to the bare quark mass by
\begin{equation}\label{hop}
\kappa = \frac{1}{2(4r+m_{q}a)}.
\end{equation} 
$\kappa$ is known as the hopping parameter. Equation (\ref{hop}) can be solved for the quark mass $m_{q}$ in terms of lattice parameters $\kappa$ and $r$. The quark mass then is
\begin{eqnarray}\label{qm}
m_{q}a &=& \frac{1}{2\kappa} - 4r \\
       &=& \frac{1}{2}(\frac{1}{\kappa} - \frac{1}{\kappa_{c}}). 
\end{eqnarray}
$\kappa_{c}=1/8$ for the non-interaction theory. The same formula holds for the interaction case.

This discrete fermion action, also known as the Wilson action, is written 
\begin{eqnarray}\label{WilsonAction1}
S^{W}_{F} &=& \kappa \sum_{x}\bar \psi(x) \psi(x) \\
          &+& \frac{1}{2a}\sum_{\mu}[\bar\psi(x) (\gamma_{\mu} - r)U_{\mu}(x)\psi(x + \mu) - \bar\psi(x) (\gamma_{mu}+r)U^{\dagger}_{\mu} (x-\mu )\psi(x-\mu) ]. \nonumber
\end{eqnarray}
In the free field limit, when $r=1$ the doubling problem is resolved. The matrices $(\gamma_{\mu} - 1)U_{\mu}(x)$ and $(\gamma_{\mu} + 1)U^{\dagger}_{\mu}$ $\space$ in (\ref{WilsonAction1}) are the forward and backward quark hopping terms, respectively.

We can rescale the fields in the Wilson action by letting $\psi \rightarrow \sqrt{2\kappa}\psi$, giving a convenient form of the Wilson action 

\begin{eqnarray}\label{WilsonAction}
S^{W}_{F} &=& \sum_{x}\bar \psi(x) \psi(x) \\
          &+& \kappa \sum_{\mu}[\bar\psi(x) (\gamma_{\mu} - r)U_{\mu}(x)\psi(x + \mu) - \bar\psi(x) (\gamma_{\mu}+r)U^{\dagger}_{\mu} (x-\mu )\psi(x-\mu) ]. \nonumber
\end{eqnarray}

It is known that for small quark mass, $m^{2}_{\pi} \approx m_{q} \approx \kappa_{c} - \kappa$. By definition, $\kappa_{c}$ in (\ref{qm}) is the value which causes the pion mass to be zero. The calculation of $\kappa_{c}$ is statistical in nature and is determined by the limit $m_{\pi} \rightarrow 0$. When a zero mode occurs at a value of $\kappa < \kappa_{c}$ $\space$ for a given configuration, the quark propagator becomes singular in a physical region. These unphysical modes are called ``exceptional configurations", and are a large concern for the Wilson action in the quenched approximation (see section \ref{qa}). Dealing with this problem is a major focus of this thesis.  

A consequence of the ``r" term in the Wilson action is that it breaks chiral symmetry at $O(a)$ in lattice spacing. Consequently, an additive mass renormalization is required. The loss in chiral symmetry results in operator mixing and additional field renormalizations. 

Even though the Wilson action introduces ``exceptional configurations" and breaks chiral symmetry, it does preserve a one-to-one correspondence between the Dirac and flavor degrees of freedom and the continuum theory. This is a huge advantage because it allows the interpolating field operators to be constructed in the same manner as in the continuum limit. For example, 
$\bar\psi(x)\psi(x)$ $\space$ (scalar) and $\bar\psi(x)\gamma_{\mu}\psi(x)$ $\space$ (vector) have the same form on the lattice as in the continuum.  

An alternative formalism that is closely related to the Wilson action is the twisted mass action. In this formalism an additional term is added to the Wilson action that removes the unphysical quark modes. This formalism was proposed by Frezzotti and Rossi in 2001 ~\cite{frezzotti-2001-0108}. Twisted mass LQCD is the new frontier for lattice calculations and will be discussed in depth in future chapters. 

\subsection{Staggered Fermions}

Staggered Fermions reduce the number of fermion species by using one component ``staggered" fermion fields rather than the usual four component Dirac spinors and by employing a spin diagonalization of the spin components of the fermion fields ~\cite{Kogut:1974ag,Banks:1975gq,Susskind:1976jm}. Each of the staggered flavor and spin fields is placed on a corner of the lattice.  The diagonalization of the fermion fields removes the 16-fold doubling problem of the n$\ddot{a}$ive fermion action. This discretization of the action also preserves chiral symmetry when $m_{q} \rightarrow 0$, because there is no rotation under the subgroup $U(1)$ from the single spin index.  When chiral symmetry is desired, staggered fermions are preferred to Wilson fermions. The exceptional configuration problem is also greatly reduced and one can go lower in quark mass in computer simulations. 

The disadvantage of this formalism is that there is now a 4-fold degeneracy for each physical flavor in the continuum limit. The degenerate states are called ``tastes", to distinguish them from the physical flavors. This degeneracy breaks the flavor symmetry at $O(a)$ on the lattice which makes construction of operators with correct quantum numbers difficult. Computationally, staggered fermions save roughly a factor of 4 in computer time because they use only a single component Dirac spinor, thus saving on storage space as well.

%\subsection{Gauge Improvement}
%
%Since we don't use the Luscher-Wies action, tadpole improvement, or clover term should I have a section about it? I am not sure, I think that these are important things to be aware of and someone might ask "Why don't you do this.." type question. 
%
%Think about this.

\subsection{Lattice Errors}

In any lattice calculation there are statistical and systematic errors. The statistical errors are a result of the Monte Carlo stochastic method and fall off like $\frac{1}{\sqrt{N}}$. The systematic errors are a result of approximating a spatially and temporelly infinite problem on a finite lattice. Two well known errors that are a direct result of the  discretization of the lattice are the finite volume and finite lattice spacing effects. 

Another source of systematic error occurs when the lattice results are extrapolated to the continuum limit. One must implement a chiral perturbation theory to reach the continuum. This extrapolation carries inherent error that appears in the final lattice result. 
\enlargethispage*{2\baselineskip}
\subsection{Finite Volume Effects}

The volume of the lattice is given by
\begin{equation}
V_{lat} = L_{x} * L_{y} * L_{z} * L_{t},
\end{equation}
where $L_{i} = a n_{i}$. $n_{i}$ is the number of lattice sites in the $i^{th}$ direction. If $L_{i}$ is large, it has been shown that the finite volume errors fall off exponentially ~\cite{Lepage:1994yd}
\begin{equation}
error(m_{\pi}) = e^{-m_{\pi}L_{i}}.
\end{equation}
To avoid finite volume effects the length of the lattice must be larger than the particle cross-section. A light hadron cross section is about 2 fm in diameter. Since the lattice employs periodic boundary conditions the hadron on the lattice will also have reflections of itself in any given periodic direction. When $L_{i}$ is large enough the hadron does not overlap with its reflected image and the volume effect is small. On the other hand, if the lattice length is smaller than the hadron diameter and the hadron overlaps with it's image, the hadron mass will be large. This produces large finite volume errors. 

\subsection{Finite Lattice Spacing Effects}

Fields in quantum theories suffer from fluctuations at all length scales. In perturbation theory, these fluctuations are responsible for ultraviolet sensitivities and infinities in loop diagrams. In light of this, it is hard to understand how we might define a discrete approximation to a continuum field that is already randomly fluctuating and coarse. Fortunately, only long wavelength objects are physical on the lattice. In general, any low momentum, long-wavelength probe is only sensitive to space-averaged fields on the order of the probe itself. The averaging of the fields suppresses the quantum fluctuations on the lattice. Consequently the infrared behavior is not sensitive to a specific ultraviolet theory. There are, therefore, an infinite number of ways to construct an ultraviolet theory with the same infrared physics. 

In quantum theory the infrared modes can be affected by the quantum fluctuations of the ultraviolet mode via the mass and coupling terms. However, if we choose an ultraviolet theory that permits us to change the bare coupling and mass terms such that the infrared behavior is the same in the continuum limit up to a renormalization of $O(a)$, we can avoid quantum fluctuations ~\cite{Lepage:1994yd} . Effectively, the lattice acts as an ultraviolet cut-off that restricts the particle modes to low momenta. Ultimately, to avoid quantum fluctuations and costly renormalizations in a lattice measurement, the lattice spacing $a$ must be smaller than any important scale for the hadron calculation under investigation. 

\subsection{Chiral Extrapolations of Light Quark Masses}

The quark masses, $u$ and $d$, are too light to simulate in current lattice calculations because of the exceptional configuration problem and increased statistical fluctuations. While new methods are being formulated, the lowest pion mass that can be calculated is approximately 500 MeV for the Wilson formalism. (One can go much lower with staggered fermions, but there are interpretational problems.) The current method to determine the physical pion mass is to calculate many different pion masses and extrapolate to the physical value near 140 MeV. This extrapolation process is known as Chiral Pertrubation Theory ($\chi$PT). As with any statistical measurement, the extrapolated physical $m_{\pi}$ has an associated uncertainty. This technique has provided reliable results for many lattice calculations, however, the ultimate goal is to produce better simulations of the light quark masses so that there is less dependance on $\chi$PT.

\section{Quenched Approximation}\label{qa}

Full QCD calculations are currently unrealistic computationally. A remarkably good alternative to full QCD is the Quenched QCD (QQCD). It consists of neglecting the determinant of the quark matrix in the lattice gauge field action. Physically, the quenched approximation is equivalent to neglecting the vacuum polarization effects of quark loops in lattice calculations. Neglecting these vacuum loops only changes the relative weighting of the background for QQCD.

At short distances the only difference between quenched and full QCD is a small change in the QCD coupling constant. This is known as asymptotic freedom. The quenched approximation saves factors of $10^{2}$ - $10^{4}$ in computer time while preserving asymptotic freedom, confinement, and the chiral symmetry breaking that QCD includes. All of our calculations are performed in the quenched approximation.

\section{Gauge Field Construction}

In practice, to generate gauge fields for Lattice QCD Monte Carlo methods are employed for the numerical integration of Feynmann path integrals. Monte Carlo methods are especially useful in studying physical systems with a large number of coupled degrees of freedom in which the inputs have significant uncertainty. 

The QCD path integral is 
\begin{equation}
Z = \int DA_{\mu}D\psi D\bar{\psi}e^{-S},
\end{equation}
where the integration is over gluonic and fermionic fields. The associated QCD action with this path integral is 
\begin{equation}
S = \int d^{4}x \left\{ \frac{1}{4}F_{\mu\nu}F^{\mu\nu} - \bar{\psi}M \psi \right\},
\end{equation}
where $M$ is the fermion matrix. 

As an instructive, simple example ~\cite{davies-2005-}, consider the path integral of a particle moving in a one dimensional well
\begin{equation}
\int Dx(t)e^{-S[x]},
\end{equation} 
where the discrete action is 
\begin{equation}
S[x] = \int^{t_{f}}_{t_{i}} \sum^{N-1}_{i=0}[\frac{m}{2a}(x_{i+1} - x_{i})^{2} + aV(x_{i})].
\end{equation}
The corresponding picture of this action is in figure \ref{fig:claspart}.

{\setlength{\abovecaptionskip}{0mm}
\setlength{\belowcaptionskip}{0mm}
\begin{figure}[t!]
\centering
\includegraphics[scale=1, height=7cm, width=12cm]{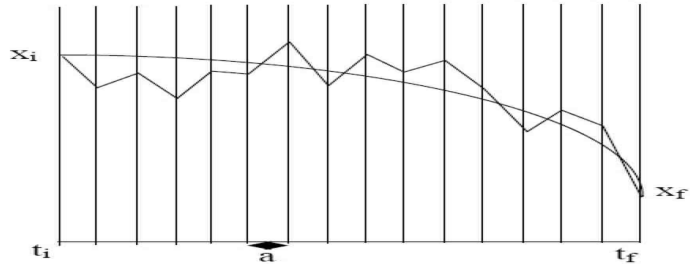}
\caption{Classical particle one-dimensional trajectory (smooth and discrete) from $x_{i} \rightarrow x_{f}$.}
\label{fig:claspart}
\end{figure}}

For large values of $N$, the path integral can be determined using a Monte Carlo method. A set of possible $\left\{x_{i}'s\right\}$ from $i=1, ..., N$ is a configuration. The exponent of the action in the path integral is analogous to the Boltzmann factor in statistical mechanics and, thus, is the weight for generating a specific configuration. To maximize the efficiency of the method, we wish to generate configurations weighted by $e^{-S}$. This process is known as importance sampling. 

A method that uses importance sampling is the Metropolis method. This method begins with an initial configuration and then slightly perturbs each $x_{i}$ of that configuration by a small, random number. This gives a small change in the action, $\Delta S$. After the perturbation, if $\Delta S < 0$ then the change to the action is accepted, otherwise another uniformly distributed random number is generated and the procedure is repeated. Each iteration of this method is known as a sweep. To insure statistical independence many sweeps occur between accepted configurations. Performing the Monte Carlo method iterations to obtain independent field configurations is called thermalization. 

A set of configurations is an ensemble. Calculations on the lattice can then be performed using the ensemble of the configurations. For the one-dimensional particle in a potential well we can calculate, for example, the quantized energy levels of the particle can be determined.  

\chapter{Twisted Mass QCD}\label{chapt:tmQCD}
As discussed in previous chapters, Wilson fermions are a good solution to the fermion doubling problem but introduce zero quark modes which correspond to massless quark flavors that produce large, unphysical statistical fluctuations in the quenched approximation. A solution was proposed by Frezzotti $et al.$ in 2001 that removes the exceptional configurations while retaining the original Wilson symmetries ~\cite{frezzotti-2001-0108}. It is called twisted mass QCD (tmQCD).

\section{Introduction to Twisted Mass}

A conceptual problem arises for Wilson fermions in the quenched approximation. As we know from field theory, the fermionic determinant contains information about the vacuum polarization loops. The quenched approximation neglects the vacuum loops and thus the fermionic determinant. When the determinant is removed, exceptional gauge field configurations occur, resulting in large statistical fluctuations leading to a corrupt ensemble average ~\cite{Luscher:1996ug}. There have been several regularization of the Wilson action schemes proposed to solve this ``exceptional problem" ~\cite{bardeen-1998-57,hoferichter-1998-63,gockeler-1999-73}. However, this problem is common to all lattice regularizations using Wilson fermions.  

One solution to the ``exceptional problem" is to add a non-standard mass term to the Wilson quark action. The lattice Dirac operator is then
\begin{equation}
D_{tmQCD} = D_{W} + m_{q} + i\mu_{q}\gamma_{5}\tau^{3},
\end{equation}
where $D_{W}$ is the massless Wilson Dirac operator, $m_{q}$ is the bare quark mass, $\mu_{q}$ is the twisted mass parameter, and $\tau^{3}$ is the third component of the Pauli matrix acting in isospin space. The lattice tmQCD action is then,
\begin{equation}\label{ltmQCD}
S_{F} = a^{4}\sum_{x} \bar \psi(x) (D_{W} + m_{q} + i\mu_{q}\gamma_{5}\tau^{3})\psi(x).
\end{equation}

The tmQCD term in (\ref{ltmQCD}) generalizes the Wilson fermion action by introducing a chiral phase between the mass and Wilson term ~\cite{abdel-rehim-2005-71}. As stated above, the twisted term protects the tmQCD action from zero quark modes.  The protection that the twisted mass action offers can be seen explicitly by manipulation of the determinant of $D_{tmQCD}$
\begin{eqnarray}\label{det}
0 &<& Det[D_{W} + m_{q} + i\mu_{q}\gamma_{5}\tau^{3}] \nonumber \\
  &=& det[(D_{W} + m_{q})^{\dagger}(D_{W} + m_{q}) + \mu^{2}_{q}],
\end{eqnarray}
where $Det$ is the determinant in two-flavor space and $det$ is the determinant in one-flavor space ~\cite{frezzotti-2002-,Aoki:1989rw}. If the twisted mass term is non-zero, the determinant in \ref{det} can not be zero thus avoiding zero quark modes. Numerical evidence is provided in reference ~\cite{Schierholz:1998bq}.

The twisted mass parameter couples to terms in flavor space and protects the Dirac operator from zero quark modes ~\cite{frezzotti-2001-0108}. Two distinct twisted mass flavors are generated from this Dirac operator corresponding to the elements of $\tau^{3}$. The twisted mass term associated with $+1$ is the ``up" flavor. Likewise, the term associated with $-1$ is the ``down" flavor. To avoid confusion with the up and down quark flavors the twisted flavors will be denoted ``tmU" and ``tmD" for clarity. 

\section{Classical Continuum Theory}

The continuum twisted mass QCD action is,
\begin{equation}\label{tmQCD}
S_{F}[\bar\psi(x) \psi(x) ] = -\int d^{4}x\bar\psi ( \slash D + m + i\mu_{q}\gamma_{5}\tau^{3} ) \psi.
\end{equation}  

The axial ($\gamma_{5}$) transformation of the fermion fields is
\begin{equation} \label{axtrans}
\psi^{\prime} = e^{i\alpha \gamma_{5} \tau^{3}/2)}\psi , \quad \bar\psi^{\prime} = \bar\psi e^{i\alpha \gamma_{5} \tau^{3}/2},
\end{equation}
which leaves the twisted action invariant ~\cite{frezzotti-2001-0108} and transforms the mass parameters 
\begin{eqnarray}\label{masses}
m^{\prime}   &=& mcos(\alpha ) + \mu_{q}sin(\alpha ), \\
\mu^{\prime} &=& -msin(\alpha ) + \mu_{q}cos(\alpha ).
\end{eqnarray}
where one defines the rotation angle of the transformation by
\begin{equation}\label{tan}
tan(\alpha ) = \frac{\mu_{q}}{m}.
\end{equation}
Notice with this definition of the twist angle the standard action is obtained when $\mu^{\prime}_{q} = 0$. 

The chiral symmetry of the massless action defines $V^{a}_{\mu}$ and $A^{a}_{\mu}$ to be
\begin{eqnarray}
A^{a}_{\mu} &=& \bar\psi\gamma_{\mu}\gamma_{5}\frac{\tau^{a}}{2}\psi , \\
V^{a}_{\mu} &=& \bar\psi\gamma_{\mu}\frac{\tau^{a}}{2}\psi .
\end{eqnarray}

It is important that the usual symmetries continue to hold in this formalism. At non-zero quark mass, the partially conserved vector and axial relations (PCVC and PCAC) take the form 

\begin{eqnarray}
\partial_{\mu}A^{a}_{\mu} &=& 2mP^{a} + i\mu_{q}\delta^{3a}S^{0}, \\
\partial_{\mu}V^{a}_{\mu} &=& -2\mu_{q}\epsilon^{3ab}P^{b}, 
\end{eqnarray}
where the pseudo-scalar and scalar densities are defined to be 
\begin{equation}
P^{a} = \bar \psi \gamma_{5} \frac{\tau^{2}}{2}\psi , \space S^{0} = \bar \psi \psi.
\end{equation}

The transformation of the quark and anti-quark to the primed basis results in a transformation of the usual Wilson operators. Useful examples of this transformation are seen in ~\cite{frezzotti-2001-0108}. The axial and vector currents in the primed basis that utilize fields from (\ref{axtrans}) are
\begin{eqnarray}\label{axmix}
A^{\prime a}_{\mu} \equiv \bar \psi \gamma_{\mu}\gamma_{5} \frac{\tau^{a}}{2}\psi^{\prime} &=& cos(\alpha )A^{a}_{\mu} + \epsilon^{3ab}sin(\alpha )V^{b}_{\mu} \\
V^{\prime a}_{\mu} \equiv \bar \psi \gamma_{\mu} \frac{\tau^{a}}{2}\psi^{\prime} &=& cos(\alpha )V^{a}_{\mu} + \epsilon^{3ab}sin(\alpha )A^{b}_{\mu},
\end{eqnarray}
for $a = 1,2$. When $a=3$ these currents have the form
\begin{eqnarray}\label{axmix3}
A^{\prime a}_{\mu} \equiv \bar \psi \gamma_{\mu}\gamma_{5} \frac{\tau^{a}}{2}\psi^{\prime} &=& A^{3}_{\mu} \\
V^{\prime a}_{\mu} \equiv \bar \psi \gamma_{\mu} \frac{\tau^{a}}{2}\psi^{\prime} &=& V^{3}_{\mu}.
\end{eqnarray}
Similarly, the pseudo-scalar and scalar operators in the primed basis are 
\begin{eqnarray}
P^{\prime a} &=& P^{a}, \space (a=1,2) \\
P^{\prime 0} &=& cos(\alpha) P^{3} + \frac{i}{2}sin(\alpha )S^{0}, \space (a = 3) \\
S^{\prime 0} &=& cos(\alpha) S^{0} + 2isin(\alpha )P^{3}, \space (a=1,2,3).
\end{eqnarray}
It is important to notice that in general there is mixing between the axial and vector currents as well as the pseudo-scalar and scalar densities. Using the rotated masses defined in (\ref{masses}) it can be shown that PCAC and PCVC relations take their usual form in the primed basis,
\begin{eqnarray}
\partial_{\mu}A^{\prime a}_{\mu} &=& 2m^{\prime}P^{\prime a} \\
\partial_{\mu}V^{\prime a}_{\mu} &=& 0, 
\end{eqnarray}
with the requirement that the rotation angle is defined as in \ref{tan}.

\section{Symmetries of the Bare Theory and Renormalizability}
%The twisted mass lattice action is ~\cite{frezzotti-2001-0108}
%\begin{equation}\label{ltmQCD}
%S_{F} = a^{4}\sum_{x} \bar \psi(x) (D_{W} + m_{q} + i\mu_{q}\gamma_{5}\tau^{3})\psi(x).
%\end{equation}
The massless Wilson Dirac operator in equation (\ref{ltmQCD}) is
\begin{equation}\label{mlessDW}
D_{W} = \frac{1}{2}\sum^{3}_{\mu=0}(\gamma_{\mu}(\nabla_{\mu} + \nabla^{*}_{\mu}) - a\nabla^{*}_{\mu}\nabla_{\mu}). 
\end{equation}

The massless Wilson Dirac operator is not invariant under a left multiplication of the axial rotation in (\ref{axtrans}) and therefore the Dirac operators are different when $\mu_{q} \not=0$ and $\mu_{q}=0$. This is a welcomed consequence because the twisted mass term in the axial rotation protects the action from zero quark modes. If this were not the case, the tmQCD theory would still suffer from ``exceptional configurations". 

It has been shown that in tmQCD there is a $U(1)$ flavor symmetry that leads to conservation of fermion number. A vectorial $U(1)$ isospin symmetry also exists which is generated by $\frac{\tau^{3}}{2}$. 

The twisted mass lattice action is invariant under axis permutations. However, reflection symmetries, such as parity, are a good symmetry only in combination with a flavor exchange between ``tmU" and ``tmD" 

\begin{equation}
\bar \psi \rightarrow \bar \psi \tau^{1}, \space \psi \rightarrow \tau^{1}\psi ,
\end{equation}
which is the equivalent to changing the sign of the twisted mass parameter $ \mu_{q} \rightarrow -  \mu_{q} $. This is a $P \times  \tau_{1,2}$ symmetry of the twisted action.

Lattice symmetries and power counting prove that the tmQCD model is renormalizable ~\cite{frezzotti-2003-}. The $P \times  \tau_{1,2}$ symmetry rules out odd parity, pure gauge terms proportional to $tr[F \tilde F]$ as $a \rightarrow 0$ to contribute to the action ~\cite{frezzotti-2004-}. While the coupling constant $g^{2}_{o}$ and the twisted mass parameter $\mu_{q}$ only require a multiplicative renormalization, the bare quark mass $m$ needs an additive and a multiplicative renormalization.

The relationship between the bare and renormalized action parameters are 
\begin{eqnarray}
 g^{2}_{R} &=& Z_{g}(g^{2}_{o},am_{q},a\mu_{q};a\mu )g^{2}_{o}, \\
 \mu_{R} &=& Z_{\mu}(g^{2}_{o},am_{q},a\mu_{q};a\mu )\mu_{q}, \\
 m_{R} &=& Z_{m}(g^{2}_{o},am_{q},a\mu_{q};a\mu )m_{q},
\end{eqnarray}
where the $Z's$ are the renormalization factors. The renormalization factors can be written in a mass-independent scheme and can be chosen to be independent of $am_{q}$ and $a\mu_{q}$ ~\cite{frezzotti-2002-}. The mass-independent renormalization parameters are obtained by renormalizing in the chiral limit ~\cite{Weinberg:1951ss,frezzotti-2002-}.
\begin{eqnarray}\label{massind}
 g^{2}_{R} &=& Z_{g}(g^{2}_{o};a\mu )g^{2}_{o}, \\
 \mu_{R} &=& Z_{\mu}(g^{2}_{o};a\mu )\mu_{q}, \\
 m_{R} &=& Z_{m}(g^{2}_{o};a\mu )m_{q}.
\end{eqnarray}

Assuming that the massless Dirac operator in (\ref{mlessDW}) is of $O(a)$, then the $O(a)$ improved bare parameters of the action are
\begin{eqnarray}
g^{2}_{o} &\rightarrow & g^{2}_{o}(1 + b_{g}am_{q}), \\
m_{q}     &\rightarrow & m_{q} + b_{m}am^{2}_{q} + \tilde b_{m}a\mu^{2}_{q}, \\
\mu_{q} &\rightarrow & \mu_{q}(1 + b_{\mu}am_{q}), 
\end{eqnarray}
where $m_{q}$ is the difference between the bare mass and the critical mass, $m_{q} = m_{o} - m_{critical}$. The improvement coefficients for the renormalization $b_{\mu}, \space \tilde b_{m}, \space b_{m}, \space b_{g}$ are determined by perturbation theory as well as the $PCVC$ and $PCAC$ relations for tmQCD ~\cite{frezzotti-2002-106}.  

\section{TmQCD at Maximal Twist}

Recall that the twist angle defined by the field transformation is defined to be
\begin{equation}
tan(\alpha) = \frac{\mu_{q}}{m}.
\end{equation}
Two interesting choices of the twist angle are $\alpha=0$ and $\alpha=\frac{\pi}{2}$. Assignment of a zero twist angle returns the standard Wilson lattice action. Choosing a twist angle $\alpha=\frac{\pi}{2}$ causes the mass, $m$, to vanish and is referred to as a maximal twist value. 

As seen in \ref{axmix} a generic rotation by $\alpha$ mixes the axial and vector currents. However, when we choose the maximal twist value, there is no mixing but the role of the vector and axial currents are exchanged. 

There are many possible definitions of the maximal twist value. One possibility is the Wilson definition of maximal twist. The twist parameter is determined by the standard Wilson action when $\alpha=0$. The pseudoscalar meson (pion) is calculated as a function of the critical mass (hopping parameter $\kappa_{c}$) and then extrapolated to vanishing pion mass. The critical mass parameter is ~\cite{abdel-rehim-2005-71}
\begin{equation}\label{Wilsonmaxtwist}
am_{c} = \frac{1}{2\kappa_{c}} - 4.
\end{equation}
The Wilson definition of maximal twist has been used in previous calculations ~\cite{Jansen:2003ir,Abdel-Rehim:2004gx}. 

The tmQCD action expressed in terms of the twisted fields (\ref{axtrans}) has a parity violating mass term. This mass term may be removed by a field redefinition where the parity violation is now associated with the Wilson term. The resulting action is said to be in the physical basis ~\cite{Frezzotti:2003ni}. The parity conservation definition of the twist angle is found by enforcing the physical property that there should be no mixing of the charged psuedoscalar and vector current in the physical basis ~\cite{Farchioni:2004fs,Sharpe:2004ny,abdel-rehim-2005-71},
\begin{equation}
\sum_{x}<V^{-}_{\nu}(\vec x,t)P^{+}(0)> = 0,
\end{equation}
where the charged pseudoscalar is
\begin{equation}
P^{+}(x) = \bar d(x)\gamma_{5} u(x).
\end{equation}
Employing the vector transformation in (\ref{axmix}) and with the understanding that the charged pseudoscalar is invariant under (\ref{axtrans}) we can write the parity definition of maximal twist as
\begin{equation}
tan(\alpha) = \frac{i\sum_{\vec x}<\tilde V^{-}_{\nu}(\vec x,t)P^{+}(0)>}{\sum_{\vec x}<\tilde A^{-}_{\nu}(\vec x,t)P^{+}(0)>},
\end{equation}
where again the currents with a tilde are constructed in the twisted basis. 

In reference ~\cite{abdel-rehim-2005-71} a comparative numerical study between the Wilson and parity definitions of maximal twist was performed. Their study showed that there are no significant lattice spacing effects on the nucleon or vector meson masses for either definition of maximal twist. However, the pion decay constant was found to be independent of lattice spacing for the parity maximal twist while the Wilson was not. 

For a fixed value of the twisted mass parameter the parity maximal twist yielded smaller pion masses than the Wilson definition. It is desired that the square of the pion mass be minimized at maximal twist. The present results imply that the parity conserving definition of maximal twist is better for this observable. For this reason, the set of $(\kappa, \mu_{q})$ pairs found in ~\cite{abdel-rehim-2005-71} will be used in this thesis. 

%The parity conservation maximal twist parameters from reference ~\cite{abdel-rehim-2005-71} are listed in the table below,
%
%\begin{table}[h!]
%\begin{center}
%\begin{tabular} {ccccccc}
%$\beta$   & &    lattice size    & &    am         & &    a$\mu$ \\
%6.0       & &    $20^{3}x32$     & &   -0.8110     & &    0.030 \\
%          & &                    & &   -0.8170     & &    0.015 \\
%          & &                    & &   -0.8195     & &    0.010 \\
%          & &                    & &   -0.8210     & &    0.005 \\
%\end{tabular}
%\caption[TMP]{Twisted Mass Parameters for Parity Conservation at Maximal Twist}
%\label{TMP}
%\end{center}
%\end{table}
%
%The associated $\kappa$ value for each $a\mu$ in table \ref{TMP} is determined using equation \ref{Wilsonmaxtwist}. 

\section{Continuum and Chiral Limit}

In tmQCD the lattice cut-off effects resulting from the chiral violating twisted mass term may change dramatically as a function of the quark mass. This fact is important when chiral symmetry is spontaneously broken. During spontaneous symmetry breaking, the chiral phase of the vacuum state in the continuum theory is driven by the quark mass term. This is also true in the lattice formalism, therefore the continuum limit is taken before the twisted mass $\mu_{q} \rightarrow 0$ ~\cite{frezzotti-2004-}. 

Even with the advancements in computational technologies, lattice techniques are not able to compute physical quark masses. Therefore, in the continuum limit, a lattice chiral perturbation method is used to reach physical results. Lattice chiral perturbation theory (ChPT) is an expansion in powers of the quark mass and the lattice spacing parameter that provides estimates of physical observalables in terms of a few low energy constants ~\cite{Bar:2004xp}. When ChPT is applied in the tmQCD ~\cite{Munster:2004dj,Munster:2004wt}, ChPT involves the renormalized quark mass $m_{R}$ and the rescaled twist angle 
\begin{equation}
\alpha_{R} = tan^{-1}[Z tan( \alpha )],
\end{equation}
where $Z$ is a renormalization constants of the operators $\bar \psi^{\prime}\gamma_{5}\tau_{a} \psi^{\prime}$ and $\bar \psi^{\prime}\psi^{\prime}$ in the mass independent scheme described in equation (\ref{massind}) in reference ~\cite{Frezzotti:2001ea}. 

$O(a)$ cutoff effects of the pion mass and the pion decay constants are automatically absent when the  twist angle is $90^{o}$. However, there are lattice artifacts of $O(\frac{a^{2}}{m_{R}})$ that remain from the chiral Lagrangian density in the pion mass ~\cite{Scorzato:2004da}.

\chapter{Lattice Techniques}
In this chapter a brief review of lattice strategies to extract information from lattice calculations is presented. The purpose of this chapter will be to present a review of two-point Green function source techniques and correlation functions. We will also discuss the strange matrix elements of the nucleon.   

\section{Grassmann Integration}
Grassmann integration is a useful tool to evaluate fermionic integrals in two and three point functions. A brief summary of the properties for Grassmann variables is presented here. Let the Grassmann variables and it's conjugate by $\zeta$ and $\zeta^{*}$. If these are to be Grassmann variables they must obey the anti-commutation relations
\begin{equation}
[\zeta_{i}, \zeta_{j} ]_{+} \space = \space [\zeta^{*}_{i}, \zeta_{j}]_{+} \space = \space [ \zeta^{*}_{i}, \zeta^{*}_{j}]_{+} \space = \space 0.
\end{equation}
%The inverse operation of a Grassmann variable is 
%\begin{equation}
%(\zeta^{*})^{*} \space = \space -\zeta.
%\end{equation}

Integration over Grassmann variables can be defined as
\begin{eqnarray}\label{Grassint}
\int d\zeta \space &=& \space \int d\zeta^{*} \space = \space 0, \nonumber \\
\int d\zeta \zeta  &=& \int d\zeta^{*} \zeta^{*} \space = \space 1.                   
\end{eqnarray}
From equation \ref{Grassint} we can deduce the property
\begin{equation}
\int \Pi _{m}d\zeta^{*}_{m} d\zeta_{m}exp[-\sum_{ij}\zeta^{*}_{i}M_{ij}\zeta_{j}] \space = \space det(M).
\end{equation}
This integral differs from the corresponding integral over commuting variables by resulting in the $det(M)$ rather than $det(M)^{-1}$. 

Suppose now that the Grassmann variables represent a quark field. Then, for example, using Wick contractions between quark and anti-quark fields then the integral in (\ref{Grassex}) results in a quark propagator 
\begin{equation}\label{Grassex}
\int d\bar \zeta d\zeta \zeta_{\alpha} \bar \zeta_{\beta}e^{-\bar \zeta M \zeta} \space = \space det(M)S_{\alpha \beta}.
\end{equation}
We set $det(M) \space = \space 1$ for these types of integrals in the quenched approximation. A similar expression can be determined for tmQCD. 
In chapter \ref{chapt:tmQCD}, the field transformations at maximal twist was expressed as 
\begin{equation} \label{fields}
\psi_{tm} = \frac{1}{\sqrt{2}}(1 \pm  i\gamma_{5})\psi \space , \space \bar\psi_{tm} = \frac{1}{\sqrt{2}}\bar\psi (1 \pm i\gamma_{5}),
\end{equation}
where the $+$ and $-$ represents ``tmU" and ``tmD", respectively.

We are interested in how Grassmann integration behaves using twisted fields. Our example from equation (\ref{Grassex}) using maximally twisted fields can be expressed as
\begin{equation}
\int d\bar \zeta d\zeta (1 \pm i\gamma_{5})\zeta_{\alpha} \bar \zeta_{\beta}(1 \pm i\gamma_{5}) e^{-\bar \zeta M^{\prime} \zeta} \space = \space det(M^{\prime})(1 \pm i\gamma_{5})S_{\alpha \beta}(1 \pm i\gamma_{5}).
\end{equation}
with $M^{\prime} \space = \space (1 \pm i\gamma_{5})M(1 \pm i\gamma_{5})$ and  
%Now the integral is
%\begin{equation}
%\int d\bar \zeta d\zeta (1 \pm i\gamma_{5})\zeta_{\alpha} \bar \zeta_{\beta}(1 \pm i\gamma_{5}) e^{-\bar \zeta M^{\prime} \zeta} \space = \space (1 \pm i\gamma_{5})S_{\alpha \beta}(1 \pm i\gamma_{5}),
%\end{equation}
where the propagator, $S_{\alpha \beta}$, is in the physical basis. Again, let $det(M^{\prime}) \space = \space 1$. This instructive, simple example shows how to create quark propagators in the twisted basis and return to the physical basis by twisting the ends of the propagator. This strategy was employed to determine hadron masses in reference ~\cite{Abdel-Rehim:2005qv}. 

\section{Green Function Methods for Proton/Neutron}

In this section we will review the proton two and three point function method presented in reference ~\cite{Wilcox:1991cq} as well as the twisted mass representation. The twisted interpolation fields used for the proton two point function are
\begin{eqnarray}\label{interpolate}
\chi_{\alpha}(x)_{tm} &=& \epsilon^{abc} \psi^{(u)a}_{\alpha}(x)_{tm}\psi^{(u)b}_{\beta}(x)_{tm}(\tilde C)_{\beta \gamma}\psi^{(d)c}_{\gamma}(x)_{tm},\\
\bar \chi_{\alpha^{\prime}}(x)_{tm} &=& -\epsilon^{a^{\prime}b^{\prime}c^{\prime}} \bar \psi^{(d)c^{\prime}}_{\gamma^{\prime}}(x)_{tm}(\tilde C)_{\gamma^{\prime} \beta^{\prime}} \bar \psi^{(u)b^{\prime}}_{\beta^{\prime}}(x)_{tm} \bar \psi^{(u)a^{\prime}}_{\alpha^{\prime}}(x)_{tm}. \nonumber
\end{eqnarray}
The Greek and Latin indices represent Dirac and color indices, respectively, in equation (\ref{interpolate}). The interpolation fields for the neutron are given by a $u \rightarrow d$ field exchange. 

The proton two point function for forward time $(t > 0)$ can be written in terms of the interpolation fields as follows:

\begin{eqnarray}\label{G2pt}
G_{pp}(t;\vec p,\Gamma^{\prime}) &\equiv & \sum_{\vec x}e^{-i\vec p \cdot \vec x}\Gamma^{\prime}_{\alpha^{\prime}\alpha}<vac|T (\chi_{\alpha} (x)_{tm} \bar \chi_{\alpha^{\prime}}(0)_{tm})|vac> \\
                        &=& \sum_{\vec x}e^{-i\vec p \cdot \vec x}\Gamma^{\prime}_{\alpha^{\prime}\alpha} \epsilon^{abc} (-\epsilon^{a^{\prime}b^{\prime}c^{\prime}})(\tilde C)_{\beta \gamma}(\tilde C)_{\gamma^{\prime} \beta^{\prime}}\\
                        & &<vac|\psi^{(u)a}_{\alpha}(x)_{tm}\psi^{(u)b}_{\beta}(x)_{tm} \psi^{(d)c}_{\gamma}(x)_{tm}\bar \psi^{(d)c^{\prime}}_{\gamma^{\prime}}(0)_{tm} \bar \psi^{(u)b^{\prime}}_{\beta^{\prime}}(0)_{tm} \psi^{(u)a^{\prime}}_{\alpha^{\prime}}(0)_{tm}|vac>. \nonumber
\end{eqnarray}
The $4 \times 4$ $\Gamma^{\prime}$ matrix determines which correlation function is to be evaluated and is generic until specified. A similar function can be written for the neutron using the correct interpolation fields; however, we will focus on the proton here for clarity. 

We have defined the charge conjugation matrix $C = \gamma_{2}$ and $\tilde C = C\gamma_{5}$, which satisfies the relation $\tilde C \gamma_{\mu} \tilde C^{-1} = \gamma^{*}_{\mu}$. A general transformation can be constructed for a general matrix $Q$ such that $\underline Q \equiv (\tilde C Q \tilde C^{-1})^{T}$.

In Euclidean space the integration formula for the time ordered N-point function is defined to be
\begin{equation}
<vac|T(\psi_{\alpha}(-it_{A})\bar \psi_{\beta}(-it_{B})...)|vac> = Z^{-1} \int dU d\bar \zeta d\zeta e^{-S_{G}-S_{F}[\bar \zeta , \zeta ]}\zeta_{\alpha}(t_{A})\bar \zeta_{\beta}(t_{B})
\end{equation} 
where $S_{G}$ and $S_{F}[\bar \zeta,\zeta]$ are the Euclidean gluonic and fermionic actions respectively. 

%The fermionic Wilson action in Euclidean space is $S_{F} = \sum_{I,J} \bar \zeta_{I} M_{IJ} \zeta_{J}$. 

Using Grassmann integration, we may write the proton two point function in the physical basis as  
\begin{eqnarray}
G_{pp} &=& \sum_{\vec x} e^{-i\vec p \cdot \vec x} \epsilon^{abc} \epsilon^{a^{\prime}b^{\prime}c^{\prime}}(tr[\Gamma^{\prime} \frac{(1+i\gamma_{5})}{\sqrt{2}}S^{(u)aa^{\prime}}(x,0) \frac{(1+i\gamma_{5})}{\sqrt{2}}\nonumber \\
       &\times &  \frac{(1-i\gamma_{5})}{\sqrt{2}}\underline S^{(d)bb^{\prime}}(x,0) \frac{(1-i\gamma_{5})}{\sqrt{2}} \frac{(1+i\gamma_{5})}{\sqrt{2}}S^{(u)cc^{\prime}}(x,0) \frac{(1+i\gamma_{5})}{\sqrt{2}}] \nonumber \\
           &+& tr[\Gamma^{\prime} \frac{(1+i\gamma_{5})}{\sqrt{2}}S^{(u)aa^{\prime}}(x,0) \frac{(1+i\gamma_{5})}{\sqrt{2}}]tr[ \frac{(1-i\gamma_{5})}{\sqrt{2}}\underline  S^{(d)bb^{\prime}}(x,0) \frac{(1-i\gamma_{5})}{\sqrt{2}}\nonumber \\
           &\times & \frac{(1+i\gamma_{5})}{\sqrt{2}}S^{(u)cc^{\prime}}(x,0) \frac{(1+i\gamma_{5})}{\sqrt{2}}]),
\end{eqnarray}
where a configuration average is understood and the trace is only over Dirac indices.
Using the property of traces we can rearrange the multiplications such that
\begin{eqnarray}
G_{pp} &=& \sum_{\vec x} e^{-i\vec p \cdot \vec x} \epsilon^{abc} \epsilon^{a^{\prime}b^{\prime}c^{\prime}}(tr[ \frac{(1+i\gamma_{5})}{\sqrt{2}}\Gamma^{\prime} \frac{(1+i\gamma_{5})}{\sqrt{2}}S^{(u)aa^{\prime}}(x,0)\underline S^{(d)bb^{\prime}}(x,0)S^{(u)cc^{\prime}}(x,0)] \nonumber \\
           &+& tr[\frac{(1+i\gamma_{5})}{\sqrt{2}}\Gamma^{\prime} \frac{(1+i\gamma_{5})}{\sqrt{2}}S^{(u)aa^{\prime}}(x,0)]tr[\underline  S^{(d)bb^{\prime}}(x,0)S^{(u)cc^{\prime}}(x,0)]).
\end{eqnarray}
If we define a new gamma matrix, $\Gamma^{tw} \space = \space \frac{1}{2}(1+i\gamma_{5})\Gamma^{\prime} (1+i\gamma_{5})$ it is possible to write the proton two point function as
\begin{eqnarray}\label{Gpptm}
G_{pp} &=& \sum_{\vec x} e^{-i\vec p \cdot \vec x} \epsilon^{abc} \epsilon^{a^{\prime}b^{\prime}c^{\prime}}(tr[\Gamma^{tw} S^{(u)aa^{\prime}}(x,0) \underline S^{(d)bb^{\prime}}(x,0)S^{(u)cc^{\prime}}(x,0)] \nonumber \\
           &+& tr[\Gamma^{tw} S^{(u)aa^{\prime}}(x,0)]tr[\underline  S^{(d)bb^{\prime}}(x,0)S^{(u)cc^{\prime}}(x,0)]).
\end{eqnarray}
The proton two point function presented in the Wilson formalism is ~\cite{Wilcox:1989yi}
\begin{eqnarray}\label{GppW}
G_{pp} &=& \sum_{\vec x} e^{-i\vec p \cdot \vec x} \epsilon^{abc} \epsilon^{a^{\prime}b^{\prime}c^{\prime}}(tr[\Gamma S^{(u)aa^{\prime}}(x,0) \underline S^{(d)bb^{\prime}}(x,0)S^{(u)cc^{\prime}}(x,0)] \nonumber \\
           &+& tr[\Gamma S^{(u)aa^{\prime}}(x,0)]tr[\underline  S^{(d)bb^{\prime}}(x,0)S^{(u)cc^{\prime}}(x,0)].
\end{eqnarray}
The form of the two point function is the same in equations (\ref{Gpptm}) and (\ref{GppW}) if $\Gamma^{tw} \space = \space \Gamma$.
This discussion shows that the same techniques can be employed as in the original Wilson case with the exchange of $\Gamma \space \rightarrow \Gamma^{\prime}$. 

As suggested by ~\cite{Abdel-Rehim:2005qv}, in practice the ends of the propagator are twisted upon creation of the quark propagators so that calculations can be done in the usual way in the physical basis. Since the usual hadronic two-point functions may be used, the rest of this chapter will assume we are doing the calculating in the physical basis. 

\section{Correlation Functions}

Properties of correlation functions are a fundamental concept for analysis of hadron structure ~\cite{Wilcox:1991cq,Draper:1989pi,Woloshyn:1989xw,Wilcox:1989yi,Leinweber:1990dv}. A review of correlation functions is given in this section.

In the large time limit $(t >> 1)$, the proton two point function is 
\begin{equation}\label{ltlimit}
G_{pp}(t;\vec p, \Gamma) \rightarrow N_{v}\sum_{s} e^{-Et}\Gamma_{\alpha^{\prime}\alpha}<vac|\chi_{\alpha}(0)|\vec p,s><s,\vec p|\bar \chi_{\alpha^{\prime}}(0)|vac>,
\end{equation}
where $N_{v}$ is the number of spatial lattice points. In the two-point function we have used the fermionic lattice completeness relation 
\begin{equation}
\sum_{n,\vec p,s}|n,\vec p, s><n,\vec p, s| = I.
\end{equation}
The corresponding continuum completeness relation is
\begin{equation}
\sum_{n,s}\int \frac{d^{3}p}{(2\pi)^{3}} \frac{m}{E}|n,\vec p,s><n, \vec p,s| = I.
\end{equation}
Thus, the correspondence between lattice and continuum states is
\begin{equation}
|n,\vec p, s>_{lattice} \rightarrow (\frac{m}{VE})^{1/2}|n,\vec p,s>_{cont.}
\end{equation}
where the volume of lattice sites is $V =N_{v}a^{3}$. With this relation and the continuum field relation, $\psi_{lat} \rightarrow \frac{1}{\sqrt{2 \kappa}}a^{3/2}\psi_{cont}$ it is possible to determine the matrix elements of the interpolation fields in (\ref{ltlimit}). These are
\begin{eqnarray}
<vac|\chi_{\alpha}(0)|\vec p,s>_{lat} &\rightarrow & \frac{a^{3}}{(2\kappa)^{3/2}}(\frac{m}{N_{v}E})^{1/2}<vac|\chi_{\alpha}(0)|\vec p,s>_{cont} \\
<\vec p,s|\bar \chi_{\alpha}(0)|vac>_{lat} &\rightarrow & \frac{a^{3}}{(2\kappa)^{3/2}}(\frac{m}{N_{v}E})^{1/2}<\vec p,s|\bar \chi_{\alpha}(0)|vac>_{cont}. \\
\end{eqnarray}
The lattice matrix elements are related to the continuum free spinors $u_{\alpha}(\vec p,s)$ and $\bar u_{\alpha^{\prime}}(\vec p,s)$ by
\begin{eqnarray}
<vac|\chi_{\alpha}(0)|\vec p,s>_{lat} &=& Au_{\alpha}(\vec p,s), \\
<\vec p,s|\bar \chi_{\alpha^{\prime}}(0)|vac>_{lat} &=& A^{*}\bar u_{\alpha^{\prime}}(\vec p,s),
\end{eqnarray}
where A is a complex scalar in general. Now we are prepared to determine the large time limit of the proton two-point function as a function of the momentum and $\Gamma$ ~\cite{Wilcox:1991cq}. 
\begin{equation}
G_{pp} \rightarrow \frac{|A|^{2}a^{6}m}{(2\kappa)^{3}E}e^{-Et}tr[\Gamma (\frac{-i\slash{p} + m}{2m})],
\end{equation}
where the usual relation for free spinor fields has been employed,
\begin{equation}
\sum_{s}u(\vec p,s)\bar u(\vec p,s) = \frac {-i \slash p + m}{2m}.
\end{equation}
A similar argument is proposed for the proton three point function. The three point function is constructed with a current insertion between the interpolation fields in equation \ref{G2pt}. The large time limit of the three point function is then
\begin{eqnarray}
G_{pJ_{\mu}p}(t_{2},t_{1};\vec p,\vec p^{\prime},\Gamma ) &\rightarrow &-iN^{2}_{v}\sum_{s,s^{\prime}}e^{-E_{p}(t_{2}-t_{1})}e^{-E_{p^{\prime}}t_{1}} \times \\
 & & \Gamma_{\alpha,\alpha^{\prime}}<vac|\chi_{\alpha}(0)_{tm}|\vec p,s> \times \nonumber \\
 & & <\vec p,s|J_{\mu}(0)|\vec p^{\prime},s^{\prime}><\vec p^{\prime},s^{\prime}|\bar \chi_{\alpha^{\prime}}(0)_{tm}|vac> \nonumber
\end{eqnarray}
where $t_{2}$ is a time after the current insertion and $t_{1}$ is a time before. Pictorially, the two and three-point functions are seen in Figure \ref{fig:23point}. 
$t$ is the final time index and $t^{\prime}$ is the time step at which the current is inserted in this picture. 

{\setlength{\abovecaptionskip}{0mm}
\setlength{\belowcaptionskip}{0mm}
\begin{figure}[h!]
\centering
\includegraphics[scale=1, height=7cm, width=14cm]{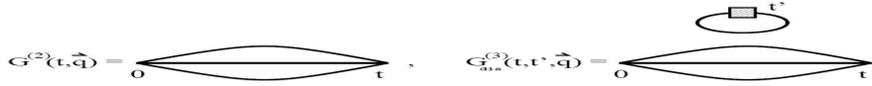}
\caption{Two and Three Point correlators. The solid lines represent quark propagators and the shaded box is a current insertion.}
\label{fig:23point}
\end{figure}}

The lattice, continuum relation for the current expectation value above is
\begin{equation}
<\vec p,s|J_{\mu}(0)|\vec p^{\prime},s^{\prime}>_{lat} \rightarrow \frac{1}{N_{v}}(\frac{m^{2}}{E_{p}E_{p^{\prime}}})^{1/2}<\vec p,s|J_{\mu}(0)|\vec p^{\prime},s^{\prime}>_{cont},
\end{equation}
where the continuum state is
\begin{equation}
<\vec p,s|J_{\mu}(0)|\vec p^{\prime},s^{\prime}>_{cont} = i\bar u(\vec p,s)(\gamma_{\mu}F_{1} - \sigma_{\mu \nu} \frac{q_{\nu}}{2m}F_{2})u(\vec p,s). 
\end{equation}
$F_{1}$ and $F_{2}$ are real functions and $\sigma_{\mu \nu} \space = \space \frac{1}{2i}[\gamma_{\mu},\gamma_{\nu}]$.

%It is an instructive review to evaluate the two and three point function with different $\Gamma^{\prime}$ matrices. If we define $\Gamma^{\prime} \space = \space \frac{(1 \pm i\gamma_{5})}{\sqrt{2}}\Gamma_{4}\frac{(1 \pm i\gamma_{5})}{\sqrt{2}}$ where   
Given,
\\
\begin{displaymath}\
\Gamma_{4} =
\frac{1}{2}
\left(
\begin{array}{rr}
 I & 0\\
 0 & 0\\
\end{array}
\right) ,
\end{displaymath}
\\
and we choose the zero momentum $(\vec p =0)$ charge density $(\mu=0)$ as the current, the proton three-point function becomes
\begin{equation}
G_{pJ_{\mu}p}(t_{2},t_{1};0,-\vec q,\Gamma^{\prime}) \rightarrow Be^{-m(t_{2} - t_{1})}e^{-Et_{1}}(\frac{E+m}{2E})(F_{1} - \frac{q^{2}_{\mu}}{(2m)^{2}}F_{2}).
\end{equation}
Here we identify $G_{e}(q^{2}) \equiv (F_{1} -\frac{q^{2}_{\mu}}{(2m)^{2}}F_{2})$ as the electric form factor of the nucleon.
% In a similar fashion, with a  choice of $\Gamma^{\prime}$ to be $\Gamma^{\prime} \space = \space \frac{(1 \pm i\gamma_{5})}{\sqrt{2}}\Gamma_{i}\frac{(1 \pm i\gamma_{5})}{\sqrt{2}}$ where
Similarly, with
\\
\begin{displaymath}\
\Gamma_{i} =
\frac{1}{2}
\left(
\begin{array}{rr}
 \sigma_{k} & 0\\
 0 & 0\\
\end{array}
\right) ,
\end{displaymath}
\\
the zero momentum, space-like $(\mu=i)$ three point function becomes 
\begin{equation}
G_{pJ_{j}p}(t_{2},t_{1};0,-\vec q,\Gamma^{\prime}_{i}) \rightarrow \frac{B}{2E}e^{-m(t_{2} - t_{1})}e^{-Et_{1}}\epsilon_{jkl}q_{l}(F_{1} + F_{2}).
\end{equation}
For this choice of $\Gamma$ we find the magnetic form factor $G_{m}(q^{2}) \equiv (F_{1}+F_{2})$. 

\section{Strange Matrix Elements}
Once the two and three-point functions are calculated methods are employed to extract the electric, magnetic, and strange matrix elements from the correlators. 

A common technique is to create a ratio of the correlators and then sum over the time insertion index. The ratio itself is
\begin{equation}\label{ratio}
R_{X}(t,t^{\prime},\vec q) \equiv \frac{G^{(3)}_{X}(t,t^{\prime},\vec q)G^{(2)}(t^{\prime},\vec 0)}{G^{(2)}(t,\vec 0)G^{(2)}(t^{\prime},\vec q)},
\end{equation}
where the index $X=\left\{E,M,S\right\}$ are the electric, magnetic, and scalar ratios respectively. The indices $t$ and $t^{\prime}$ are the sink (final time) and current insertion time values ~\cite{lewis-2002-}. The three point function in equation (\ref{ratio}) is constructed from the correlation of the two point function with the loop data when a disconnected part is evaluated. 

Define the Fourier transform of the self contracted disconnected lattice current,
$J(\vec{x},t)$, to be 
\begin{equation}
J^{\prime}(\vec{q},t) = \sum_{\vec{x}}e^{-i\vec{q} \cdot \vec{x}}J(\vec{x},t).
\end{equation}
The disconnected three-point function can then be written generically as ~\cite{Wilcox:2000qa}
\begin{eqnarray}
G^{(3)}(t,t^{\prime},\vec{q}) &=& <G^{(2)}(t,0) \space J^{\prime}(\vec{q},t^{\prime})> \\
                               &-&<G^{(2)}(t,0)>\space <J^{\prime}(\vec{q},t^{\prime})>. \nonumber
\end{eqnarray}

Strange matrix elements are extracted from equation (\ref{ratio}). The extracted matrix elements are related to the form factors by 
\begin{eqnarray}
M_{(E,M,S)} &=& \left\{ G_{S},\frac{\epsilon_{ijk}q_{k}G_{M}}{E_{q}+m},G_{E} \right\}.
\end{eqnarray}
For the magnetic case, $i,j,k$ are indices over the spatial directions. All other indices for the magnetic form factor are suppressed for simplicity. 

There are many ways to extract the matrix elements from the form factors. One way to acquire the matrix element is to sum over the contributions of the inserted strange quark currents ~\cite{Viehoff:1997wi}
\begin{equation}
\sum^{t}_{t^{\prime}=1}R_{X}(t,t^{\prime},\vec q) \rightarrow constant \space + \space tM_{X}(t,\vec q).
\end{equation}
A disadvantage of this method is that it depends on a linear fit of the data, which may only be accurate in a specific temporal region ~\cite{lewis-2002-}. An alternative method employeed by reference ~\cite{Mathur:2000cf} is
\begin{equation}\label{KTway}
\sum^{t_{fixed}}_{t^{\prime}=1}R_{X}(t,t^{\prime},\vec{q}) \rightarrow constant \space + \space tM_{X}(t,\vec{q}),
\end{equation}
where $t_{fixed} \space > \space t$. 

In both of the previous methods a linear temporal fit of the data is require to measure the matrix element. In practice, the fit is restricted to a limited set of time slices. To remove this linear dependance, a differential method can be employed ~\cite{Wilcox:2000qa,lewis-2002-}. Using the form
\begin{equation}
\sum^{t+1}_{t^{\prime}=1}[R_{X}(t,t^{\prime},\vec{q}) \space - \space R_{X}(t-1,t^{\prime},\vec{q})] \rightarrow M_{X}(t,\vec{q}).
\end{equation}
The resulting matrix element is constant over a larger range of time slices and is not subject to a linear fit of the data. This method was employed in the high statistics study of these matrix elements conducted in reference ~\cite{lewis-2002-}. 
\chapter{Linear Equations Solution Techniques}
For either the Wilson or Twisted Mass approach to LQCD, we are faced with solving large, sparse systems of linear equations to determine the respective quark propagators. This chapter focuses on improving iterative methods for solving these systems of linear equations, which often involve multiple right-hand sides and multiple shifts. New Krylov iterative methods to solvie these systems of equations will be presented in this chapter.   

\section{Projection Methods}
\vspace{12pt}
\subsection{Eigenvalue Projections}

There are two general types of projection methods used to evaluate eigenvalue equations. These two are oblique and orthogonal projection methods. In this thesis, we consider only orthogonal projections. Orthogonal projection methods approximate an eigenvector $z$ by a vector $\tilde z$.

Let $M$ be an n $\times$ n complex matrix and $K$ be an $m-dimensional$ subspace of the space $C^{n}$. Our goal is to determine the eigenvalues, $\lambda$, and eigenvectors, $z$, of the eigenvalue equation 
\begin{equation}\label{eeqn}
Mz = \lambda z,
\end{equation}
where $z$ belongs to $C^{n}$ and $\lambda$ belongs to $C$. 

To determine the projection operator we must find the appropriate eigenpair ($\tilde \lambda,\tilde z$) for equation (\ref{eeqn}), with $\lambda$ in $C$ and $\tilde z$ in $K$, such that the Galerkin condition is satisfied. The Galerkin condition is the requirement that the vector $M \tilde z - \tilde\lambda \tilde z$ in $K$ is orthogonal to all other vectors $v \in K$,
\begin{equation}
M \tilde z - \tilde\lambda \tilde z \perp K,
\end{equation} 
which can be written as
\begin{equation}\label{diff}
(M \tilde z - \tilde \lambda \tilde z,v) = 0, \quad \forall v \in K.
\end{equation}
When this condition is true, the approximate eigenvector $\tilde z$ is completely contained in $K$ and therefore is exact. 

Assume that an orthonormal basis $\{v_{1},v_{2}, ... ,v_{m}\}$ of $K$ exists and that the matrix $V$ is constructed with the vectors $v_{1}, v_{2}, ... ,v_{m}$ as columns.

In this chapter, $(a,b)$ denotes an inner product between two vectors $a$ and $b$. 
Let 
\begin{equation}
\tilde z = Vy,
\end{equation} 
so that equation (\ref{diff}) becomes
\begin{equation}
(MVy - \tilde\lambda Vy,v_{j}) = 0, \space j=1, ... ,m.
\end{equation}

If we identify the matrix $B_{m}=V^{\dagger}MV$, $y$ and $\tilde\lambda$ must satisfy
\begin{equation}
B_{m}y = \tilde\lambda y.
\end{equation}
This provides a numerical method to determine approximate eigenvalues and eigenvectors of $M$ using the Galerkin condition in equation (\ref{diff}). This is known as the Rayleigh-Ritz procedure and can be summarized in Table (\ref{RR}).

%This procedure for calcualting the approximate eigenvalues and eigenvectors employing the Galerkin condition of the matrix $M$ is refered to as the Rayleigh-Ritz procedure ~\cite{Saadeval}. The Rayleigh-Ritz procedure can be summarized 

It is possible to reformulate orthogonal projections in an operator language. Consider again the Galerkin condition in (\ref{diff}). Define the projection operator $P_{K}=V^{\dagger}V$. The Galerkin condition becomes

\begin{equation}\label{Gap}
P_{K}(M\tilde z - \tilde \lambda \tilde z) = 0 \space , \space \tilde \lambda \in C \space , \space \tilde z \in K.
\end{equation}
Since the operation of the projection operator on the approximate eigenvector   $\tilde z$ is invariant, the operation of $P_{K}$ on equation (\ref{diff}) can be viewed as a linear transformation from $K$ to $K$ ~\cite{Saadeval}. Another way to write the operator expression of the Galerkin condition is
\begin{equation}
P_{K}MP_{K}\tilde z = \tilde \lambda \tilde z \space , \space \tilde \lambda \in C \space , \space \tilde z \in C^{n}
\end{equation}
which explicitly shows the linear operator $A_{m}=P_{K}AP_{K}$ for the whole space $C^{n}$. If we are restricted to an orthogonal space $K$, this is the matrix $B_{m}$. Equation (\ref{Gap}) is known as the Galerkin approximate problem.

\begin{table}[t!]
\caption[Rayleigh-Ritz Procedure]{Rayleigh-Ritz Procedure}
\vspace{-12pt}
\begin{center}
\begin{tabular}{cl}\hline \\
1. & Compute an orthonormal basis $\{v_{i}\}_{i=1,...,m}$ of the subspace $K$.  \\
    & $Let$ $V=[v_{1},v_{2},...,v_{m}]$ whose columns span $K$. \\
2. & Compute $B_{m} = V^{\dagger}MV;$ \\
3. & Compute the eigenvalues of $B_{m}$ and select the $k$ desired \\
   & $\tilde \lambda$, $i =1,2,...,j$ where $k\leq m$ \\
4. & Compute the eigenvectors $y_{i}$, $i =1,2,...,k,$ of $B_{m}$ associated \\
   & with $\tilde \lambda$, $i =1,2,...,k$ \\
   & and the corresponding approximate eigenvectors of $M$, \\
   & $\tilde z_{i}=Vy_{i}$,$\space i =1,2,...,k,$ \\
   \\
\hline
\vspace{-34pt}
\end{tabular}
\label{RR}
\end{center}
\end{table}

A useful property for estimating the convergence of projection methods for eigenvalue equations is the distance $\parallel (I-P_{K})z\parallel_{2}$ of the exact eigenvector $z$ from the subspace $K$. For this distance we have the inequality ~\cite{Saadeval}
\begin{equation}
\parallel \tilde z - z\parallel_{2} \space \geq \space \parallel (I-P_{K})z\parallel_{2},
\end{equation}
such that a good approximation of the eigenvector $z$ from $K$ results when $\parallel (I-P_{K})z\parallel_{2}$ is small. 
%When this distance is small our approximate eigenvector is approaching the exact value. It can also be interpreted as the sine of the acute angle between the exact eigenvector $z$ and the subspace $K$.

\subsection{Harmonic Rayleigh-Ritz Procedure}

While Rayleigh-Ritz values do a good job of determining approximate eigenvalues (Ritz Values) on the exterior of the eigenvalue spectrum, problems can occur when interior Ritz values are calculated. When a Ritz value is on the exterior of the spectrum, the associated Ritz vector usually has some significance. In contrast, the Ritz vector in the interior may be a combination of many eigenvectors in the subspace giving an interior Ritz value with little meaning ~\cite{Harmonic}. These are known as Spurious Ritz Values (SRV). Spurious Ritz values can have adverse effects on existing Ritz values of significance. When a SRV is near a ``good Ritz value" the corresponding eigenvectors blend together. In this situation, it is necessary to determine the residual norm to distinguish which of the Ritz values is of significance. 

A solution to eliminate the SRV problem is to convert interior Ritz values to exterior Ritz values. A modified Rayleigh-Ritz procedure called the `Interior' or `Harmonic' Rayleigh-Ritz procedure is presented ~\cite{Harmonic,Freund,interior}. The Harmonic Rayleigh-Ritz procedure presents a solution to the SRV problem by shifting the interior values to the exterior of the eigenvalue spectrum. 

Consider the eigenvalue problem 
\begin{equation}
 M z = \lambda z.
\end{equation}
Let $K$ be a j-dimensional subspace of $C^{n}$. It is from this subspace that we wish to extract the approximate eigenvectors. To extract an interior eigenvalue the shifted matrix $(M - \sigma I)^{-1}$ should be used in the Rayleigh-Ritz method. This matrix shifts the eigenvalues to the exterior of the spectrum for this operator. The analysis of the procedure will make use of this operator, but in practice this shifted, inverted matrix is never calculated. Creating this matrix is impractical because of the additional computational cost of finding solutions of linear equations.   

Applying the generalized Rayleigh-Ritz procedure to the shifted interior problem, we find
\begin{equation}\label{genRR}
Q^{\dagger}(M - \sigma I)^{-1}Qd = \frac{1}{\theta - \sigma I}Q^{\dagger}Qd
\end{equation}
where $(\theta,Qd)$ is the approximate eigenpair of the matrix $M$. The matrix $Q$ should span the columns of the subspace $K$. Instead, to avoid having to calculate the inverted, shifted matrix, let $Q=(M - \sigma I)P$. Equation (\ref{genRR}) becomes
\begin{equation}
P^{\dagger}(M - \sigma I)^{\dagger}Pd = \frac{1}{\theta - \sigma}P^{\dagger}(M - \sigma I)^{\dagger}(M - \sigma I)Pd.
\end{equation}

Solving this generalized shifted and inverted Rayleigh-Ritz equation yields the eigenpair $(\frac{1}{\theta-\sigma},Qd)$. This is the corresponding eigenpair for the matrix $M$. However, since we are trying to extract an interior eigenvalue with the shifted, inverted matrix $(M - \sigma I)^{-1}$, a better choice for the approximate eigenpair of $M$ is $(\rho,Pd)$ where $\rho$ is the Rayleigh quotient with respect to $M$. $Pd$ is a better approximation for the interior eigenvector than $Qd$ since we have shifted the problem. Likewise, the Rayleigh quotient $\rho$ is a better approximate eigenvalue of $M$ than $\theta$. 

This analysis has led us to expect that if $z$ is approximately in $K$, then the harmonic Rayleigh-Ritz method will produce a good approximation to $z$ and an associated eigenvalue near $\sigma$. If we let the approximate eigenvalue of the shifted system be $\theta \space = \space \sigma + \delta$, then we may write the harmonic Rayleigh-Ritz equation as
\begin{equation}\label{StRR}
P^{\dagger}(M - \sigma I)^{\dagger}Pd = \frac{1}{\delta}P^{\dagger}(M - \sigma I)^{\dagger}(M - \sigma I)Pd,
\end{equation}
By multiplying by the vector $d^{*}$ and determining the two-norm we find that equation (\ref{StRR}) yields 
\begin{eqnarray}
||(M - \sigma I)Qd||^{2}_{2} &\leq & |\delta | ||(M - \sigma I)Qd||_{2} \\ 
||(M - \sigma I)Qd||_{2}     &\leq & |\delta |. 
\end{eqnarray}
Therefore, if the harmonic Ritz value is within $\delta$ of the shift $\sigma$, the residual norm must be bounded by $|\delta|$ ~\cite{Stewart}. For a harmonic Ritz value close to $\sigma$ and in the limit $\delta \rightarrow 0$, the harmonic Ritz vector cannot be spurious.

\section{Projections for Linear Equations}\label{lineareq}

Projection methods are useful for solving systems of linear equations as well as eigenvalue problems. Most practical iterative methods for solving a large system of equations employ a projection process at some stage of the algorithm. A few good projection techniques that are used are the Galerkin, MinRes, and Left-Right projections. 

\subsection{General Projection Method for Linear Equations}

Consider the linear system of equations
\begin{equation}
M(x - x_{o}) = r_{o},
\end{equation}
where the n$\times$n matrix $M$ is a complex. Projection techniques are designed to extract an approximate solution of the set of linear equations from a subspace of $C^{n}$. Let $K$ be the m-dimensional search subspace. There must be m constraint equations to extract a solution from the subspace $K$. The usual way to determine the $m$ constraints is to enforce $m$ independent   orthogonality conditions. Specifically, we require the residual vector $r = b - Mx$ to be orthogonal to $m$ linearly independent vectors. This set of $m$ linearly independent vectors defines another subspace $L$ which is referred to as the constraint subspace or left subspace ~\cite{Saad}. This general structure is known as the Petrov-Galerkin conditions. 

Let the column vectors of the matrix $V_{nxm} = [v_{1}, v_{2}, ... , v_{m}]$ form a basis for $K$. Likewise, let the columns of $W_{nxm}=[w_{1},w_{2}, ... ,w_{m}]$ form a basis for $L$. If the approximate solution vector extracted from the search space is 
\begin{equation}
x = x_{o} + Vy,
\end{equation} 
where $x_{o}$ is the initial guess, then the orthogonality condition requires that the system of equations for the solution vector y must be
\begin{equation}
W^{\dagger}MVy = W^{\dagger}r_{o}.
\end{equation}
$r_{o}$ is the residual vector associated with the initial solution vector $x_{o}$. If the assumption is made that the $mxm$ matrix $W^{\dagger}MV$ is non-singular, then the approximate solution vector is
\begin{equation}
\tilde x = x_{o} + V(W^{\dagger}MV)^{-1}W^{\dagger}r_{o}.
\end{equation}
%A useful algorithm in iterative methods is the prototype projection method. 
%The prototype projection method is
The procedure just described is known as the prototype projection method and is summarized in Table (\ref{Prototype}).
%\vspace{12pt}
\begin{table}[t!]
\caption[Prototype Projection Method]{ALGORITHM:: Prototype Projection Method}
\vspace{-12pt}
\begin{center}
\begin{tabular} {cl} \hline \\
1. & Until convergence, Do \\
2. & Select a pair of subspaces $\kappa$ and $L$ \\
3. & Choose bases $V = [v_{1}, v_{2}, ... ,v_{m}]$ and $W = [w_{1},w_{2}, ..., w_{m}]$ for $\kappa$ and $L$ \\
4. & $r := b - Mx$ \\
5. & $y := (W^{\dagger}MV)^{-1}W^{\dagger}r$ \\
6. & $x := x + Vy$ \\
7. & Enddo. \\
\\
\hline
\vspace{-34pt}
\end{tabular}
\label{Prototype}
\end{center}
\end{table}
%\vspace{12pt
The approximate solution vector, $x$, extracted from the prototype projection method is only valid if the matrix $B = W^{\dagger}MV$ is non-singular. 

The matrix $B$ can be singular even when the matrix $M$ is non-singular. If either of the following conditions in Table (\ref{criteria}) hold, then $B$ is non-singular for any bases $V$ and $W$ of $K$ and $L$ and the prototype projection method solution exist ~\cite{Saad}. The conditions that need to be satisfied  are in the Table (\ref{criteria}).
\begin{table}[t!]
\caption[Subspace Criteria]{Existence Criteria of the solution x.}
\vspace{-12pt}
\begin{center}
\begin{tabular} {cl} \hline \\
1. & M is positive definite and the left subspace $L = \kappa$ or \\
2. & M is non-singular and $L = M\kappa$. \\
\\
\hline
\vspace{-34pt}
\end{tabular}
\label{criteria}
\end{center}
\end{table}

Specific projection methods are determined by choosing specific vectors that form a basis for the search and left subspaces $K$ and $L$, respectively. Two common projection methods are the Minimal Residual and Galerkin Projection methods. 
\subsection{Specific Projections for Linear Equations}

The Minimum Residual projection method (MinRes) is created with a specific choice for the spaces $K$ and $L$. For a MinRes projection we choose the left subspace to be $L = MK$. The basis vectors for the left subspace are then $W = MV$. Therefore, the MinRes projection can be written as
\begin{eqnarray}
(MV)^{\dagger}MVy &=&  (MV)^{\dagger}r_{o} \\
                             y &=&  ((MV)^{\dagger}MV)^{-1}(MV)^{\dagger}r_{o} .
\end{eqnarray}
The approximate solution vector is constructed out of the search space as before $x = x_{o} + Vy$. 

The Galerkin projection can be constructed with the choice for the left subspace to be $L=K$. The basis vectors of the left space are $W = V$. The projected system of equations that we wish to solve now is
\begin{eqnarray}
V^{\dagger}MVy &=&  (V)^{\dagger}r_{o} \\
                             y &=&  (V^{\dagger}MV)^{-1}V^{\dagger}r_{o} ,
\end{eqnarray}
where the solution vector is constructed in the same manner as with the MinRes projection. Projection methods are incredibly useful in that they project large problems of dimension-n into smaller, more manageable problems of dimension-m. This property is valuable for many iterative methods discussed in this thesis.  

\section{Orthogonal Matrices}

In many algorithms an orthogonal basis for the search subspace is needed to find a solution to a system of linear equations. A few common methods to create the basis vectors of the search space are standard Gram-Schmidt, Householder reflectors, and Fast Givens Rotations. We will discuss the numerical advantages and disadvantages of these orthonormal rotations in this section. 

\subsection{Gram-Schmidt}

The set of vectors $G = (g_{1}, g_{2}, ...,g_{r})$ is said to be an orthogonal set if the inner product of all the elements of $G$ are zero when $i\not=j$. This same set of vectors is said to be orthonormal if in addition $ \parallel g_{i} \parallel_{2} = 1 $, $\forall i$. A vector that is orthogonal to all the vectors in the set G is said to be the orthogonal complement of G and denoted by $G_{\perp}$. A unique vector $x_{i}$ can be written as the sum of vectors from G and $G_{\perp }$. The Gram-Schmidt process takes any vector $ x_{j}$ ($x_{j} \in G_{\perp}$) and orthogonalizes that vector with respect to all previous vectors $x_{i}$ ($x_{i} \in G$) to form an orthonormal set of bases vectors. The Gram-Schmidt algorithm is given in Table \ref{Gram-Schmidt Procedure}.
\begin{table}[t!]
\caption[Gram-Schmidt Procedure]{ALGORITHM :: Gram-Schmidt Procedure}
\vspace{-12pt}
\begin{center}
\begin{tabular} {cl} \hline \\
1. & Compute  $r_{11} = \parallel x_{1}\parallel _{2}$. If $r_{11} = 0$ Stop, else compute $q_{1}=x_{1}/r_{11} $  \\
2. & $For j = 2, ..,r, Do$ \\
3. &      Compute $r_{ij} = (x_{j},q_{i})$ $for$ $i =1,2,...,j-1$ \\
4. &      $\hat q = x_{j} - \Sigma_{i=1}^{j-1} r_{ij}q_{i}$ \\
5. &      $r_{j,j} = \parallel \hat q \parallel _{2}$ \\
6. &      If  $r_{j,j}=0$ then Stop, else $q_{j} = \hat q/r_{j,j}$ \\
7. & $Enddo$ \\
\\
\hline
\vspace{-34pt}
\end{tabular}
\label{Gram-Schmidt Procedure}
\end{center}
\end{table}

Here $\hat q$ is an orthogonal normalization measure in the context of the convergence of the algorithm.

Notice in steps 4 and 5 of the Gram-Schmidt algorithm that the vectors $\hat{q}$ and $r_{j,j}$ are generated with a QR decomposition. A QR decomposition exists whenever the column vectors form a linearly independent set. The Gram-Schmidt algorithm is a common orthogonalization method, but is known not to be as numerically stable as other algorithms.

\subsection{Householder Matrices}

An alternative approach to the Gram-Schmidt procedure is to use the Householder algorithm. This technique uses Householder reflectors to build an orthogonal matrix. A reflector is a matrix of the form
\begin{equation}
Q = I - 2ww^{T}
\end{equation}
where $w$ is a normalized work vector. A reflection matrix that leaves the first $k-1$ columns unchanged while zeroing the $k^{th}$ column is of the form
\begin{equation}
Q_{k} = I - 2w_{k}w^{T}_{k}.
\end{equation}
These reflectors geometrically represent the reflection of a vector $x_{i}$ into some plane. To obtain an orthogonal set of vectors using Householder reflectors we construct
\begin{equation}
X = Q^{T}R
\end{equation}
where $Q^{T} = (Q_{m-1}...Q_{1})^{T}$ and R is an upper triangular matrix generated from $m-1$ Householder transformations onto $X$. Householder reflectors have the advantage of being more stable than standard Gram-Schmidt but have an additional overhead expense due to the multiplication of the work vector $w$ on to itself to form the Householder reflectors. For large matrices, the additional cost of the Householder matrices can make the overhead of this algorithm large. 

\subsection{Givens Rotations}

A fast method that can be invoked to determine orthogonal matrices is the fast Givens Rotations ~\cite{Golub}. In contrast to Householder Reflectors that eliminate all the elements but the first in a given vector, a Givens Rotation eliminates each element individually. In a parallel computing environment  (such as MPI), both fast Givens Rotations and the Householder algorithm can have a significant speed advantage relative to Gram-Schmidt. The Householder Method requires $O(nlog(n))$ steps and $(n-1)$ square roots using $n(n-1)$ processes while the fast Givens Rotations require $O(n)$ steps to create an orthogonal matrix ~\cite{Kuck}. 
An example of a Givens Rotation matrix is 

\begin{displaymath}
 G_{i} = 
\left(
\begin{array}{rrrrrrr}
1 & ..& 0 & .. & 0 &.. &0\\
: & : & : & : & : &: &: \\
0 & ..& c & .. & s & .. &0 \\
: & : & : & : & : &: &: \\
0 & .. & -s & .. & c &.. &0\\
: & : & : & : & : &: &: \\
0 &.. & 0 & .. & 0& .. &  1\\
\end{array}
\right) ,
\end{displaymath}
\\
where $c$ and $s$ are the Givens cosine and sine, respectively. These trigonometric functions can be determined explicitly. For example, to annihilate the bottom element of a $2x1$ vector we have
\\
\begin{displaymath}
\left(
\begin{array}{rr}
 c & s\\
-s & c\\
\end{array}
\right)^{T}
\left(
\begin{array}{r}
 a\\
 b\\
\end{array}
\right)=
\left(
\begin{array}{rr}
r\\
0\\
\end{array}
\right) ,
\end{displaymath}
\\
which gives the constraints on the cosine and sine. The constraints are $sa + cb = 0$ and $c^{2} + s^{2}=1$, which result in the following algebraic form of the Givens cosine and sine:
\begin{equation}
c = a/ \sqrt[2]{a^{2} + b^{2}}, s = -b/ \sqrt[2]{a^{2} + b^{2}}.
\end{equation}
The factorization is then determined by
\begin{equation} 
Q = G_{1}G_{2}...G_{g},
\end{equation}
where there are $g = (2m + n + 1)/2$ Givens matrices for a generic m $\times$ n matrix M. This method to determine orthogonal matrices is preferred when solving large systems of equations due to the reduction in overhead of the calculation in comparison with the aforementioned algorithms. This method is stable while producing reliable results. 

\section{Krylov Subspace Methods}

A general projection method extracts an approximate solution vector $x_{m}$ from the system of equations
\begin{equation}
Mx=b
\end{equation}
by employing the Petrov-Galerkin condition that requires the residual vector, $r=b-Mx$, to be orthogonal to the left space $L$. A Krylov method is a method in which the test space is a Krylov subspace
\begin{equation}\label{KSr}
K_{m}(M,r_{o}) = span\left\{ r_{o}, Mr_{o}, M^{2}r_{o},...,M^{m-1}r_{o}\right\},
\end{equation}
where $x_{o}$ is the initial guess and $r_{o} = b-Mx_{o}$. This condition is true for all Krylov methods. Krylov methods differ in their choices of the left space $L_{m}$ and by how the problem is preconditioned. It is clear that the approximate solution vectors extracted from the Krylov subspace is of the form
\begin{eqnarray}
M^{-1}b &\approx & x_{m} \nonumber \\
        &=&        x_{o}+q_{m-1}(M)r_{o},
\end{eqnarray}
where $q_{m-1}(M)$ is a polynomial in $M$ generated by $K_{m}(M,r_{o})$. The choice of the left space, which is generated by the constraints used to build these approximations, will have an important effect on the particular iterative method. Two examples of $L_{m}$ choices are for the MinRes projection in which $L_{M}=MK_{m}$ and the Galerkin projection with $L_{m}=K_{m}$.

\section{Arnoldi Method}
%
%In this section we consider the projection methods on Krylov subspaces of the form
%\begin{section}
%K_{m}(M,v) \equiv span{v,Mv,M^{2}v,...,M^{m-1}v},
%\end{section}
%where $v$ are the first vector of the matrix $V_{m}$. 

Arnoldi's method is an orthogonal projection method onto a Krylov subspace of $dim(m)$ for general non-Hermitian matrices. The Arnoldi procedure can be used both to compute eigenvalues and to solve systems of linear equations. The Arnoldi procedure to build an orthogonal basis is listed in Table \ref{Arnoldi}.
\begin{table}[t!]
\caption[Arnoldi Algorithm]{Arnoldi Algorithm}
\vspace{-12pt}
\begin{center}
\begin{tabular} {cl} \hline \\
1. & Choose a vector $v_{1}$ such that  $\parallel v_{1}\parallel_2$ $=1$   \\
2. & $For j = 1,2, ..,m, Do$ \\
3. & Compute $h_{ij} = (Mv_{ij},v_{i})$ $for$ $i =1,2,...,j$ \\
4. & Compute $w_{ij} := Mv_{j} - \Sigma_{i=1}^{j}h_{ij}v_{i}$ \\
5. & $h_{j+1,j} = \parallel w_{j} \parallel _{2}$ \\
6. & $If h{j+1,j}=0$ $then$ $stop$ \\
7. & $v_{j+1} = w_{j}/h_{j+1,j}$ \\
8. & Enddo \\
\\
\hline
\vspace{-34pt}
\end{tabular}
\label{Arnoldi}
\end{center}
\end{table}
At any step in the algorithm the previous Arnoldi vector, $v_{j}$, is multiplied by the matrix M to form $v_{j+1}$. This vector is orthonormalized against all previous $v_{i}$ vectors with a standard Gram-Schmidt procedure. The set of vectors,$\left\{ v_{1}...v_{m} \right\}$ form an orthonormal basis of the Krylov subspace. Let $V_{m}$ be the n $\times$ m matrix whose columns are $\left\{ v_{1}...v_{m} \right\}$. Let $H_{m}$ be the m $\times$ m upper-Hessenberg matrix formed by the $h_{ij}$ values from the algorithm. Then the Arnoldi iteration gives the recurrence ~\cite{Saad}
\begin{eqnarray}\label{Arnoldi}
 MV_{m} = V_{m} \bar H_{m} + h_{m+1,m}v_{m+1}e_{m}^{T}
        = V_{m+1} \bar H_{m},
\end{eqnarray}
giving,
\begin{equation}
V_{m}^{T}MV_{m} = H_{m}.
\end{equation}
Pictorially we can see the action of M on the basis vectors $V_{m}$ in Figure (\ref{fig:Arnoldi2}).
{\setlength{\abovecaptionskip}{0mm}
\setlength{\belowcaptionskip}{0mm}
\begin{figure}[b!]
\centering
\includegraphics[scale=1, height=5cm, width=13cm]{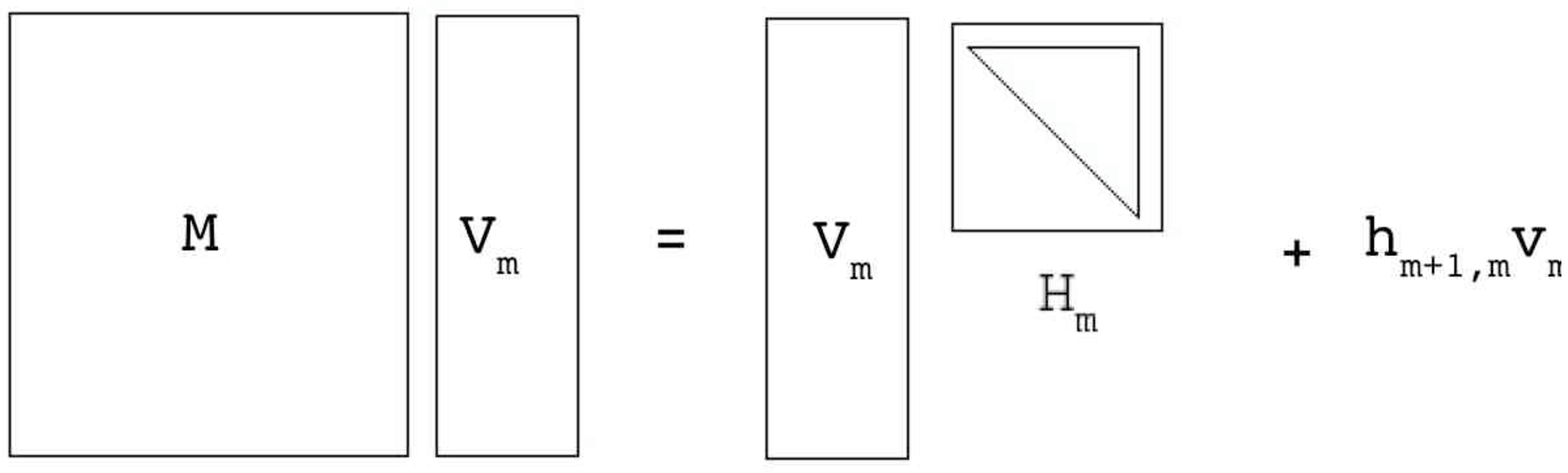}
\caption{The action of M on $V_{m}$ gives $V_{m}H_{m}$ plus a vector}
\label{fig:Arnoldi2}
\end{figure}}
We first consider how the Arnoldi recurrence can be used for eigenvalue computations. Essentially the Arnoldi algorithm combines use of a Krylov subspace with the Rayleigh-Ritz projection.
Arnoldi concluded that the eigenvalues of a Hessenberg matrix smaller than the dimension of the original matrix can provide accurate approximations to some eigenvalues of the original n $\times$ n matrix ~\cite{Saad}. Once these approximate eigenvalues are known, an approximate solution to the original problem can be determined. 

As a result of the projection onto $K_{m}$ we gain the approximate eigenvalues $\lambda_{i}^{(m)}$ of the Hessenberg matrix $H_{m}$ ~\cite{Saadeval}. The approximate eigenvector associated with the the eigenvalue $\lambda_{i}^{(m)}$ is defined to be
\begin{equation}
d_{i}^{(m)} = V_{m}y_{i}^{(m)}.
\end{equation}
Using the eigenvalue equation, the small $i^{th}$ eigenvalue problem is then
\begin{equation}
V^{T}MVd_{i} = \theta_{i}d_{i},
\end{equation}
where $\theta_{i}$ is the $i^{th}$ approximate eigenvalue. The associated Rayleigh-Ritz approximate eigenvector is $y_{i} = Vd_{i}$. The eigenvectors and eigenvalues form Rayleigh-Ritz pairs $(\theta_{i},y_{i})$ where $y_{i}^{(m)}$ is the associated eigenvector of length $m$.

For a moderately sized Krylov subspace, a few of the approximate eigenvalues are usually good approximations to the true eigenvalues of the original matrix $M$. As the dimension of the Krylov subspace increases, the quality of these approximate eigenvalues improves until all of the desired eigenvalues of $M$ are found. Obviously it is not practical to have a large Krylov subspace due to storage and computational cost. However, with a reasonable dimensioned subspace, the Ritz eigenvalues can play an important role in deflated Krylov methods.  
It is important to be able to cost-effectively estimate the residual norm during Krylov method iterations. A cheap way to determine the residual norm makes use of the expression ~\cite{Saadeval},
\begin{equation}\label{stuff}
(M - \lambda_{i}^{(m)} I)u_{i}^{(m)} = h_{m+1,m}e_{m}^{H}y_{i}^{(m)}v_{m+1}.
\end{equation}
The two norm of equation (\ref{stuff}) is
\begin{equation}\label{resnorm}
\parallel (M - \lambda_{i}^{(m)} I)u_{i}^{(m)} \parallel_{2} = h_{m+1,m}\mid e_{m}^{H}y_{i}^{(m)}\mid .
\end{equation}
So, the residual norm is equal to the last component of the eigenvector $y_{i}^{(m)}$ multiplied by $h_{m+1,m}$ ~\cite{Saadeval}. 

When multiple shifts are desired with Krylov methods, the Arnoldi iteration can be modified to handle these shifted systems of equations ~\cite{Harmonic}. 

The new shifted operator $(M - \sigma I)$ gives the eigenvalue equation
\begin{equation}\label{shiftedArnoldi0}
V^{T}(M - \sigma I)^{H}(M - \sigma I)V \tilde d = (\tilde \theta - \sigma I)V^{T}(M - \sigma I)^{H}V \tilde d,
\end{equation}
where $\tilde \theta_{i}$ is a Harmonic Ritz value. The harmonic Rayleigh-Ritz pairs are $(\tilde{\theta_{i}},\tilde{y_{i}})$ where we have used the relation $\tilde y_{i} = V \tilde d_{i}$. When the harmonic Rayleigh-Ritz procedure is applied to the Arnoldi iteration we have the eigenvalue equation ~\cite{Harmonic,Harmonic1,Harmonic2,Harmonic3,Harmonic4} 
\begin{equation}\label{HRitzeq}
(H_{m} + h_{m+1,m}^{2}fe_{m}^{T})\tilde d = \tilde \theta \tilde d,
\end{equation}
where $f = (H_{m} - \sigma I)^{-H}e_{m}$. The problem has now been altered from finding eigenvalues and eigenvectors of $H_{m}$ to finding the eigenpairs of equation (~\ref{HRitzeq}). 

\section{Arnoldi Methods for Linear Equations}
We next consider using the Arnoldi recurrence for solving linear equations. These are ways of applying the projection techniques from section \ref{lineareq} to a Krylov subspace. The choice of the left subspace determines the iterative technique. In the next section we will introduce popular methods that are widely used for different choices of the left subspace $L_{m}$.

\subsection{Full Orthogonalization Method}

We consider an orthogonal projection method for a system of equations $Mx=r_{o}$ which uses the left space $L_{m} \space = \space K_{m} \space = \space K_{m}(M,r_{o})$, with $K_{m}$ defined in equation (\ref{KSr}). This method finds an approximate solution vector $x_{m}$ from the affine subspace $x_{o}+K_{m}(M,r_{o})$ by imposing the Petrov-Galerkin condition
\begin{equation}
b-Mx_{m} \perp K_{m}.
\end{equation}

If the first basis vector of the Krylov subspace in Arnoldi's method is $v_{1} \space = \space r_{o}/||r_{o}||_{2}$, then
\begin{equation}
V^{T}_{m}MV_{m} = H_{m}
\end{equation} 
holds with $\beta=||r_{o}||_{2}$. If we then employ (\ref{Arnoldi}) we may write 
\begin{equation}
V^{T}_{m}r_{o} \space = \space V^{T}_{m}(\beta v_{1}) \space = \space \beta e_{1}.
\end{equation}
This results in the approximate solution vector
\begin{eqnarray}
y_{m} &=& H^{-1}_{m}(\beta e_{1}), \\
x_{m} &=& x_{o} + V_{m}y_{m}. 
\end{eqnarray}

The Arnoldi method for linear equations with a Galerkin projection is referred to as the Full Orthogonalization Method (FOM) ~\cite{Saad}. The FOM algorithm is described in Table \ref{FOM}.
\begin{table}[t!]
\caption[Full Orthogonalization Algorithm]{FOM Algorithm}
\vspace{-12pt}
\begin{center}
\begin{tabular} {cl} \hline \\
1. & Compute the residual vector $r_{o}$, with $\beta := ||r_{o}||_{2}$ and \\
   & $v_{1} := \frac{r_{o}}{\beta}$  \\
2. & Define the m$\times$ m Hessenberg matrix $H_{m} \space = \space h_{i,j=1,m}$ \\
   &  and initialize it to zero.\\
3. & For $j=1,m$, Do\\
4. & $Compute$ $w_{j} := Mv_{j}$ \\
5. & For $i=1,j$, Do\\
6. & $h_{ij} = (w_{j},v_{i})$ \\
7. & $w_{j}  =  w_{j} - h_{i,j}v_{i}$ \\
8. & Enddo \\
9. & Compute $h_{j+1,j}=||w_{j}||_{2}$. If $h_{j+1,j}=0$, set $m \space = \space j$ \\
   & and compute the solution vector. \\
10.& Compute $v_{j+1}=\frac{w_{j}}{h_{j+1,j}}$ \\
11.& Enddo \\
12.& Compute $y_{m}=H^{-1}(\beta e_{1})$ and $x_{m} = x_{o} + V_{m}y_{m}$. \\
\\
\hline
\vspace{-34pt}
\end{tabular}
\label{FOM}
\end{center}
\end{table}

\section{GMRES Methods}
\vspace{12pt}
\subsection{Standard GMRES}
The General Minimal Residual method (GMRES) is the MinRes projection applied to a Krylov subspace. As with FOM, it uses the Arnoldi iteration to generate an orthogonal basis for the Krylov subspace. Since GMRES is a Krylov method, any vector $x$ in the subspace $K_{m}$ can be written as 
\begin{equation}
\tilde{x} = \tilde{x}_{0} + V_{m}y
\end{equation}
where $\tilde{x}_{0}$ is an approximate initial guess and $\tilde{x}$ is the approximate solution to the system of linear equations. A residual vector is a measure of the accuracy of the approximate solution vector for a system of linear equations. The residual vector is defined to be $r = b - M\tilde{x}$ where $b$ is the right-hand side vector. The residual norm is the two-norm of the residual vector. It can be written as
\begin{equation}\label{res1}
||r||_{2} = \parallel b - Mx \parallel_{2} = \parallel b - M(x_{0} + V_{m}y) \parallel_{2}
\end{equation}
Using the definition of the residual vector and Arnoldi iteration we can write
\begin{eqnarray}
r &=& r_{0} - MV_{m}y \nonumber \\
  &=& \beta v_{1} - V_{m+1} \bar H_{m}y \\ 
  &=& V_{m+1}(\beta e_{1} - \bar H_{m}y). \nonumber 
\end{eqnarray}
Recall that in the discussion of orthogonal rotations, $V$ is an orthonormal matrix. The residual norm is then
\begin{eqnarray}
\parallel r \parallel_{2}&=& \parallel b - M(x_{0} + V_{m}y \parallel_{2} \\
                         &=& \parallel \beta e_{1} - \bar H_{m}y \parallel_{2}.
\end{eqnarray}
The solution that GMRES produces, $x$, minimizes the residual norm. This can be found by finding the minimum residual solution with the vector $y_{m}$. This vector is the minimizer of the residual norm in equation (~\ref{res1}). The minimizer
\begin{equation}
y_{m} = min \parallel \beta e_{1} - \bar H_{m}y \parallel_{2}
\end{equation}
is computed by an inexpensive (m+1) $\times$ m least-squares problem. $m$ is small for a practical LQCD application. The Arnoldi iteration used above minimizes the solution vector of the system of linear equations ~\cite{Saad}. This gives the GMRES(m) algorithm listed in Talble \ref{GMRES}.
\begin{table}[t!]
\caption[GMRES(m) Algorithm]{ALGORITHM :: GMRES(m)}
\vspace{-12pt}
\begin{center}
\begin{tabular} {cl} \hline \\
1. & $Compute$  $r_{0} = b - Mx_{0}$, $\beta = \parallel r_{0} \parallel_{2}$ and $v_{1} = r_{0}/\beta$ \\
2. & $For j = 1,2, ..,m, Do$ \\
3. & $Compute$ $w_{j} = Mv_{j}$\\
4. & For $i=1,...,j$,Do\\
5. & $h_{i,j} =(w_{j},v_{i})$ \\
6. & $w_{j} = w_{j} - h_{i,j}v_{i}$ \\
7. & $Enddo$ \\
8. & $h_{j+1,j} = \parallel w_{j} \parallel_{2}$ If $h_{j+1,j}=0$ set $m=j$ and goto step 11. \\
9. & $v_{j+1} = w_{j}/h_{j+1,j}$\\
10.& Enddo \\
11.& Define the $(m+1) x m$ Hessenberg matrix $\bar H_{m} =(h_{ij})_{1 \leq i \leq m+1,1 \leq j \leq m}$ \\
12.& Compute $y_{m}$, the minimizer of $\parallel \beta e_{1} - \bar H_{m}y \parallel_{2}$, and $x_{m} = x_{0} + V_{m}y_{m}$ \\
\\
\hline
\vspace{-34pt}
\end{tabular}
\label{GMRES}
\end{center}
\end{table}
%\vspace*{4.0in}
In the GMRES(m) algorithm, Givens rotations are employed in practice to determine the matrix elements $h_{ij}$. 
\subsection{Restarted GMRES}
In practice, the GMRES algorithm becomes impractical due to growth of memory and computational resources when the dimension of the Krylov subspace becomes large. As the dimension $m$ increases, the computational cost increases at least as $O(m^{2}n)$ per cycle because of the orthogonalization of the elements of $\bar{H}$. The memory cost increases as $O(mn)$. ~\cite{Saad}  A solution to eliminate the high computational cost is to restart the Arnoldi iteration. The restarted GMRES(m) algorithm is listed in Table ~\cite{Saad}
\begin{table}[t!]
\caption[Restarted GMRES(m) Algorithm]{ALGORITHM :: Restarted GMRES(m)}
\vspace{-12pt}
\begin{center}
\begin{tabular} {cl} \hline \\
1. & Compute $r_{0} = b - Mx_{0}$, $\beta = \parallel r_{0} \parallel_{2}$ ,and, $v_{1} = r_{0}/\beta$ \\
2. & Generate the Arnoldi basis and the matrix $\bar H_{m}$ using the Arnoldi algorithm \\
   & starting with $v_{1}$ \\
3. & Compute $y_{m}$, which minimizes $\parallel \beta e_{1} - \bar H_{m} \parallel_{2}$, and $x_{m} = x_{0} + V_{m}y_{m}$ \\
4. & If satisfied then Stop, else set $x_{0} = x_{m}$ and go to Step 1. \\
\\
\hline
\vspace{-34pt}
\end{tabular}
\label{Restarted}
\end{center}
\end{table}
The algorithm has the ability to exit when the desired residual norm is reached for a given subspace size. If the residual norm is not satisfactory, the old Krylov subspace is replaced with a new subspace generated with the restarted initial guess $x_{m}$. The restarted GMRES(m)  method is the basis for many algorithms. One variation of this algorithm that we will consider is a Restarted Deflated GMRES method.

\subsection{Deflated GMRES}
For large, sparse matrices new GMRES techniques are required when the matrix eigenvalue spectrum contains small eigenvalues. For example, the Wilson matrix in LQCD contains small eigenvalues that give rise to exceptional configurations and need to be addressed to give sensible results. Techniques have been developed to solve problems of this nature for LQCD ~\cite{GMRESDR,evalDR}. One method is GMRES with Deflated Restarting. This is referred to as GMRES-DR(m,k) where m is the dimension of the subspace and k is the number deflated eigenvalues for the spectrum. 

\subsection{An Invariant Krylov Subspace}
For GMRES to remain effective, augmentation of the subspace by Rayleigh-Ritz vectors should return a Krylov subspace as well. In this subsection we verify that the Krylov subspace is still a Krylov subspace under the augmentation of approximate eigenvectors ~\cite{GMRESDR}. Since we are using restarted methods, each pass through the Arnoldi iteration (equation~\ref{Arnoldi}) between restarts is referred to as a ``cycle". It was shown by Sorensen ~\cite{Sorensen} that if the implicitly restarted Arnoldi method is restarted with approximate eigenvalues (Ritz vector), the new initial vector is a combination of the desired Ritz vectors that generated these eigenvalues. So the subspace 
\begin{equation}\label{augArnoldi}
Span(y_{1},y_{2},...,y_{k},v_{m+1},Mv_{m+1},M^{2}v_{m+1},...,M^{m-k}v_{m+1})
\end{equation}
is the implicitly restarted Arnoldi space in ~\cite{implicit}. The vector $v_{m+1}$ is the last Arnoldi vector from the previously-generated Arnoldi cycle. This vector is now the starting vector for the newly restarted Arnoldi cycle. It can be shown that equation (~\ref{augArnoldi}) is equivalent to
\begin{equation}
Span(y_{1},y_{2},...,y_{k},My_{i},M^{2}y_{i},...,M^{m-k}y_{i}),  1\leq i \leq k,
\end{equation}
where we have used the Arnoldi iteration from equation (~\ref{Arnoldi}). Equation (~\ref{augArnoldi}) is a Krylov subspace generated by a Ritz vector for each cycle. Similarly, in a restarted GMRES method, let $r_{0}$ be the residual vector from the previous cycle. Then, the subspace is 
\begin{equation}\label{subspace}
Span(r_{0},Mr_{0},M^{2}r_{0},...,M^{m-k-1}r_{0},\tilde y_{1},\tilde y_{2},...,\tilde y_{k}),
\end{equation}
where $\tilde{y_{i}}$ are harmonic Ritz vectors. As shown in ~\cite{implicit,Eiermann}, this subspace is equivalent to a subspace with the Harmonic Ritz vectors at the front of the subspace
\begin{equation}\label{equivspace}
Span(\tilde y_{1},\tilde y_{2},...,\tilde y_{k},M \tilde y_{i},M^{2} \tilde y_{i},...,M^{m-k} \tilde y_{i}),
\end{equation}
for $1 \leq i \leq k$. The span of these vectors is a Krylov subspace including the harmonic Ritz vectors, $\tilde y_{i}$. By the preceding arguments we find that a Krylov subspace is still a Krylov subspace under augmentation of approximate eigenvectors. 

GMRES-DR(m,k) begins with a cycle of GMRES(m) which computes the solution vector and the matrix $\bar{H}_{m}$. When the first cycle is finished, k-Harmonic Ritz vectors have been computed along with the matrix $V_{m}$ using the Arnoldi recurrence. $V_{m}$ is constructed by the vectors that span the subspace in equation (~\ref{subspace}). For the second cycle of GMRES-DR(m,k), as seen in equation (~\ref{equivspace}), the first k-columns of the new matrix $V_{k}$ consist of the orthonormalizied harmonic Ritz vectors. The vector $v_{k+1}$ must be generated by orthogonalizing the residual vector from the first cycle with respect to the columns of $V_{m}$. Now that we have all the vectors needed to use the Arnoldi iteration we can form the rest of the Krylov subspace in (~\ref{equivspace}). The GMRES-DR(m,k) algorithm ~\cite{GMRESDR} is summarized in table (\ref{GMRES-DR(m,k)}).
\begin{table}[t!]
\caption[GMRES-DR(m,k) Algorithm]{ALGORITHM :: GMRES-DR(m,k)}
\vspace{-12pt}
\begin{center}
\begin{tabular} {cl} \hline \\
1. & $\fontshape{it}Start$: Choose $m$, the maximum size of the subspace \\
   &  and the desired number of  \\
   & approximate eigenvectors. Choose an initial guess, $x_{0}$, and \\
   & compute $r_{0} = b - Mx_{0}$  \\
   & The new problem is $M(x - x_{0}) = r_{0}$. Let $v_{1} = r_{0}/ \parallel r_{0} \parallel$ \\
   & and $ \beta = \parallel r_{0} \parallel$ \\
2. & $\fontshape{it}First$ $cycle$: Apply standard GMRES(m): use the Arnoldi \\
   & iteration to generate $v_{m+1}$ and $\bar H_{m}$. Then solve the small min. res.\\
   &  problem $min \parallel c - \bar H_{m}d \parallel$ \\
   & for $d$, where $c = \beta e_{1}$. Form the new solution vector  \\
   & $ x_{m} = x_{0} + V_{m}d.$ Let $\beta = h_{m+1,m}$, $x_{0} = x_{m}$, and $r_{0} = b - Mx_{m}$. \\
   & Compute the smallest k eigenpairs $(\tilde \theta_{i}, \tilde g_{i})$ \\
   & of $H_{m} + \beta^{2}H^{-T}_{m}e_{m}e^{T}_{m}$. \\
3. & $\fontshape{it}Orthonormaliztion$ $of$ $first$ $k$ $vectors$: \\
   & Orthonormalize the Harmonic Ritz vectors, $\tilde g_{i}$ \\
   & and form an $ m \times k$ matrix $P_{k}$. \\
4. & $\fontshape{it}Orthonormaliztion$ $of$ $k+1$ $vectors$: Append a zero entry to each vector  \\
   & in the matrix $P_{k}$ to make them length $m+1$. Then orthonormalize the   \\
   & short residual vector, $c - \bar H_{m}d$ against all the vectors in $P_{k}$ to form $p_{k+1}$. \\
5. & $\fontshape{it} Form$ $portions$ $of$ $new$ $H$ $and$ $new$ $V$ \\
   & $using$ $old$ $H$ $and$ $old$ $V$: Let \\
   & $\bar H^{new}_{k} = P^{T}_{k+1}\bar H_{m}P_{k}$ and $V^{new}_{k+1} = V_{m+1}P_{k+1}$. Then \\
   & let $\bar H_{k} = \bar H^{new}_{k}$ and $V_{k+1} = V^{new}_{k+1}$. \\
6. & $\fontshape{it} Reorthogonaliztion$ $of$ $k+1$ $vector$: Orthonogalize $v_{k+1}$ against the \\
   & earlier columns of the new $V_{k+1}$ matrix. \\
7. & $\fontshape{it} Arnoldi$ $iteration$: Apply the Arnoldi iteration from this point to \\
   & form the remaining columns of $V_{k+1}$ and $\bar H_{m}$. Let $\beta = h_{m+1,m}$. \\
%   \end{tabular}
%\caption[GMRES-DR(m,k)]{GMRES-DR(m,k) Algorithm}
%\label{GMRES-DR(m,k)}
%\end{center}
%\end{table}
%
%\begin{table}[h!]
%\begin{center}
%\begin{tabular} {cl}
8. & $\fontshape{it} Form $ $the$ $approximate$ $solution$: Let $c = V^{T}_{m+1}r_{0}$ and \\
   & solve $min \parallel c - \bar H_{m}d \parallel$ for $d$. Let the new solution vector \\
   & be $x_{m} = x_{0} + V_{m}d$. Compute the residual vector \\
   & $r = b - Mx_{m} = V_{m+1}(c - \bar H_{m}d)$. \\
   & Check $\parallel r \parallel = \parallel c - \bar H_{m}d \parallel$ for convergence, \\
   & and proceed if not satisfied. \\
9. & $\fontshape{it} Eigenvalue$ $computations$: Compute the k smallest eigenpair \\
   & $(\tilde \theta_{i}, \tilde g_{i})$ of $H_{m} + \beta^{2}H^{-T}_{m}e_{m}e^{T}_{m}$. \\
10.& $\fontshape{it} Restart$: Let $x_{0} = x_{m}$ and $r_{0} = r$. Proceed to Step 3. \\
\\
\hline
\end{tabular}
\label{GMRES-DR(m,k)}
\end{center}
\end{table}
It is important to realize that after the first cycle the Arnoldi iteration has changed slightly. The matrix $\bar H_{m}$ is upper Hessenberg except for a full leading $k+1$ by $k+1$ portion  from the augmented eigenvectors.

Computationally, it is reasonable to generate Schur vectors instead of eigenvectors. It is known that for any square matrix $M$, there exists a unitary matrix $Q$ such that
\begin{equation}
Q^{\dagger}MQ = R.
\end{equation}
The Schur decomposition is then
\begin{equation}\label{Schur}
MQ_{k} = Q_{k}R_{k},
\end{equation}
where the matrix $R =R_{1}R_{2}...R_{k}$ is triangular and similar to $M$. The matrix $Q$ is $Q = \{q_{1},q_{2},...,q_{k}\}$. This is known as the Schur decomposition of $M$. Recall that if $R$ is triangular and similar to $M$, then the diagonal elements of $R$ are the eigenvalues of $M$. The columns of $Q_{k}$ are the Schur vectors of $M$, and they will be used as the approximate eigenvectors in this algorithm ~\cite{GMRESDR}. 

A simple example is useful to see how deflation is beneficial to solving a system of linear equations. In this example, we will augment the Krylov subspace with one approximate eigenvector. Let the source vector $b=\beta_{1}z_{1} + \beta_{2}z_{2} + ... + \beta_{n}z_{n}$. After the first cycle of standard GMRES the subspace is
\begin{equation}
K = Span \{z_{1}, r_{0}, Mr_{0}, M^{2}r_{0},...,M^{m}r_{0}\}
\end{equation}
where $r_{0}$ is the new starting vector from the first cycle and $z_{1}$ is an exact eigenvector. The residual vector of this second cycle is generated by this Krylov subspace. The solution vector that is spanned by this space is $x = \gamma z_{1} + q(M)r_{0}$ where $\gamma$ is a free parameter. The residual vector for this cycle is
\begin{eqnarray}
r &=& (\beta_{1}z_{1} + ... +\beta_{n}z_{n}) - M(\gamma z_{1} +q(M)r_{0}) \nonumber \\
  &=& (\beta_{1}z_{1} + ... +\beta_{n}z_{n}) - \lambda_{1} \gamma z_{1} - Mq(M)r_{0} \\
  &=& (\beta_{1} - \lambda_{1} \gamma z_{1}) - Mq(M)r_{0} \nonumber
\end{eqnarray}
Notice that if we choose $\gamma = \beta_{1}/\lambda_{1}$ the $z_{1}$ component does not contribute to the residual vector. By this choice of $\gamma$ the polynomial can ``focus" on the rest of the spectrum. This approach is an alternative method to similar algorithms that use matrix preconditioning built of approximate eigenvectors to speed up the convergence of the residual vector ~\cite{Baglama,Burrage,Ehrel,Kharchenko}. 

\subsection{Lanczos Method}

Krylov subspace methods rely on some form of orthogonalization of the Krylov vectors in order to compute an approximate solution to a system of equations. Another class of Krylov methods are based on a biorthogonalization of a set of basis vectors. These projection methods are not orthogonal. The algorithm proposed by Lanczos ~\cite{Saad} for non-symmetric matrices builds a pair of bases for the two subspaces
\begin{equation}
K_{m}(M,v_{1}) = span(v_{1},Mv_{1}, ... ,M^{m-1}v_{1})
\end{equation}
and
\begin{equation}
K_{m}(M^{T},w_{1}) = span(w_{1},M^{T}w_{1}, ... ,(M^{T})^{m-1}w_{1}).
\end{equation}

The pair of bases are built by the algorithm in Table \ref{Lanczos}.
\begin{table}[t!]
\caption[The Lanczos Biorthogonalization Procedure]{The Lanczos Biorthogonalization Procedure}
\vspace{-12pt}
\begin{center}
\begin{tabular} {cl} \hline \\
1. & Choose two vectors $v_{1}$ and $w_{1}$ that are parallel such that $(v_{1},w_{1})=1$. \\
2. & Set $\beta_{1} \space = \space \delta_{1} \space = \space 1$, $w_{o} \space = \space v_{o} \space = \space 0$.\\
3. & For $j=1, ..., m$, Do \\
4. & $\alpha_{j} \space = \space (Mv_{j},w_{j})$ \\
5. & $\hat v_{j+1} \space = \space Mv_{j} - \alpha_{j}v_{j}-\beta_{j}v_{j-1}$ \\
6. & $hat w_{j+1} \space = \space M^{T}w_{j}-\alpha_{j}w_{j} -\delta_{j}w_{j-1}$ \\
7. & $\delta_{j+1} \space = \space |(\hat v_{j+1},w_{j+1})|^\frac{1}{2}$. If $\delta_{j+1}\approx 0$ Stop. \\
8. & $\beta_{j+1} \space = \space (\hat v_{j+1},\hat w_{j+1})/\delta_{j+1}$ \\
9. & $w_{j+1} \space = \space \hat w_{j+1}/\beta_{j+1}$ \\
10.& $v_{j+1} \space = \space \hat v_{j+1}/\delta_{j+1}$ \\
\\
\hline
\vspace{-34pt}
\end{tabular}
\label{Lanczos}
\end{center}
\end{table}
The scalars $\delta_{j+1}$ and $\beta_{j+1}$ are scaling factors for the bases vectors $w_{j+1}$ and $v_{j+1}$ respectively. If these scalar values tend toward a zero value in Steps 7 and 8, the algorithm will cease to converge and should exit in line 7 of the algorithm. 

As a result of lines 9 and 10, it is necessary to impose the constraint that
\begin{equation}\label{Tcons}
\delta_{j+1}\beta_{j+1} = (\hat v_{j+1}, \hat w_{j+1}).
\end{equation}
If equation (\ref{Tcons}) is satisfied we can write the tridiagonal matrix
\begin{displaymath}
 T_{m} = 
\left(
\begin{array}{rrrrr}
\alpha_{1} & \beta_{2}  &             &               &  \\
\delta_{2} & \alpha_{2} & \beta_{3}   &               &  \\
           &            & ...         &               &  \\
           &            & \delta_{m-1}& \alpha_{m-1}  & \beta_{m} \\
           &            &             & \delta_{m}    & \alpha_{m} \\
\end{array}
\right).
\end{displaymath}
\\
Notice that the $\delta_{j}$ is determined by the two norm of $v_{j}$ and $w_{j}$ and therefore are always positive. The $\beta_{j}$ scaling parameter is then $\pm \delta_{j}$. 

It has been shown that if the Lanczos Biorthogonalization algorithm has not broken down by the $m^{th}$ step and $\left\{ v_{i}\right\}_{i=1,..,m}$ is a basis of $K_{m}(M,v_{1})$ and $\left\{w_{i}\right\}_{i=1,..,m}$ is a basis of $K_{m}(M^{T},v_{1})$, then the following equations hold ~\cite{Saad}
\begin{eqnarray}
MV_{m}          &=& V_{m}T_{m} + \delta_{m+1}v_{m+1}e^{T}_{m}, \\
M^{T}W_{m}      &=& W_{m}T^{T}_{m} + \beta_{m+1}w_{m+1}e^{T}_{m}, \\
W^{T}_{m}MV_{m} &=& T_{m}.
\end{eqnarray}
The $T_{m}$ and $T^{T}_{m}$ matrices can be interpreted as the projection matrices of $M$ and $M^{T}$ onto the subspace $K_{m}(M,v_{1}
)$ and its orthogonal space $K_{m}(M^{T},v_{1})$. In practice, there are many techniques which do not use the matrix $M^{T}$, thus reducing the overhead of the Lanczos algorithm. These are referred to as transpose-free methods.

\subsection{Biconjugate Gradient Method}
The Biconjugate Gradient method (BiCG) is a non-symmetric Lanczos method. Implicitly, BiCG solves not only the original system of equations, $Mx=b$, but also the dual linear system of equations $M^{T}x^{*}=b^{*}$. 
% The BiCG algorithm is a Lanczos projection process onto the subspace
% \begin{equation}
% K_{m} = span( v_{1},Mv_{1}, ... ,M^{m-1}v_{1})
% \end{equation}
% which is orthogonal to 
% \begin{equation}
% L_{m} = span( w_{1},M^{T}w_{1}, ... ,(M^{T})^{m-1}w_{1}). 
% \end{equation}
The vectors $w_{1}$ and $v_{1}$ are not orthogonal to each other such that $(v_{1},w_{1})\not=0$. $w_{1}$ is obtained from the initial residual vector $b^{*} - M^{T}x^{*}_{o}$. The approximate solution vector that is obtained from the BiCG method has the form $x \space = \space V_{m}d_{m}$, where $d_{m} \space = \space T^{-1}_{m}(\beta e_{1})$. To find the inverse of the tridiagonal matrix $T_{m}$ we employ a LU factorization giving
\begin{equation}
T^{-1}_{m} = (L_{m}U_{m})^{-1}.
\end{equation}
Now define the matrix $P_{m} \space = \space V_{m}U^{-1}_{m}$ such that the solution vector is written

\begin{eqnarray}\label{Tinvx}
x_{m} &=& x_{o} + V_{m}T^{-1}_{m}(\beta e_{1}) \\
      &=& x_{o} + P_{m}L^{-1}_{m}(\beta e_{1}).
\end{eqnarray}
The residual vectors for both the linear system of equations and its dual are denoted by $r_{j}$ and $r^{*}_{j}$ and are in the same direction as $v_{j+1}$ and $w_{j+1}$ respectively. 

For the dual system we define the matrix
\begin{equation}\label{pstar}
P^{*}_{m} = W_{m}L^{-T}_{m}.
\end{equation}
Using the LU factorizations and the definitions in equations \ref{pstar} and $P_{m}$ we can show
\begin{eqnarray}\label{Psimilar}
(P^{*}_{m})^{T}MP_{m} &=& L^{-1}_{m}W^{T}_{m}MV_{m}U^{-1}_{m} \\
                      &=& L^{-1}_{m}T_{m}U^{-1}_{m} \\
                      &=& I. 
\end{eqnarray}
When equation \ref{Psimilar} is true we say that the columns of $P^{*}_{m}$ and $P_{m}$ are M-conjugate. We now have all the pieces to construct the BiCG algorithm for the system of equations of $M$. This algorithm is found in Table \ref{BiCG}.
\begin{table}[t!]
\caption[BiConjugate Gradient (BICG) Algorithm]{ALGORITHM :: BiConjugate Gradient Method}
\vspace{-12pt}
\begin{center}
\begin{tabular} {cl} \hline \\
1. & Compute the residual vector, $r_{o}\space=\space b -Mx_{o}$ and \\
   & choose the dual residual such that $(r_{o},r^{*}_{o}\not=0)$ \\
2. & Set $p_{o} = r_{o}$ and $p^{*}_{o} = r^{*}_{o}$\\
3. & Do $j=0,1,...$ convergence \\
4. & $\alpha_{j} \space =\space (r_{j},r^{*}_{j})/(Mp_{j},p^{*}_{j})$ \\
5. & $x_{j+1} \space=\space x_{j} + \alpha_{j}p_{j}$\\
6. & $r_{j+1} \space=\space r_{j} - \alpha_{j}Mp_{j}$\\
7. & $r^{*}_{j+1} \space=\space r^{*}_{j} - \alpha_{j}Mp^{*}_{j}$\\
8. & $\beta_{j} \space=\space (r_{j+1},r^{*}_{j+1})/(r_{j},r^{*}_{j})$\\
9. & $p_{j+1} \space=\space r_{j+1} + \beta_{j}p_{j}$ \\
10.& $p^{*}_{j+1} \space=\space r^{*}_{j+1} + \beta_{j}p^{*}_{j}$.\\
11.& Enddo \\
\\
\hline
\vspace{-34pt}
\end{tabular}
\label{BiCG}
\end{center}
\end{table}
To solve the dual system, the residual vector in line 1 is replaced by $r^{*}_{o} \space=\space b^{*} - M^{T}x^{*}_{o}$ and the update to the solution vector in line 5 is $x^{*}_{j+1} \space=\space x^{*}_{j} + \alpha_{j}p^{*}_{j}$. 

\section{GMRES Projection Method for Multiple Right Hand Sides}\label{sec:proj}

In many physical applications, including lattice QCD, it is desirable to solve the same matrix equation for multiple right hand sides. It is important to solve these systems of equations together and take advantage of the fact that each right hand side shares the same matrix. We next describe the GMRES Projection method for Multiple Right-Hand Sides (RHS).   

This GMRES method employees GMRES-DR(m,k) to solve the initial system of equations (first RHS) and then uses the eigenvector information for the subsequent right-hand sides ~\cite{MRHS}. Specifically, projections over the approximate eigenvectors generated in the GMRES-DR(m,k) algorithm are alternated with cycles of GMRES(m). This method is called GMRES(m)-Proj(k) where $m$ is the dimension of the Krylov subspace and $k$ is the number of approximate eigenvectors used in the projection cycles. 

The approximate eigenvectors from GMRES-DR(m,k) span a small Krylov subspace. These eigenvectors are generated by the ``Arnoldi-like" recurrence
\begin{equation}\label{Vkproj}
MV_{k} = V_{k+1}\bar{H_{k}},
\end{equation}
where $V_{k}$ is a n $\times$ k orthonormal matrix, $V_{k+1}$ is similar to $V_{k}$ with the inclusion of one extra row, and $\bar{H_{k}}$ is a full k+1 $\times$ k matrix. The columns of $V_{k}$ span a Krylov subspace as well as the subspace of approximate eigenvectors. 

A projection method is required over the approximate eigenvectors for the GMRES(m)-Proj(k) algorithm. A MinRes Projection projection over the subspace spanned by the columns of $V_{k}$ is presented below (see Table \ref{minresVk}).
\begin{table}[t!]
\caption[MinRes Projection for $V_{k}$]{ALGORITHM :: MinRes Projection for $V_{k}$}
\vspace{-12pt}
\begin{center}
\begin{tabular} {cl} \hline \\
1. & Let the current approximate solution vector be $x_{o}$ with the associated \\
   & system of equations $M(x-x_{o})=r_{o}$. Let $V_{k+1}$ and $\bar{H_{k}}$ \\
   & come from equation \ref{Vkproj}. \\
2. & Solve the least squares problem min$||c - \bar{H_{k}}d||$, where\\
   & $c = V_{k+1}^{T}r_{o}$. \\
3. & Form the new approximate solution vector $x = x_{o} + V_{k}d$. \\
4. & Form the associated residual vector $r = r_{o} - MV_{k}d = r_{o}-V_{k+1}\bar{H_{k}}d$. \\
\\
\hline
\vspace{-34pt}
\end{tabular}
\label{minresVk}
\end{center}
\end{table}
This projection is relatively inexpensive requiring $3k + 2$ vector operations (dot products and vector additions) of length $n$.

The GMRES(m)-Proj(k) method is applied to all right hand sides except for the initial system of equations that is solved with GMRES-DR(m,k). The GMRES(m)-Proj(k) algorithm is presented in ~\cite{MRHS}. GMRES(m)-Proj(k) is summarized in Table \ref{GMRESProj}
\begin{table}[t!]
\caption[GMRES(m)-Proj(k) Algorithm]{ALGORITHM :: GMRES(m) - Proj(k)}
\vspace{-12pt}
\begin{center}
\begin{tabular} {cl} \hline \\
1. & For the $i^{th}$ system of equations let $M(x^{(i)} - x^{(i)_{o}}) = r^{i}_{o}$.\\
2. & If the right hand sides are related, project over the previous solution \\
   & vectors. \\
3. & Apply the MinRes projection for $V_{k}$.\\
4. & Apply one cycle of GMRES(m).  \\
5. & Test the convergence of the current residual norm. If not satisfied, go to \\
   & step 3. \\
\\
\hline
\vspace{-34pt}
\end{tabular}
\label{GMRESProj}
\end{center}
\end{table}
The superscript on the solution and residual vectors indicates the current right hand side that is being solved. Notice that the projection in Step 3 adds very little overhead to the overall GMRES(m)-Proj(k) algorithm. A cycle of GMRES(m) requires $m^{2}+2m$ length $n$ vector operations as well as the cost of $m$ matrix-vector products. Step 3 requires $3k+2$ vector operations with no matrix-vector products.   

\section{Multiple Shift Krylov Methods}
Recall our system of equations,
\begin{equation}\label{recall}
Mx=b.
\end{equation}
Using a Krylov subspace as a basis has many advantages for solving linear equations. One of these advantages is that it allows for multiple shifted systems of equations of the form
\begin{equation}
(M - \sigma_{i})x = r_{o}
\end{equation}
to be solved simultaneously. There are more than three well-known methods that have been developed to solve this type of problem. This section is a review of Krylov methods for multiply-shifted problems.

Let the Krylov subspace of dimension $m$ generated by $M$ and $b$ be
\begin{equation}
K_{m}(M,b) = span\left\{ b, Mb, ... , M^{m-1}b\right\}.
\end{equation}
The Krylov subspace is invariant under the shifts $\sigma_{i}$,
\begin{equation}\label{shiftK}
K_{m}(M,b) = K_{m}(M - \sigma_{i},b) \space \space i=1,...,ns,
\end{equation}
where $ns$ is the number of shifted systems.

Krylov methods are iterative techniques for solving equation \ref{recall}, where the $m^{th}$ iterate $x_{m}$ satisfies $x_{m} = x_{o} + K_{m}(M,r_{o})$ with $r_{o} = b - Mx_{o}$. If $x_{o}=0$, then $r_{o} =b$ and we conclude from equation \ref{shiftK} that solution vectors of the shifted system of equations can be obtained from the same Krylov subspace as the unshifted system ~\cite{frommer-bicgstabl}. An important result of the shifted methods is that there is no added overhead to the calculation by adding the shifted systems, since the original subspace can be employed. Therefore, we can simultaneously solve the shifted system of equations for free. 

It was shown in reference ~\cite{frommer-bicgstabl} that to solve simultaneous shifted systems, residual vectors of each system have to be collinear such that $r_{shift} = \alpha r_{i}$, where $r_{i}$ is the initial system and $\alpha$ is a constant to be determined. Once $\alpha$ is determined, the residual of the shifted system is obviously a multiple of the initial residual. Therefore, the residuals come from the same Krylov subspace. Effectively, this corresponds to keeping the subspaces $K(M,b)$ and $K(M-\sigma_{i},b)$ identical. 
 
\subsection{Shifted BiCG}  

BiCG employees a coupled two-term recurrence formalism. The two-term recurrence computes the polynomials $p_{k}$, solution vectors $x_{m}$, and their residuals $r_{m}$ as seen in algorithm \ref{BiCG}. If we demand that the polynomials of the shifted and unshifted system of equations are parallel, we must insist that the scalars $\beta$ and $\alpha$ are collinear such that $\beta_{shift} =C_{1}\beta$ and $\alpha_{shift}=C_{2}\alpha$. $C_{1}$ and $C_{2}$ are determined via the constraint that the two-term recurrence polynomials are collinear for all shifted systems of equations. The details of how to determine the collinearity coefficients for shifted BiCG are in ~\cite{frommer-bicgstabl}. The algorithm is summarized in Table \ref{ShiftedBiCG}
\begin{table}[t!]
\caption[Shifted BICG Algorithm]{ALGORITHM :: Shifted BiCG}
\vspace{-12pt}
\begin{center}
\begin{tabular} {cl} \hline \\
1. & Initialize all solution vectors $x=x_{shift}=0$ and $r=b$.\\
2. & Do $k=1,2,...$ \\
3. & Employee the BiCG algorithm for the unshifted system of equations. \\
4. & For the shifted system, calculate $C_{1}$ and $C_{2}$  for $\beta_{shift}$ and\\
   & $\alpha_{shift}$, respectively. \\
5. & Determine the shifted polynomial $p_{shift} = p_{k}^{-1}(\sigma)r - \beta_{shift}p_{shift}$ \\
   & where $p_{k}^{-1}(\sigma)$ is determined by the two term recurrence. \\
6. & Determine $x_{shift} = x_{shift} + \alpha_{shift}p_{shift}$ \\
7. & Update the residual norms, $r = r - \alpha Mp$ and $r^{*} = r^{*} - \alpha Mp^{*}$. \\
8. & EndDo \\
\\
\hline
\vspace{-34pt}
\end{tabular}
\label{ShiftedBiCG}
\end{center}
\end{table}
This Lanczos method has the advantage of relying upon short recurrence relationships. However, when the matrix is non-Hermitian, the computation of each basis vector used in the method requires a multiplication with $M$ and $M^{\dagger}$ which results in added overhead to the algorithm. 

\subsection{Shifted FOM}

The standard FOM method uses the Arnoldi algorithm to construct a Krylov subspace. Recall that $V_{m}$ and $H_{m}$ are formed so that the first column of $V_{m}$ is $v_{1}=r_{o}/||r_{o}||$, and that the Arnoldi iterate is
\begin{equation}
MV_{m} = V_{m}\bar{H_{m}} + h_{m+1,m}v_{m+1}e^{T}_{m}.
\end{equation}
For the shifted system of equations, the Arnoldi algorithm differs by a shift $\sigma_{i}$ as in equation \ref{shiftedArnoldi0} such that
\begin{equation}\label{shiftedArnoldi}
(M - \sigma_{i}I)V_{m} = V_{m}(\bar{H_{m}} - \sigma I_{m}) + h_{m+1,m}v_{m+1}e^{T}_{m},
\end{equation}
where $I_{m}$ is the identity matrix of dimension $m$. According to equation \ref{shiftedArnoldi}, the only modification to the original FOM method is that the small solution vector $y_{m}$ is solved via the system of equations $(\bar{H}_{m} - \sigma I_{m})y_{m} = \beta e_{1}$ ~\cite{Simoncini-2003}. 

It was also shown in reference ~\cite{Simoncini-2003} that shifted FOM can be restarted because $r_{m}$ is a multiple of the basis vector $v_{m+1}$. (ie~ $r_{m} = -h_{m+1,m}v_{m+1}(y_{m})_{m}$) The restarted shifted FOM algorithm is below.
\begin{table}[t!]
\caption[Restarted Shifted FOM Alogorithm]{ALGORITHM :: Restarted Shifted FOM}
\vspace{-12pt}
\begin{center}
\begin{tabular} {cl} \hline \\
1. & Set $r_{o}=b$, $\beta^{i}_{m} = ||r_{o}||$, and $x^{i}_{m}=x_{o}$. Set $v_{1}=r_{o}/\beta^{i}_{m}$.\\
   & $i$ is the current shift index. \\
2. & Construct $V_{m}$ and $\bar{H}_{m}$ for $K_{m}(M,v_{1})$. \\
3. & For all shifts construct\\
   & $y^{i}_{m} = (\bar{H}_{m} -\sigma_{i}I_{m})^{-1}e_{1}\beta^{i}_{m}$ \\
   & Update the solution vector $x^{i}_{m}=x^{i}_{m}+V_{m}y^{i}_{m}$. \\
4. & Exit if last shift has been computed to convergence criteria. \\
5. & Set $\beta^{i}_{m}=-h_{m+1,m}(y^{i}_{m})_{m}$ for each shift.\\
6. & Set $v_{1} \leftarrow v_{m+1}$ to restart.  \\
7. & EndDo \\
\\ 
\hline
\vspace{-34pt}
\end{tabular}
\label{ShiftedFOM}
\end{center}
\end{table}
Similar to Shifted BiCG, restarted Shifted FOM only generates one basis $\left\{v_{1}, ... ,v_{m}\right\}$ for all shifted systems to be solved simultaneously. 

\subsection{Shifted GMRES}

Similar to both shifted BiCG and FOM, the same basis vectors for $K_{m}(M,b)$ and $K_{m}(\hat{M},b)$ can be used for the initial and shifted systems for GMRES so that the systems can be solved simultaneously where for convience we have defined $\hat{M} = (M - \sigma I)$. However, for the restarted shifted GMRES method, $r_{i}$ and $\hat{r}_{i}$ may not be parallel. This differs from the previous methods in that the matrix multiplications can not be saved upon the restart of the algorithm. A solution to this problem is presented in ~\cite{Frommer}. 

Any vector from the affine Krylov subspace can be written
\begin{equation}
x_{m} = x_{o} + p_{m-1}(M)r_{o},
\end{equation} 
where $p_{m-1}$ is a polynomial of degree $\leq m-1$. The corresponding residual is $r_{m} = b - Mx_{m}$. The residual can also be written in terms of the polynomial $p_{m-1}(M)$: 
\begin{eqnarray}
r_{m} &=& r_{o} - Mp_{m-1}(M)r_{o}, \\
      &=& q_{m}(M)r_{o}, \\
\end{eqnarray}
where $q_{m}(M) = I - Mp_{m-1}(M)$ with the initial condition that 
$q(0) = 1$. Similarly, we can define the residual norm and solution vector of the shifted system in terms of this polynomial,
\begin{eqnarray}
\hat{x}_{m} &=& \hat{x}_{o} + \hat{q}_{m-1}(\hat{M})\hat{r_{o}}, \\
\hat{r}_{m} &=& \hat{q}_{m}(\hat{M})\hat{r}_{o}, 
\end{eqnarray}
where $\hat{q}$ is similar to the polynomial above but the identity has been shifted by $\sigma$. If we assume that the initial residual vectors are collinear, then $\hat{r}_{o} = \alpha_{o}r_{o}$ where $\alpha_{o} \in C$. With this initial condition, the constraints to keep the shifted systems parallel are
\begin{eqnarray}\label{colinearity}
\hat{r}_{m} &=& \alpha_{m}r_{m},
\end{eqnarray}
which yields
\begin{equation}\label{colinearpoly}
\alpha_{o}\hat{q}_{m}(\hat{M})r_{o} = \alpha_{m}q_{m}(\hat{M}-\sigma I)r_{o}.
\end{equation}
Equations (\ref{colinearity}) and (\ref{colinearpoly}) are the defining equations for $\hat{q}_{m}$ and $\alpha_{m}$.

The Arnoldi equation for the initial and shifted system of equations are 
\begin{equation}
MV_{m} = V_{m+1}\bar{H}_{m}
\end{equation}
and
\begin{equation}\label{modArn}
\hat{M}V_{m} = V_{m+1}\hat{\bar{H}}_{m},
\end{equation}
with $\hat{\bar{H}}_{m} = \bar{H}_{m} - \sigma_{i}I_{m+1,m}$ where the last row of the m+1 $\times$ m identity matrix is full of zeros. 

The collinearity condition for the residual norms in equation \ref{colinearity}, along with the definition of the shifted Arnoldi recurrence in equation \ref{modArn} yield the underdetermined equation ~\cite{Frommer}
\begin{equation}\label{2unknown}
\hat{\bar{H}}_{m}\hat{y}_{m} + (\beta e_{1} - \bar{H}_{m}y_{m})\alpha_{m} = \alpha_{o}||r_{o}||_{2}e_{1}.
\end{equation}
There are two unknown variables, $\hat{y}_{m}$ and $\alpha_{m}$, in equation \ref{2unknown}. In practice, a $QR$ factorization can be used to solve this equation to determine $\alpha_{m}$. Once the collinearity parameter is determined, the solution vector $\hat{y}_{m}$ is determined, which allows the shifted GMRES algorithm to be restarted. 

For a fixed $m$ value in GMRES(m), the Shifted GMRES(m) algorithm is written in Table \ref{ShiftedGMRES}
\begin{table}[t!]
\caption[Shifted GMRES(m)]{ALGORITHM :: Shifted GMRES(m)}
\vspace{-12pt}
\begin{center}
\begin{tabular} {cl} \hline \\
1. & Set the initial guesses for $x_{o}$, $\hat{x}_{o}$ such that $\hat{r}_{o}=\alpha_{o}r_{o}$.\\
2. & Employee the Arnoldi algorithm to determine $V_{m}$ and $\bar{H}_{m}$. \\
3. & Use GMRES(m) for the initial system of equations to determine\\
   & $y_{m}$, the solution $x_{m}=x_{o} + V_{m}y_{m}$ and \\
   & $(\beta e_{1} - \bar{H}_{m}y_{m})$. \\
4. & Using the output of the initial GMRES(m) algorithm, \\
   & solve equation \ref{2unknown} for $\alpha_{m}$ and $\hat{y}_{m}$.\\
5. & Determine the solution of the shifted system, $\hat{x}_{m}=\hat{x}_{o}+V_{m}\hat{y}_{m}$.\\
\\
\hline
\vspace{-34pt}
\end{tabular}
\label{ShiftedGMRES}
\end{center}
\end{table}
The restarted shifted GMRES(m) algorithm is in Table \ref{resShiftedGMRES}.
\begin{table}[t!]
\caption[Restarted Shifted GMRES(m) Algorithm]{ALGORITHM :: Restarted Shifted GMRES(m)}
\vspace{-12pt}
\begin{center}
\begin{tabular} {cl} \hline \\
1. & Set the initial guesses for $x_{o}$, $\hat{x}_{o}$, $\alpha_{o}$ as in\\
   & the shifted GMRES(m) algorithm. Set the restart value, $k$. \\
2. & Do $j=1,2, ... ,j_{max}$ \\
3. & Use Shifted GMRES(m) to determine $x^{j+1}_{k}$, $\hat{x}^{j+1}_{k}$,\\
   & and $\alpha^{j+1}_{k}$ such that $\hat{r}^{j+1}_{k} = \alpha^{j+1}_{k}r^{j+1}_{k}$ \\
   & via equation \ref{2unknown}. \\
4. & Stop if the residual norm of the initial and shifted systems \\
   & has reached convergence criteria.\\
\\
\hline
\vspace{-34pt}
\end{tabular}
\label{resShiftedGMRES}
\end{center}
\end{table}
An example comparing restarted, shifted GMRES(m) and other shifted methods is shown in the next section in Figures (\ref{fig:gmresdrshift}) and (\ref{fig:shiftfomgmres}).
 
\section{Krylov Methods with Multiple Shifts and Multiple Right Hand Sides}

We have already discovered that GMRES-DR(m,k) can be implemented in applications where it is desired to solve multiple shifts simultaneously $or$ multiple right-hand sides. Sometimes both of these methods are needed for the same application. One specific application is Wilson LQCD. In the Wilson formalism the different shifts represent different quark masses and the right hand sides are quark sources on the lattice. A new deflated GMRES technique for $both$ multiple RHS and multiple shifts is provided in this thesis. 

Consider the large system of linear equations that not only has multiple right hand sides, but also has multiple shifts for each right hand side. Let $nrhs$ be the number of right hand sides and $ns$ be the number of shifts. The problem to solve is then
\begin{equation}
(M - \sigma_{i})x^{j}_{i} = b_{j},
\end{equation} 
with $j \space = \space 1,...,nrhs$ and $i \space = \space 1,...,ns$. $M$ is a large matrix which may be nonsymmetric or complex non-Hermitian. In this section $\sigma_{1}$ is referred to as the base shift. 

\subsection{Shifted GMRES-DR(m,k)}

To achieve our goal of an algorithm to solve multiple shifts simultaneously for multiple RHS, another new shifted GMRES(m) algorithm was needed which would take advantage of the deflation of eigenvalue spectrum to speed convergence. Such a method has been developed in this work and is referred to as GMRES-DRS(m,k), which stands for shifted GMRES-DR(m,k) where $m$ and $k$ have the same meaning as in the GMRES-DR(m,k) algorithm. 

For the deflated GMRES(m) method we can solve multiple shifted systems concurrently. The derivation is the same except that we employee the ``Arnoldi-like" recurrence that differs from the Arnoldi recurrence in that $\bar{H}_{m}$ is an upper Hessenberg m+1 $\times$ m matrix except for a full $k+1 \times k$ leading portion that contains approximate eigenvector information needed by the deflation technique. The shifted GMRES-DR(m,k) algorithm is listed in Table \ref{ShiftedGMRESDR}.
\begin{table}[t!]
\caption[GMRES-DRS(m,k) Algorithm]{ALGORITHM :: Shifted GMRES-DRS(m,k)}
\vspace{-12pt}
\begin{center}
\begin{tabular} {cl} \hline \\
1. & At the begging of a cycle of GMRES-DRS(m,k), assume the current\\
   & problem is $(M - \sigma_{i}I)(x_{i} - \tilde{x}_{o,i}) \space = \space \beta_{i}r_{o,i}$, \\
   & with $\beta_{1}=1$ and where $\tilde{x}_{o,i}$ is the current \\
   & approximate solution to the $i^{th}$ shifted system. \\
2. & Apply GMRES-DR(m,k) to $M$ and generate the ``Arnoldi-like" recurrence \\
   & $MV_{m} = V_{m+1}\bar{H}_{m}$. \\
3. & For the base system, solve the minimum residual problem \\
   & min$||c-(\bar{H}_{m}-\sigma_{1}\bar{I})$, where\\
   & $c = V^{T}_{m+1}r_{o,1}$ and $\bar{I}$ is a $m+1 \times m$ identity matrix.  \\
   & The new approximate solution vector is $\tilde{x}_{1} = \tilde{x}_{o,1} + V_{m}d_{1}$. The new \\
   & residual vector is $r_{1} = r_{o,1} - MV_{m}d_{1} = r_{o,1} - V_{m+1}\bar{H}_{m}d_{1}$. \\
4. & For the other shifted systems $i=1,...,ns$ form $s=c-(\bar{H}_{m} -\sigma_{1}\bar{I})d_{1}$. Apply\\
   & a QR factorization: $\bar{H}_{m} - \sigma_{i}\bar{I} = QR$. Solve $Rd_{i}=\beta_{i}Q^{T}c+\alpha_{i}Q^{T}s$,\\
   & using the last row to solve for $\alpha_{i}$ and the first $m$ rows for $d_{i}$. \\
5. & The new approximate solution vector of the $i^{th}$ system is $\tilde{x}=x_{o,i}V_{m}d_{i}$, and \\
   & the new residual is $r_{i} = \alpha_{i}r_{o}$. \\
6. & Test the residual norm for convergence. If not satisfied, for $i=2,...,ns$ set \\
   & $\beta_{i} = \alpha_{i}$ and for $i=1,...,ns$, set $\tilde{x}_{o,i}=\tilde{x}_{i}$ and $r_{o,i}=r_{i}$.\\
   & Then go back to step 1. \\
 \\
 \hline
 \vspace{-34pt}
\end{tabular}
\label{ShiftedGMRESDR}
\end{center}
\end{table}
GMRES-DRS(m,k) is related to the method GMRES-E ~\cite{morgan-1995}. In practice, the QR factorization in this algorithm uses Givens rotations to solve the system of equations in Step 4. 

A comparative study between shifted GMRES(25) and the new GMRES-DRS(25,10) algorithm was carried out to study the convergence of each method. The matrix had $n \space =\space 1000$ and is bidiagonal with $0.1,1,2,3,...,998,998$ on the main diagonal and 1's on the superdiagonal. Thus, $M$ has the form:
\begin{displaymath}
 M = 
\left(
\begin{array}{rrrrr}
.1 & 1& 0 & .. & 0 \\
0 & 1 & 1 & : & :  \\
0 & 0& 2 & 1 & 0  \\
: & : & : & : & : \\
0 & .. & 0 & .. &\\
\end{array}
\right) ,
\end{displaymath}
\\

The associated right hand side is randomly generated. The shifts are $\sigma=1,-0.4,-2$. This matrix has a small eigenvalue that slows down the convergence of the residual norm for restarted, shifted GMRES(m), especially for the base system. Restarted, shifted GMRES(m) is compared to GMRES-DRS(m,k). The results are in Figure (\ref{fig:gmresdrshift}). \enlargethispage{\baselineskip}

{\setlength{\abovecaptionskip}{0mm}
\setlength{\belowcaptionskip}{0mm}
\begin{figure}[h!]
\centering
\includegraphics[scale=1, height=8cm, width=12cm]{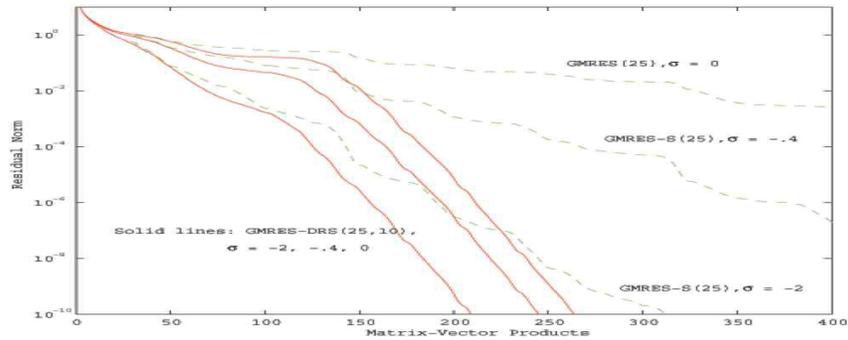}
\caption{Comparative results for restarted, shifted GMRES(m) and GMRES-DRS(m,k)}
\label{fig:gmresdrshift}
\end{figure}}

Shifting the matrix by 0.4 improves the convergence of GMRES(25) because the smallest eigenvalue in the spectrum is moved from 0.1 to 0.5. GMRES-DRS(25,10) converges rapidly once it generates approximate eigenvectors corresponding to these eigenvalues. The convergence for all three shifted systems is similar because once the small eigenvalues are removed by the deflation, shifting has little effect on convergence. 

In this example the second and third shifted systems converge faster than the first base system. However, in some situations, there can be convergence problems for the non-base systems ~\cite{Simoncini-2003}. This author compare multiply-shifted GMRES(m) and FOM. Since FOM has parallel residuals for all shifted systems, it is argued that it is a more natural approach to the shifted problem ~\cite{Simoncini-2003}. However, convergence depends on the roots of the polynomial, which correspond to eigenvalues of $M$ and where the roots fall in relation to the shift. The next example demonstrates this dependence. 
\enlargethispage{-\baselineskip}
Recall that shifted GMRES(m) uses the same polynomial for all shifted systems. Likewise, shifted FOM also uses one polynomial for all shifts. This polynomial is subject to the condition $q(0) = 1$. We may view this as the polynomial being 1 at zero and the spectrum shifted, or the polynomial has the value 1 at the shift and the spectrum fixed as that of $M$. We prefer the second view. So, for shifted GMRES(m) with the base system $(M - \sigma_{i}I)$, we view the polynomial chosen by this method as being 1 at $\sigma_{1}$ and needing to be small over the spectrum of $M$. 

In the next two examples, plots are given of the roots of the polynomials. Because the polynomial must be small over the spectrum of $M$, a small root of the polynomial at the shift can cause a convergence problem. So for Krylov methods to be effective, the roots of the polynomial need to generally stay away from the shift value. 

For the bidiagonal matrix used in the first example, we apply shifted GMRES(40) and shifted FOM(40) with shifts $\sigma=0.4,0$. The results are in Figure (\ref{fig:shiftfomgmres}).
{\setlength{\abovecaptionskip}{0mm}
\setlength{\belowcaptionskip}{0mm}
\begin{figure}[b!]
\centering
\includegraphics[scale=1, height=8cm, width=12cm]{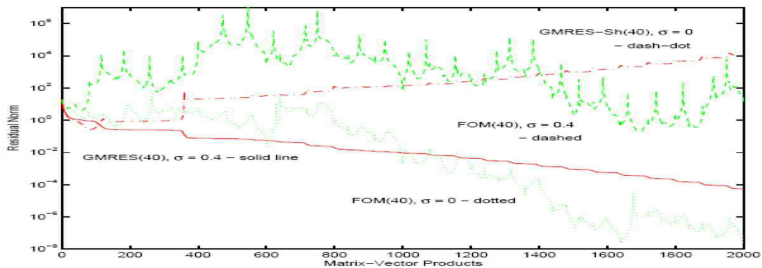}
\caption{Comparative results for shifted GMRES(40) and shifted FOM(40)}
\label{fig:shiftfomgmres}
\end{figure}}
For the base shift of $\sigma=0.4$, GMRES(40) works better than FOM(40). Alternatively, FOM performs better than GMRES for the shift value $\sigma=0$. Figure (\ref{fig:ritzvalues}) shows the harmonic Ritz values nearest the shift for 25 cycles of GMRES. These are the roots of the GMRES polynomial for the system $(M - 0.4I)x_{1} = b$, shifted so they correspond to the spectrum of $M$. \enlargethispage{\baselineskip}

{\setlength{\abovecaptionskip}{0mm}
\setlength{\belowcaptionskip}{0mm}
\begin{figure}[h!]
\centering
\includegraphics[scale=1, height=8cm, width=12cm]{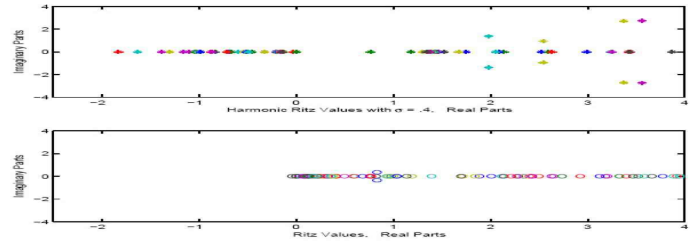}
\caption{Distribution of Smallest Harmonic and Regular Ritz values, m=40, 25 cycles.}
\label{fig:ritzvalues}
\end{figure}}

The harmonic Ritz values avoid the region about 0.4. This is a good result since the shifted GMRES(40) polynomial for $\sigma=0.4$ is defined to have a value of 1 near this value and be small over the spectrum of $M$. This polynomial cannot be effective if there is a root near 0.4. GMRES(40) converges slowly  for this fairly difficult problem. In contrast to GMRES(40), FOM(40)  with $\sigma=0.4$ is not able to converge because the roots of the FOM polynomial are not separated from 0.4 as seen in the bottom plot in figure (\ref{fig:ritzvalues}). For the second shift value of $\sigma=0$ shifted GMRES(40) is not effective. There are too many harmonic Ritz values at and surrounding 0. Shifted FOM(40) gives erratic convergence because of some Ritz values near zero, but it does manage to converge. See ~\cite{Simoncini-2003} for more comparative examples between shifted FOM and GMRES. We next give an example where GMRES is more effective than FOM.

The matrix is the same as in the previous two examples. The base shift is $\sigma=0$ and a second shift of $\sigma=1.4$ is used. Shifted GMRES(80) is compared with shifted FOM(80). The results are shown in Figure (\ref{fig:gmresfom80}).

{\setlength{\abovecaptionskip}{0mm}
\setlength{\belowcaptionskip}{0mm}
\begin{figure}[h!]
\centering
\includegraphics[scale=1, height=8cm, width=12cm]{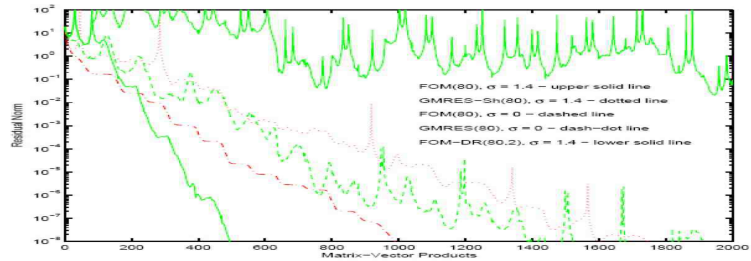}
\caption{Comparative results for shifted GMRES(80) and shifted FOM(80)}
\label{fig:gmresfom80}
\end{figure}}

Some of the Ritz values in the FOM(80) spectrum fall around 1.4, while there is a gap around these values in the harmonic Ritz values. These approximate eigenvalue distributions are shown in Figure (\ref{fig:ritzvalues80}). \enlargethispage{\baselineskip}

{\setlength{\abovecaptionskip}{0mm}
\setlength{\belowcaptionskip}{0mm}
\begin{figure}[h!]
\centering
\includegraphics[scale=1, height=8cm, width=12cm]{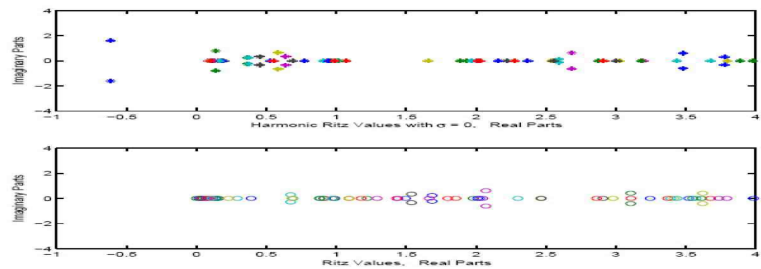}
\caption{Distribution of Smallest Harmonic and Regular Ritz values, m=80, 25 cycles.}
\label{fig:ritzvalues80}
\end{figure}}

In Figure (\ref{fig:ritzvalues80}) we see that shifted GMRES(80) works much better for the second system. Therefore, we conclude that methods for shifted GMRES(m) and shifted FOM must both be used with caution. However, deflating eigenvalues can help this problem. Figure (\ref{fig:gmresfom80}) also has a plot of FOM-DRS(80,2) for $\sigma=1.4$ and we see that deflating only two approximate eigenvalues fixes the convergence problem. Deflated methods in general will eliminate the ``small eigenvalue" problem. For the rest of the examples in this thesis the matrices and shifts are such that the base systems have the slowest convergence.

\section{Deflated GMRES for Multiple Right-Hand Sides and Multiple Shifts}

We now consider solving multiply-shifted systems that also have multiple right-hand sides.  It is important to reuse information or share information among the right-hand sides. Block methods~\cite{Saad} share information between right-hand sides.  It is possible to design multi-shifted versions of both Block-GMRES~\cite{Saad} and Block-GMRES-DR~\cite{bgdr}.  However, here we will concentrate on a non-block approach.  The right-hand sides are solved separately, and eigenvector information from the solution of the first right-hand side is used to assist the subsequent ones.  More specifically, we will generalize for multiple shifts the GMRES-Proj approach mentioned in Section \ref{sec:proj}.  See~\cite{MRHS} for more on this method, including comparison with block methods.

First, some of the difficulties of deflating for subsequent right-hand sides will be discussed.  Suppose the first right-hand side has been solved and approximate eigenvectors have been generated.  Then for the non-shifted case, there are several ways to deflate eigenvalues.  Some of these are given in~\cite{MRHS}.  However, generally they do not work for multiply-shifted systems.  For example, if the deflation involves building a preconditioner from the approximate eigenvectors~\cite{BaCaGoRe,BuEr,MRHS}, then as mentioned earlier, the differently shifted preconditioned systems cannot be solved together.  
\enlargethispage{\baselineskip}

For the GMRES(m)-Proj(k) method, there is trouble with one of the two phases.  We know the GMRES portion can be adapted to keep right-hand sides parallel for multiple shifts.  However, the phase with projection over approximate eigenvectors generally fails to produce parallel residual vectors.  Even though this projection is over a Krylov subspace of dimension $k$ spanned by the columns of $V_k$, this subspace does not contain the current right-hand side (the residual vector); so the derivation in \ref{sec:proj} does not work.  Specifically, the residual vectors between adjacent right-hand side vectors are not parallel since $r_{0,i}$ for $i=1,...nrhs$ and $r_{0,1}$ are not in the span of the columns of $V_{k+1}$. One case in which the projection does keep the residual vectors parallel is exact eigenvectors, as shown below.

%\begin{theorem}
Theorem ::
Assume that before the minres projection, the shifted systems are \[M(x-x_i) = r_{0,i}\,\] 
with $r_{0,i} = \beta_i r_{0,1}$ for $i = 2, \ldots, ns$.
Then after the minres projection over the subspace $Span\{z_1, z_2, \ldots, z_k\}$, where $z_1$ through $z_k$ are eigenvectors of $M$, the residual vectors are still parallel.
%\end{theorem}
{\em Proof}.  Let $Z_k$ be the matrix with $z_1, \ldots, z_k$ as columns.  For exact eigenvectors, the minres projection is equivalent to Galerkin.  The Galerkin orthogonality condition gives for the $i^{th}$-system
\[Z_k^T(\beta_i r_{0,1} - (M-\sigma_i I) Z_k d_i) = 0. \]
Solving gives 
\[d_i = \beta_i (\Lambda_k - \sigma_i I_k)^{-1} (Z_k^TZ_k)^{-1}Z_k^T r_{0,1}, \]
where $\Lambda_k$ is the k $\times$ k diagonal matrix with diagonal entries $\lambda_1$ through $\lambda_k$ and $I_k$ is the k $\times$ k identity matrix.
The residual vector after projecting is then 
\begin{eqnarray*}
r_{0,i}^{new} &=& \beta_i r_{0,1} - (M-\sigma_i I) Z_k d_i \\
&=& \beta_i r_{0,1} - \beta_i (M-\sigma_i I)Z_k (\Lambda_k - \sigma_i I_k)^{-1} (Z_k^TZ_k)^{-1}Z_k^T r_{0,1} \\
&=& \beta_i r_{0,1} - \beta_i Z_k(\Lambda_k - \sigma_i I_k) (\Lambda_k - \sigma_i I_k)^{-1} (Z_k^TZ_k)^{-1}Z_k^T r_{0,1}  \\
&=& \beta_i (I - Z_k (Z_k^TZ_k)^{-1}Z_k^T) r_{0,1}.  
\end{eqnarray*}
This shows that all $r_{0,i}$ are multiples of each other.
\endproof

So, one option for using GMRES-Proj with multiple shifts is to use only fairly accurate eigenvectors.  We could sort through the approximate eigenvectors computed by GMRES-DR and apply only ones with acceptable accuracy to the projection in GMRES-Proj. This is now tested.  

{\it Example 1.} For the same matrix, we first solve the $\sigma_1 = 0$ system to an accuracy relative to the residual norm below $10^{-10}$ or 250 matrix-vector products.  Table \ref{tbl:evec} shows the accuracy of the eigenvectors thus produced in its second column.  The worst residual norm of the 10 approximate eigenvectors is $3.3\times 10^{-4}$.  If, for example, only the four smallest approximate eigenvectors are used, their accuracy has a residual norm of $1.0\times 10^{-6}$ or better.  Table \ref{tbl:evec} lists the number of matrix-vector products for the relative residual norm of the base shifted system with the second right-hand side to reach $1.0\times 10^{-8}$.  We see that convergence is better using all ten eigenvectors.  However, the fourth column gives the accuracy  attained by the worst of the last two shifted systems (usually it is the third shift).  It only reaches a residual norm of $3.3\times 10^{-4}$ if all 10 approximate eigenvectors are used in the projection.  With only four eigenvectors, the residual norm reaches a better level of $5.4\times 10^{-8}$, but the convergence is almost twice as slow.  
%(next mention the 385 case)
\begin{table}
\caption{Effect of projecting over different accuracies of eigenvectors} 
\vspace{-12pt}
\begin{center} \footnotesize
\begin{tabular}{cccccccc}  \hline
       & \underline{250 mvp's for 1st}   &   &         & \underline{385 mvp's  1st} & &    \\ 
$k$ & eig. res. & mvp's  & lin. eqs. res.  &  eig. res.  & mvp's  & lin. eqs. res.  \\  
\hline 
10  & 4.1e-2  & 135    & 3.3e-4  & 4.4e-1  & 135 & 1.7e-6 \\ 
8   & 3.3e-3  & 165    & 3.7e-5  & 7.1e-6  & 165 & 1.3e-9 \\ 
6   & 8.0e-5  & 180    & 2.8e-6  & 3.3e-10 & 180 & 1.3e-9 \\ 
4   & 1.0e-6  & 255    & 5.4e-8  & 2.4e-11 & 255 & 5.5e-10 \\  
2   & 5.4e-9  & 435    & 1.5e-8  & 9.9e-12 & 435 & 3.6e-10 \\ 

\hline
\vspace{-34pt}
\end{tabular} 
\end{center} 
\label{tbl:evec}
%\label{} 
\end{table}  
%(... should we switch to relative residuals in table? ....)

%(........... add figure of convergence curves (of 3rd shifted system) with different numbers of eigenvectors like on back of p.11 in notes)

The problem with this approach is that the eigenvector computation during solution of the first right-hand side needs to be done to a considerable accuracy, since we do not want to slow down convergence of the subsequent systems.  If many right-hand sides are to be solved, this extra expense might not be significant.  However, we next propose an approach that does not require eigenvectors to be accurate.  
%(? on new para....)

The key idea is that although the residual vectors cannot be kept parallel, they can be chosen so that they relate to each other.  We force the residuals of the non-base systems to be parallel to the residual of the base system except for a component in the direction of $v_{k+1}$, the last column of the $V_{k+1}$ matrix from Equation~(\ref{Vkproj}). \enlargethispage{\baselineskip}
Now we will look at the correction phase that is needed at the end of GMRES-Proj.  Assume that we have already solved shifted systems with the extra right-hand side $v_{k+1}$ (this solution will be discussed next) and have 
\begin{equation}
(M-\sigma_i I)s_i = v_{k+1}. \label{keyder3}
\end{equation}
We assume that for a particular right hand side, the systems have been solved by GMRES-Proj to the point that the residual is only in the direction of $v_{k+1}$:
\begin{equation}
b - (M - \sigma_i I) \bar x_i = r_i = \gamma_i v_{k+1}, \label{keyder4}
\end{equation}
for $i=2,\ldots,ns$ and for some scalar $\gamma_i$.  Here $\bar x_i$ is the approximate solution to $x_i$ (the superscripts for the particular right-hand sides are left off here for simplicity).  We perform a Galerkin projection for system~(\ref{keyder4}) over the subspace spanned by the single vector $s_i$ from solution of~(\ref{keyder3}): 
\[s_i^T(M-\sigma_i I)s_i \delta = \gamma_i s_i^T v_{k+1}.\]  Using Equation~(\ref{keyder3}), this becomes \[s_i^T v_{k+1} \delta = \gamma_i s_i^T v_{k+1}.\]  Then $\delta = \gamma_i$.  To determine $\gamma_i$, we start with $r = \gamma_i v_{k+1}$. Multiplying both sides by $v_{k+1}^T$ and using that $v_{k+1}$ is of unit length gives \begin{equation}
\gamma_i = v_{k+1}^T r. \label{keyder4b}
\end{equation}
The corrected solution of the system is $x_i = \bar x_i + \delta s_i$.

We need to fill in the method for solving~(\ref{keyder3}), the shifted systems with the extra right-hand side.  First, GMRES-Proj is applied until the residual is negligible except in the direction of $v_{k+1}$.  So we assume that
\begin{equation}
(M - \sigma_i I) \bar s_i = r \label{keyder5}
\end{equation}
is the current system, where
\[
r = v_{k+1} - (M - \sigma_i I) \bar s_i = \gamma_i v_{k+1},
\]
for some scalar $\gamma_i$.
Rearranging gives 
\begin{equation}
(M - \sigma_i I) \bar s_i = (1-\gamma_i) v_{k+1}.\label{keyder6}
\end{equation}
Applying Galerin projection over the subspace spanned by the single vector $\bar s_i$ to the system (\ref{keyder5}) gives
\[
\bar s_i^T (M - \sigma_i I) \bar s_i \delta = \gamma_i \bar s_i^T v_{k+1}.
\]
With Equation~(\ref{keyder6}), this becomes
\[
(1-\gamma_i) \bar s_i^T v_{k+1} \delta = \gamma_i \bar s_i^T v_{k+1},
\]
and this simplifies to
\[\delta = {\gamma_i \over {1-\gamma_i}}.\] 
So the corrected solution is 
$s_i = \bar s_i + {\gamma_i \over {1-\gamma_i}} \bar s_i.$
Now \[s_i = { 1\over {1-\gamma_i}} \bar s_i.\]
Finally, the $\gamma_i$ is determined to be $\gamma_i = v_{k+1}^T r $ as it was for (\ref{keyder4b}).

We next give the algorithms for solution of the systems with second and subsequent right-hand sides and for the extra right-hand side.  Note these are in order of how they were derived here, not in order of how they are actually used.
The algorithm for solution of the systems with second and subsequent right-hand sides is given in Table \ref{ShiftedGMRES-Proj}. 
%(Have not considered related right-hand sides and projecting over previous solutions!  Need to add this in....................maybe with comparison like Dean did..........does projection over previous solutions mess up the parallel rhs's or can we do a correction, since have solved systems already?)
%(Should bar I's also have subscript of k?)
% Should the first algorithm have "bars" like the second does?
% Mostly use bar for current approximation, but also use tilde some.  Must choose.
\begin{table}[t!]
\caption[Shifted GMRES algorithm for Multiple RHS.]{ALGORITHM :: GMRES-Proj-Sh for the second}
\centering{and subsequent right-hand sides}
\vspace{-12pt}
\begin{center}
\begin{tabular} {cl} \hline \\
1. & Consider the systems with the $j$th right-hand side (and with all $ns$ shifts).\\
   & At the beginning of a cycle of GMRES(m)-Proj(k)-Sh, assume the current \\
   & problem is $(M-\sigma_i I) (x^j_i- \tilde x^j_{0,i}) = \beta_i r_{0,i},$\\
   & with $\beta_1 = 1$, and where $\tilde x_{0,i}$ is the current \\
   & approximate solution to the i$th$ shifted system \\
2. & Apply the Minres Projection for $V_k$ to the first right-hand side. \\
   & This uses the $V_{k+1}$ and $\bar H_k$ matrices developed while solving \\
   & the first right-hand side with GMRES-DRS(m,k). \\
3. &  For shifted systems $is=2 \ldots ns$, solve $(H_k - \sigma_i I) d_i = \beta_i (H_k - \sigma_1 I)d_1$.\\
4. & Apply one cycle of GMRES(m)-Sh.\\
5. & Test the residual norms for convergence (can also test during GMRES cycles). \\
   & For the non-base systems, can ignore the error term in the direction of $v_{k+1}$.\\
6. & Correction phase: Suppose the computed solution for the $i$th shifted system \\
   & Let the solution to the system with the extra right hand side \\
   & so far is $x^j_i$. $v_{k+1}$ and shift $\sigma_i$ be $s_i$. \\
   & The corrected solution is $(x^j_i)^{corrected} = x^j_i + (v_{k+1}'*r)*xv_{i}$. \\
   & The corrected residual norm can now be calculated. \\
\\
\hline
\vspace{-34pt}
\end{tabular}
\label{ShiftedGMRES-Proj}
\end{center}
\end{table}
%\vspace{.10in}
%\begin{center}
%\textbf{GMRES-Proj-Sh for the second and subsequent right-hand sides}
%\end{center}
%\begin{enumerate}
% \item Consider the systems with the $j$th right-hand side (and with all $ns$ shifts). 
% \item At the beginning of a cycle of GMRES-Proj-Sh, assume the current problem is 
%$(A-\sigma_i I) (x^j_i- \tilde x^j_{0,i}) = \beta_i r_{0,i},$ with $\beta_1 = 1$, and where $\tilde x_{0,i}$ is the current approximate solution to the i$th$ shifted system.
% \item Apply the Minres Projection for $V_k$ to the first right-hand side.  This uses the $V_{k+1}$ and $\bar H_k$ matrices developed while solving the first right-hand side with GMRES-DR-Sh.  
% \item  For shifted systems $is=2 \ldots ns$, solve $(H_k - \sigma_i I) d_i = \beta_i (H_k - \sigma_1 I)d_1$.
% \item Apply one cycle of GMRES(m)-Sh.
% \item Test the residual norms for convergence (can also test during the GMRES cycles).   For the non-base systems, can ignore the error term in the direction of $v_{k+1}$. %need to be more specific:  this is not right yet: $||(\bar H_k -\sigma_i \bar I) d_i||  
%If not satisfied, go back to step 2.
% \item Correction phase: Suppose the computed solution for the $i$th shifted system so far is $x^j_i$.  Let the solution to the system with the extra right hand side $v_{k+1}$ and shift $\sigma_i$ be $s_i$.  The corrected solution is $(x^j_i)^{corrected} = x^j_i + (v_{k+1}^T r) s_{i}$.  The corrected residual norm can be calculated.
%\end{enumerate} 
%\vspace{.15in}

%Need to change item number 2 to 7:
The algorithm for solution of the systems with extra right-hand sides is given in Table \ref{ShiftedGMRES-ProjX}.
\begin{table}[t!]
\caption[GMRES-Proj-Sh for the Extra Right-Hand Side $v_{k+1}$]{ALGORITHM :: GMRES-Proj-Sh for the extra right-hand side $v_{k+1}$}
\vspace{-12pt}
\begin{center}
\begin{tabular} {cl} \hline \\
   & Same as for previous algorithm except for $\dots$ \\
1. & Consider the systems with right-hand side $v_{k+1}$ (and with all $ns$ shifts). \\
2. & Correction phase: Suppose the computed solution for the $i$th shifted system \\
   & so far is $\bar s_i$. The corrected solution is  \\
   & $s_i = \bar s_i + ({ 1\over {1-\gamma_i}})*\bar s_{i}$, with $\gamma_i = v_{k+1}^T r.$ \\
\\
\hline
\vspace{-34pt}
\end{tabular}
\label{ShiftedGMRES-ProjX}
\end{center}
\end{table}

{\it Example 2.}  We use the same test matrix.  All right-hand sides are generated randomly.  The systems with the first right-hand sides are solved with GMRES-DRS(25,10) as before.  Then the extra right-hand $v_{k+1}$ systems are solved (for all shifts) with GMRES(15)-Pr(10)-Sh.  Finally, the second right-hand side systems are also solved with GMRES(15)-Pr(10)-Sh.  All relative residual tolerances are $rtol=1.0 \times 10^{-6}$.  Figure (\ref{fig:gsf3}) has residual curves for only two shifts, the base shifts of zero and $\sigma = -2$.  The dotted line shows the uncorrected residual norm for the second shift, while the dash-dot line has the second shift residuals if they are corrected (actually the correction needs to be done only once at the end of each right-hand side).  The uncorrected residual norm for the second shifted system levels off at $4.0 \times 10^{-3}$, but this is fixed by the correction phase.  The convergence is faster than for GMRES-DRS, because the eigenvectors are used from the beginning to speed up the convergence.  Also the cost of GMRES(15)-Proj(10)-Sh is less than for GMRES-DRS(25,10), because it is fairly inexpensive to project over the approximate eigenvectors compared to keeping the eigenvectors in the GMRES-DR subspace.  Here the expense for the extra right-hand side is fairly significant, however it will not be if there are more right-hand sides.  Figure (\ref{fig:gsf3b}) has the case of solving a total of 10 right-hand sides.  Also, the extra right-hand side is solved only to relative residual tolerance of $1.0 \times 10^{-3}$.  Now the expense for the extra right-hand side $v_{k+1}$ is small compared to the amount saved by speeding up the solution for all the remaining right-hand sides.
{\setlength{\abovecaptionskip}{0mm}
\setlength{\belowcaptionskip}{0mm}
\begin{figure}[h!]
\centering
\includegraphics[scale=1, height=8cm, width=12cm]{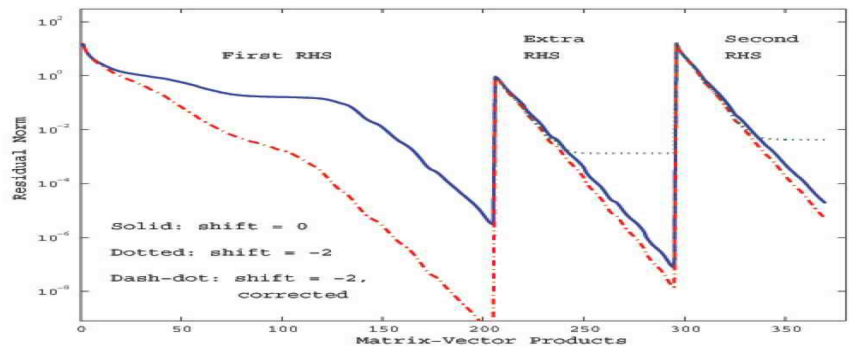}
\caption{Solution of first rhs, extra rhs and second rhs with two shifts.}
\label{fig:gsf3}
\end{figure}}
%\begin{figure}
%\vspace{.10in}
%\special{eps:c:/PCTeXv4/FIGURES/gpf3.eps x=5in y=3in}
%\centereps{4.2in}{2.8in}{gsf3.eps}
%\vspace{.10in}
%\caption{Solution of first rhs, extra rhs and second rhs with two shifts.}
%\end{figure}

%(............. Note since am using thicker lines in gsf3, need thicker in earlier ones also...............)
{\setlength{\abovecaptionskip}{0mm}
\setlength{\belowcaptionskip}{0mm}
\begin{figure}[h!]
\centering
\includegraphics[scale=1, height=8cm, width=12cm]{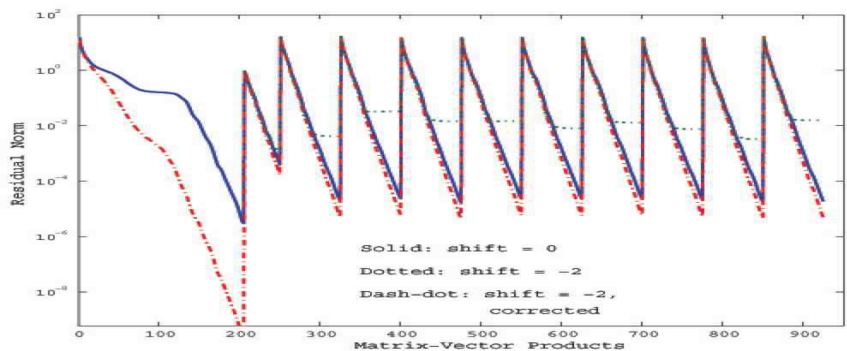}
\caption{Solution of first rhs, extra rhs and ten rhs with two shifts.}
\label{fig:gsf3b}
\end{figure}}
%\begin{figure}
%\vspace{.10in}
%\special{eps:c:/PCTeXv4/FIGURES/gpf3.eps x=5in y=3in}
%\centereps{4.2in}{2.8in}{gsf3b.eps}
%\vspace{.10in}
%\caption{Solution of first rhs, extra rhs and second rhs with two shifts.}
%\end{figure}

{\it Example 3.}
At the end of the previous example, the extra right-hand side is solved to low accuracy, but the correction for the subsequent right-hand sides is still successful.
We now experiment with solving the extra right-hand side to different levels of accuracy.  Table \ref{tbl:RHS} shows the accuracy after correction for the $\sigma=-2$ system when the extra right-hand side system is solved to relative residual tolerances ranging from $1.0 \times 10^{-6}$ down to $1.0 \times 10^{-1}$ (the tolerance is checked for termination only at the end of GMRES cycles).  The first and second right-hand side systems are solved to three different residual norm tolerances ($1.0\times 10^{-6}$, $1.0\times 10^{-8}$ and $1.0 \times 10^{-10}$) in the three rows of the table.  The conclusion of this experiment is that the extra right-hand side systems do not need to be solved very accurately.  With tolerances $1.0 \times 10^{-6}$ and $1.0\times 10^{-8}$ for the first and second right-hand sides, the extra right-hand side systems need only to be solved to a relative tolerance of $1.0\times 10^{-3}$ for essentially full accuracy.
\begin{table}
\caption{Effect of solving the extra right-hand side system to different accuracies} 
\vspace{-12pt}
\begin{center} \footnotesize
\begin{tabular}{ccccccccc}  \hline
desired rtol  	& accurracy of $2$nd    &   & & & & &  \\ 
of $1$st and $2$nd sys's	 & before correction & 1.e-6  & 1.e-5  &  1.e-4  & 1.e-3  & 1.e-2 & 1.e-1  \\   
\hline 
1.e-6 & 4.2e-3  & 4.8e-6  & 4.8e-6    & 4.8e-6  & 4.9e-6  & 6.4e-6 & 3.8e-4 \\ 
1.e-8 & 3.6e-4  & 2.4e-8  & 2.4e-8    & 2.4e-8  & 2.5e-7  & 9.4e-7 & 9.9e-5 \\ 
1.e-10 & 1.2e-3 & 1.5e-10 & 2.7e-10   & 1.0e-9  & 9.7e-8  & 2.7e-7 & 3.1e-5 \\ 

\hline
\vspace{-34pt}
\end{tabular} 
\end{center} 
\label{tbl:RHS}
%\label{} 
\end{table}  

{\it Example 4.}
We look at a Wilson-Dirac matrix from lattice QCD. The dimension is 393,216 by 393,216.  The value of $\kappa$ is 0.158 for the base shift.  This is approximately $\kappa_{critical}$.  The right-hand sides are unit vectors associated with particular space-time, Dirac and color coordinates.  Often there are a dozen or more right-hand sides associated with each matrix and perhaps seven shifts for each right-hand side.  We will just show solutions of the second right-hand side for three shifts, $\sigma = 0, -0.3, -0.5$.
The first right-hand side is solved with GMRES-DRS(50,30) to a residual tolerance of 1.e-8 and the extra right-hand side to $1.0\times 10^{-7}$.  Then for the second right-hand side, GMRES-Proj uses 30 approximate eigenvectors for the projection in between cycles of GMRES(20).  See Figure (\ref{fig:gsf5}) for the results.  GMRES(20)-Proj(30)-Sh can converge in about one-tenth of the iterations needed for GMRES(20).  To reach a residual norm of less than $10^{-7}$ for the toughest system with shift of zero takes 2680 matrix-vector products for GMRES(20)-Sh and 280 for GMRES(20)-Proj(30)-Sh.   
%(.... say something about eigenvalue distribution or refer to other paper(s)....  what else is needed?....)

%(check if figure has been changed to GMRES(20)-Sh, etc. .........................)
{\setlength{\abovecaptionskip}{0mm}
\setlength{\belowcaptionskip}{0mm}
\begin{figure}[b!]
\centering
\includegraphics[scale=1, height=8cm, width=12cm]{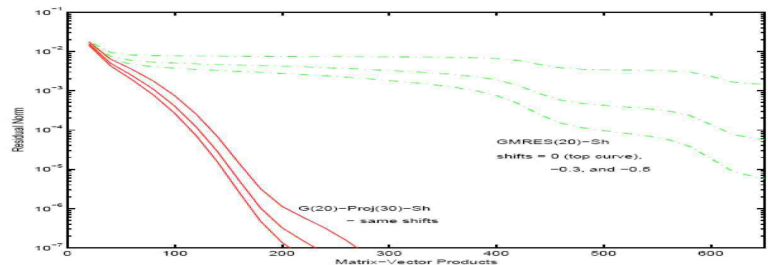}
\caption{Solution of second RHS for large QCD matrix with three shifts.}
\label{fig:gsf5}
\end{figure}}

\section{Projection Methods for tmQCD}

To solve shifted system of equations simultaneously each shifted system of equations must use the same Krylov subspace. In tmQCD at maximal twist, the matrix $M$ changes for each shift due to necessary even-odd preconditioning of the problem. Since the matrix changes for every shifted system, the Krylov subspace used to solve the base system will not work for the following shifted systems. So, simultaneous shifted Krylov methods do not work for the tmQCD formalism and the systems must be solved serially. Preliminary results of a new method that help convergence of the subsequent shifted systems using a projection over solutions are presented in this section.
\enlargethispage{\baselineskip}

The solution vectors of the shifted system of equations  $(M - \sigma_{i}I)$ for $i=1, ...,j-1$ can help the convergence of a shifted system $(M - \sigma_{j}I)$ where $j > i$. To help the convergence of the $j^{th}$ system of equations a MinRes projection over the previous $j-1$ solution vectors is used to create an initial guess $\tilde{x_{o,j}}$ for the current system. In this method, we solve the most difficult system of equations last. In doing so, we take advantage of the projection over all the previous solution vectors. 

Again, the problem referenced in Example 2 is explored. The six shifted systems in Figure (\ref{fig:tmshifts}) correspond to the shifts $\sigma_{i}=\left\{-0.05,-0.04,-0.03,-0.02,-0.01,0 \right\}$ for $i=1,...,6$.

{\setlength{\abovecaptionskip}{0mm}
\setlength{\belowcaptionskip}{0mm}
\begin{figure}[h!]
\centering
\includegraphics[scale=1, height=8cm, width=10cm]{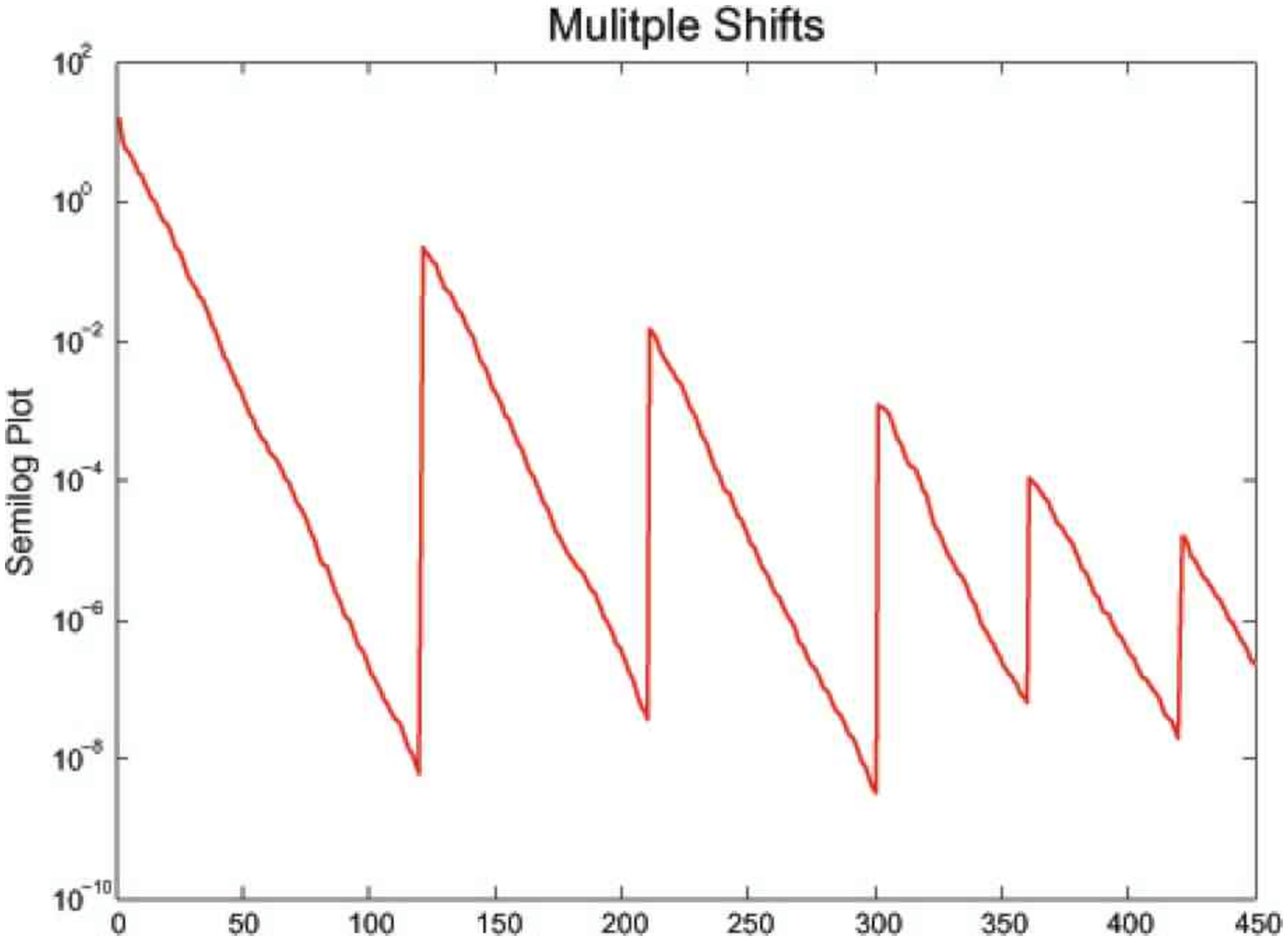}
\caption{Residual vector for serial shifted systems as a function of MVP's for GMRES(30)}
\label{fig:tmshifts}
\end{figure}}

The initial base system is solved with GMRES(30). It is obvious from Figure (\ref{fig:tmshifts}) that the convergence is helped for each successive RHS. The first system of equations required approximately 120 matrix-vector products to form a residual of $10^{-7}$. In contrast, the last system of equations (which is the most difficult shifted system) needed approximately 50 matrix-vector products. Next, the same example using the same shifts is repeated for GMRES-DR(30,10). 

{\setlength{\abovecaptionskip}{0mm}
\setlength{\belowcaptionskip}{0mm}
\begin{figure}[h!]
\centering
\includegraphics[scale=1, height=8cm, width=10cm]{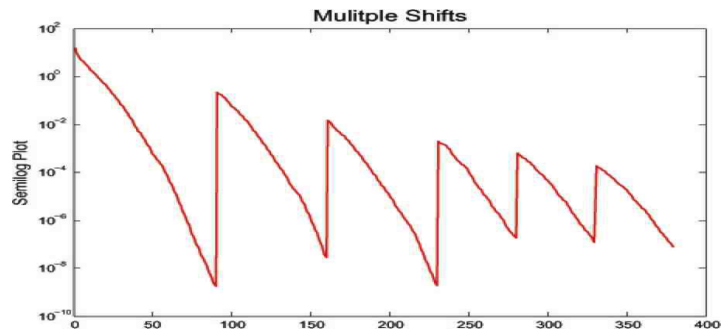}
\caption{Residual vector for serial shifted systems as a function of MVP's for GMRES-DR(30,10)}
\label{fig:tmdrshifts}
\end{figure}}

In Figure (\ref{fig:tmdrshifts}) the total number of matrix-vector products to solve all of the shifted systems is approximately 380. The total number of matrix-vector products (MVP) in Figure (\ref{fig:tmshifts}) is approximately 450. The saved MVP are a result of deflation and the projection over the previous solution vectors. In future work, this method will be extended for multiple RHS vectors using GMRES(m)-Proj(k). 
\chapter{Disconnected Sea Quarks}\label{chapt:discon}

Disconnected loop calculations have historically been a challenging problem for hadronic physics. Exact 
calculations of light quark matrix elements at each lattice point is extremely expensive computationally and currently not realistic with our current computer resources. An alternative to the exact calculation is to utilize an unbiased, stochastic estimate of the loops ~\cite{Wnoise,Liunoise,Thron1}. 

This technique uses noise theory to project out the loop operator expectation values. A continuing challenge with the noise methods is to reduce the variance of the calculation such that a stronger signal is acquired. By reducing the variance, the computational costs also decrease for the operator calculation. Higher order subtraction results are presented in this chapter as well as preliminary subtraction results for a twisted perturbative subtraction technique.
 
\section{Noise Theory}

The disconnected loops can be described by the systems of equations
\begin{equation}
 Mx = \eta 
\end{equation}
where $M$ is the quark matrix, $x$ is the solution vector and $\eta$ is a noise vector that is used to project the disconnected signal. The constraints on the system of equations are
\begin{equation}
< \eta_{i}> = 0 , <\eta_{i}\eta_{j}>=\delta_{ij},
\end{equation}
 where the average is over all noises used. Using these identities, a particular inverse matrix element $M^{-1}_{ij}$, can be determined by
\begin{eqnarray}
 <\eta_{j}x_{i}> &=& \sum_{k} M^{-1}_{ik}<\eta_{j}\eta_{k}> \\
                 &=& M^{-1}_{ij}.
\end{eqnarray}

At this time it is instructive to review the basic principals of matrix inversion using noise theory ~\cite{Wnoise,Thron1}. The expressions for the expectation value and variance of a matrix in terms of a general noise vector are found in Ref. ~\cite{Thron1}. 

Let the average of the projection of one general noise vector onto the other be 
\begin{equation}
X_{mn} \equiv \frac{1}{N} \sum_{k=1}^{N}\eta_{mk} \eta_{nk}^{\ast},
\end{equation}
for $(m,n=1, ... ,D)$, where $D$ is the dimension of the matrix and $(k=1,...,N)$, where N is the number of noises used. The matrix $X_{mn}$ is hermitian with expectation value $<X_{mn}> = \delta_{mn}$ as above. Using this notation, the variance of the measurement is defined to be

\begin{eqnarray}\label{var}
V[Tr{QX}] &\equiv & <|\sum_{m,n}q_{mn}X_{mn} - Tr{Q}|^{2}> \\
          &=& \sum_{n}<|X_{nn} - 1|^{2}><|q_{nn}|>^{2} \nonumber \\
          &+& \sum_{m \not =n}(<|X_{mn}|^{2}|q_{mn}|^{2} + q_{mn} q_{nm}^{\ast}<(X_{mn})^{2}>). \nonumber
\end{eqnarray}

\subsection{Real Z(2) Noise}

The $Z(2)$ noise constraints are
\begin{equation}
<|X_{mn}|^{2}> = \frac{1}{N}, <(X_{mn})^{2}> = \frac{1}{N}.
\end{equation}
for $m \not =n$.

Allow the matrix, $QX$, to be real. Now consider equation (\ref{var}), when we apply the constraints for $Z(2)$ noise and notice  $<|X_{nn} - 1|^{2}> = 0$. We may write the variance as 
\begin{equation}\label{varz2}
V[Tr{QX_{real}}] = \frac{1}{N} \sum_{m \not= n}(|q_{mn}|^{2} + q_{mn}q^{*}_{nm}).
\end{equation}

Therefore, by equation (\ref{varz2}), $Z(2)$ noise has the lowest variance of any real noise. This implies that the variance for $Z(2)$ noise is a result of the off-diagonal matrix elements. 

\subsection{General $Z(N)$ Noise}

For general $Z(N)$ noise ($N \geq 3$) we have a different set of constraints,
\begin{equation}
<|X_{mn}|^{2}> = \frac{1}{N}, <(X_{mn})^{2}> = 0
\end{equation}
for $m \not= n$.  Similarly to the real case, we have that  $<|X_{nn} - 1|^{2}> = 0$. Thus, the expression for the variance becomes
\begin{equation}\label{Znvariance}
V[Tr{QX_{Z(N)}}] = \frac{1}{N} \sum_{m \not=n}|q_{mn}|^{2}. 
\end{equation} 
For a general matrix $Q$, the variance relationship of Z(2) and Z(N) is not fixed. The difference in the variances is due to that fact that the square of an equally weighted distribution, as is the case for $Z(2)$, is not itself always uniformly distributed. In contrast, the square of the uniformly weighted $Z(N)$ for $N>3$ is uniformly distributed.  Even so, if the phases of $q_{mn}$ and $q_{mn}^{*}$ are not correlated, the variances for $Z(2)$ and $Z(N)$ ($N>3$) are approximately the same. For the operators that we calculate this appears to be the case ~\cite{Wnoise}.

\section{Perturbative Subtraction}
Perturbative noise subtraction gives a computationally efficient and effective way to reduce the variance of disconnected operators by using noise theory methods ~\cite{Wnoise}. A review of the methodology in reference ~\cite{Wnoise} is useful for our twisted mass formalism.

The trace of a matrix is obviously invariant under addition of another traceless matrix. Given two matrices, $Q$ and $\tilde{Q}$, the expectation values are related by
\begin{equation}
<Tr\{ (Q - \tilde Q)X \} >  = <Tr \{ QX \} >,
\end{equation}
where $ \tilde Q$ is traceless. The variance, however, is not invariant under the addition of the traceless matrix $\tilde Q$:
\begin{equation}\label{variance}
V[Tr\{ (Q - \tilde Q)X] = <|\sum_{m\not=n}(q_{mn} - \tilde q_{mn})X_{mn} - TrQ|^{2}>.
\end{equation}
The variance is completely determined by the off-diagonal elements of $Q$ and $\tilde{Q}$. The variance can be minimized if the off-diagonal elements of $Q$ and $\tilde Q$ are similar. It is important to find a $\tilde{Q}$ matrix that is traceless because we only want to reduce the variance of $Q$ not the diagonal elements that contribute to the disconnected loop expectation values. 

% For lattice QCD calculations the correct choice for $\tilde Q$ is the perturbative expansion of the quark matrix. The elements of the Wilson matrix can be written as 
The Wilson matrix  can be written as
\begin{equation}\label{Wilsonmatrix}
(M^{-1})_{IJ} = \frac{1}{\delta_{IJ} - \kappa P_{IJ}},
\end{equation}
where the capital indices are over space, color and Dirac indices ($I,J = \{ x,a,\alpha \}$). For Wilson fermions, the matrix elements $P_{IJ}$ are
\begin{equation}
P_{IJ} = \sum_{\mu}[(1-\gamma_{\mu})U_{\mu}(x)\delta_{x,y-a_{\mu}} + (1+\gamma_{\mu})U^{\dagger}_{\mu}(x-a_{\mu}\delta_{x,y+a_{\mu}})].
\end{equation} 
Expanding equation (\ref{Wilsonmatrix}) in a geometric series in the hopping parameter $\kappa$, one can write the perturbative Wilson quark matrix as
\begin{equation}\label{Mpert}
M^{-1}_{pert}(P) = I + \kappa P + \kappa^{2}P^{2} + \kappa^{3}P^{3} + ... .
\end{equation}
To reduce the variance of the weak matrix elements expectation values, $<\eta_{j}M^{-1}_{ik}\eta_{k}>$, it is natural to choose $\tilde{Q}$ to be the perturbative quark matrix $M_{pert}(P)$. According to equation (\ref{variance}), the variance is calculated from the off-diagonal elements of $Q$. With the perturbative Wilson matrix as our choice of $\tilde{Q}$, we calculate the difference of the expectation values of the Wilson quark matrix, $<\eta_{j}M^{-1}_{ik}\eta_{k}>$, and the perturbative matrix $<\eta_{j}M^{-1}_{pert}(P)_{ik}\eta_{k}>$ to reduce the variance of the disconnected loop operator. However, the perturbative quark matrix that we used to reduce the variance is, in fact, not traceless and therefore the diagonal elements of $M^{-1}_{ik}$ are changed by the subtraction of $M^{-1}_{pert}$. 

In general, the expectation value of an operator $O$ is 
\begin{equation}\label{loopop}
<\bar \psi O \psi> = -Tr(OM^{-1}). 
\end{equation}
The subtraction that changed the variance of the loop operators also changed the diagonal elements of the $M$ matrix. These values contribute to the vacuum expectation value and need to be added back in to get the full, unbiased answer. One way to calculate the diagonal elements is to explicitly construct all the gauge invariant paths that contribute to a given operator. 

%An alternative approach is to use a statistical method to determine the perturbative trace. This is the approach taken in this thesis.  This method is referred to as a ``statistically unbiased method" in reference ~\cite{Wnoise}.

Only closed loop, gauge invariant objects contribute to the trace in equation (\ref{loopop}). The local operators require a perturbative correction starting at $4^{th}$ order and non-local, vector operators need corrections starting at $3^{rd}$ order in $\kappa$. (The vector operator also requires a correction at zeroth order in $\kappa$. ) Another way to view the perturbative trace is that only closed path objects with an area $A$ contribute to the trace in equation (\ref{loopop}). A general picture of the vector and scalar operators is in Figure (\ref{fig:loopfig}). The quark lines in this figure represent all possible quark propagators that add to the operators.

{\setlength{\abovecaptionskip}{0mm}
\setlength{\belowcaptionskip}{0mm}
\begin{figure}[h!]
\centering
\includegraphics[scale=1, height=8cm, width=10cm]{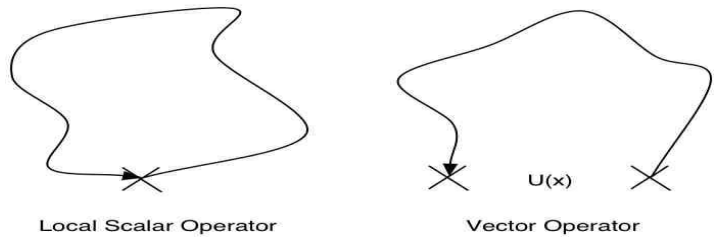}
\caption{General diagram of the quark line contributions to the flavor singlet scalar and vector operators.}
\label{fig:loopfig}
\end{figure}}
\pagebreak
The local scalar operators that contribute to the operator signal begin and end on the same lattice site. Vector operators are next-neighbor interactions that are connected with a gauge link $U(x)$. 

\subsection{Subtraction Methods}

To reduce the noise variance of the disconnected operators we need to subtract the inverse of the perturbative quark matrix $M^{-1}_{pert}$ from the quark matrix $M^{-1}_{IJ}$. In practice, we construct the perturbative matrix to a desired order in $\kappa$ with equation (\ref{Mpert}). 

The construction of the perturbative Wilson matrix in our program has a geometric interpretation. The program forms a hollow hypercube about the position of the current lattice site for each order. Table \ref{tbl:mdvalues} displays the number of steps from the original lattice site (at $O(\kappa)$) at which the hypercube is created as a function of $\kappa$. For example, for $O(\kappa^{2})$ a hollow hypercube of length one is created. The next order in $\kappa$ will make a hypercube one step farther out in all directions from the previous order, thus, expanding the size of the hypercube by one unit. This is represented in the transition between $O(\kappa^{2})$ and $O(\kappa^{3})$.
{\begin{table}[h!]
\caption{Dimension of the hollow-hypercubes as a function of $\kappa$. }
\vspace{-12pt}
\begin{center}
\begin{footnotesize}
\begin{tabular}{l cccccccc}
\hline \\
%Heading Line 1
       & $\kappa$ & $\kappa^{2}$ & $\kappa^{3}$ & $\kappa^{4}$ & $\kappa^{5}$ & $\kappa^{6}$ & $\kappa^{7}$ & $\kappa^{8}$ \\
\hline
steps  & 0        & 1            & 0            & 1            & 0            & 1            & 0            & 1            \\
       &          &              & 2            & 3            & 2            & $3^{*}$      & 2            & 3            \\
       &          &              &              &              & $4^{*}$      & $5^{*}$      & 4            & 5            \\
       &          &              &              &              &              &              & 6            & 7            \\  
\\
\hline
\vspace{-46pt}
\end{tabular}
\label{tbl:mdvalues}
\end{footnotesize}
\end{center}
\end{table}
}
Our program calculates the scalar loop value to $O(\kappa^{6})$ . According to Table \ref{tbl:mdvalues}, the corresponding hollow hypercubes that contribute are (1,3,5). For the perturbative VEV calculation the only hypercubes that contribute are those that form closed loop, gauge-invariant objects. We see that that hypercube of $O(\kappa^{5})$ that has been expanded four steps will not contribute to the sixth order VEV because it is not possible for the gauge-links on this surface to form a closed object with links that are three steps away. Therefore, for $O(\kappa^{5})$ no hypercube is constructed for the VEV calculation. Similarly,  for $O(\kappa^{6})$ hypercubes of length 3 and 5 are omitted for higher orders in kappa. (All of these values are marked with an asterisk.) However, in contrast to the VEV, in the noise calculation these contributions are retained because all objects that mimic the noise are included.
%For a specific lattice site, each perturbative order in kappa forms a hollow hyper-cubic shell about that lattice site. For example the perturbative quark matrix to first order in kappa is a hollow hypercube with sides of length $2a$, where $a$ is the lattice spacing.

Previous disconnected nucleon calculations have only used a subtraction method to reduce the variance of the vector and scalar operators to $O(\kappa^{3})$ and $O(\kappa^{4})$. In this thesis, the same calculation is done to $O(\kappa^{5})$ and $O(\kappa^{6})$, and can easily be extended to higher orders in $\kappa$.

To determine the perturbative quark propagators we solve the system of equations
\begin{equation}
M_{pert}x = b,
\end{equation}
where $b \in Z(2)$. Using a noise vector to solve this system of equations gives quark propagators of the off-diagonal (as well as diagonal) elements of $M^{-1}_{pert}$. The perturbative quark propagators from the off-diagonal elements of $M^{-1}_{pert}$ can have any open path up to a given $O(\kappa)$. These propagators are not gauge invariant and do not contribute to the operator signal. For example, non-gauge invariant propagator paths of $O(\kappa^{2})$ and $O(\kappa^{3})$ are shown in Figure \ref{fig:vector}.

{\setlength{\abovecaptionskip}{0mm}
\setlength{\belowcaptionskip}{0mm}
\begin{figure}[h!]
\centering
\includegraphics[scale=1, height=8cm, width=10cm]{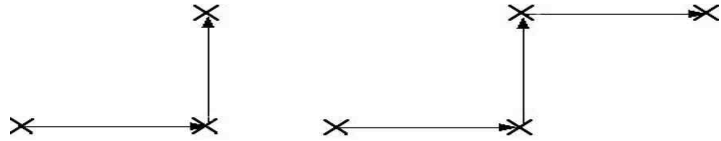}
\vspace{-48pt}
\caption{Perturbative noise contribution to $O(\kappa^{2})$ and $O(\kappa^{3})$ used to reduce the variance of the disconnected loop operators.}
\label{fig:vector}
\end{figure}}

% Physically, these non-gauge invariant objects represents quark propagators of $O(\kappa^{n})$ from surrounding lattice sites contributing to variance reduction at one specific, final lattice site. 
In practice, we find all the perturbative contributions at each lattice site and use that data to reduce the operator variance. 

\subsection{Vacuum Expectation Values}

The variance-reduction technique reduces the operator signal calculated with equation (\ref{loopop}). To correct for the loss of signal, the perturbative trace must be added back into the calculation. 
% An exact calculation of these operators would be expensive and impractical. An alternative solution is to use a Monte Carlo method to generate the "knocked down" matrix elements. Even the Monte Carlo method can become expensive computationally when higher orders of kappa are considered. 

A picture is instructive to determine which orders in the perturbative expansion contribute to the local and vector operators. The scalar and pseudoscalar operators are local to each individual lattice site. We wish to include all contributions that start and end on the same lattice site for these operators. As seen in figure \ref{fig:scalar2}, the orders in kappa which contribute to the local operators are $\kappa^{4}$ and $\kappa^{6}$. These are local gauge invariant contributions to the signal. 

{\setlength{\abovecaptionskip}{0mm}
\setlength{\belowcaptionskip}{0mm}
\begin{figure}[h!]
\centering
\includegraphics[scale=1, height=8cm, width=10cm]{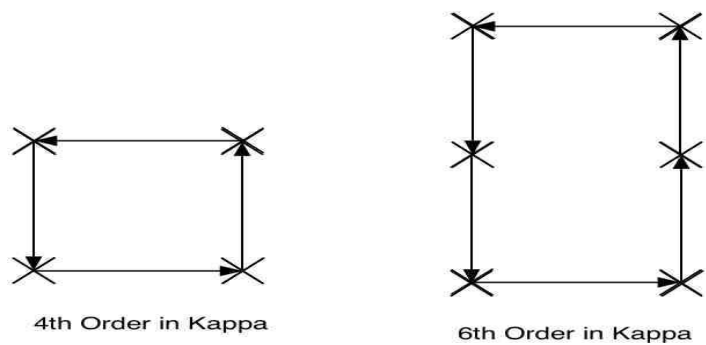}
\caption{Perturbative scalar operator contributions.}
\label{fig:scalar2}
\end{figure}}

In general, all even powers of $\kappa$ contribute to the local operators.

For the non-local operators, initial and final lattice sites are connected by a gauge-link $U(x)$. In this chapter it is understood that there is an implicit $\kappa$ multiplication in $U(x)$. Thus, in our calculations, contributions to the vector operators are of order $\kappa^{3}$ and $\kappa^{5}$. In Figure (\ref{fig:vector2}), the third order diagram is referred to as a staple. The fifth order diagram is referred to as a chair diagram. For the vectors, all odd-order terms in $\kappa$ contribute.

{\setlength{\abovecaptionskip}{0mm}
\setlength{\belowcaptionskip}{0mm}
\begin{figure}[h!]
\centering
\includegraphics[scale=1, height=8cm, width=10cm]{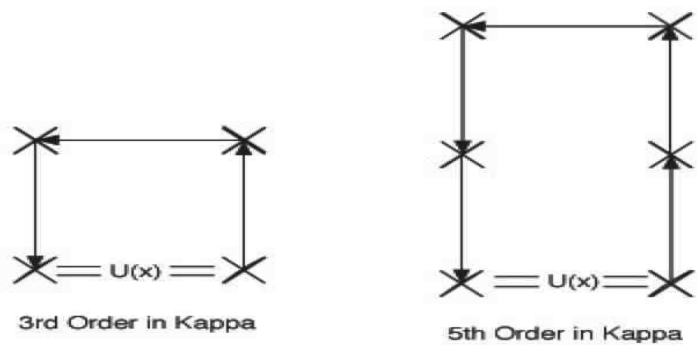}
\caption{Perturbative vector operator contribution.}
\label{fig:vector2}
\end{figure}}
\vspace{-12pt}
%For the vector operators, we gain an extra order in kappa. The Dirac projection operators are $P_{\pm}=(1 \pm \gamma_{\mu})$ where $P_{+}$ is the forward projection operator and $P_{-}$ is the backward operator.  A picture of a ``free order" chair object is in figure \ref{fig:freeorder}. Each propagator has an associated $P_{\pm}$ operator. The addition of the ``tail" in the figure does not affect the gauge invariance of the chair diagram. 
A similar chair diagram that contributes perturbatively is shown in figure \ref{fig:freeorder}. In this figure, a chair diagram has been constructed around a gauge link $U(x)$. At the final position of the object, a ``tail" is attached between this lattice site and a site that is adjacent. The ``tail" is a $O(\kappa^{6})$ contribution. This contribution is explicitly removed by our perturbative subtraction. We refer to this as reducing the variance ``for free" since there is no extra VEV calculation involved. 

{\setlength{\abovecaptionskip}{0mm}
\setlength{\belowcaptionskip}{0mm}
\begin{figure}[h!]
\centering
\includegraphics[scale=1, height=10cm, width=10cm]{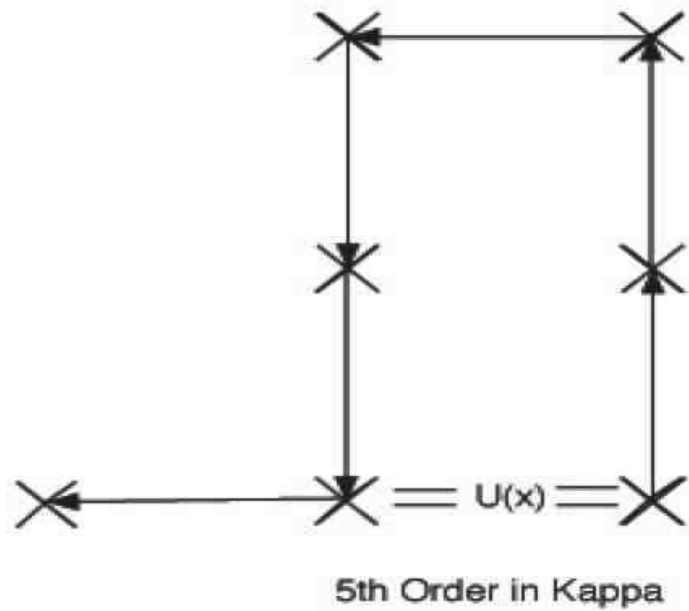}
\caption{Free Order in kappa for vector objects.}
\label{fig:freeorder}
\end{figure}}

The perturbative vacuum expectation value does not use a noise vector to solve the system of equations. Instead we solve the system
\begin{equation}
Mx = e_{i}, 
\end{equation}
where the vector $e_{i}$ is an element of the set of all unit vectors that span the Euclidean, color, and Dirac spaces. For each $e_{i}$, quark propagators are calculated that begin at this lattice site and ``spread out" to sites that are of $O(\kappa^{n})$ away. In our calculation we determine all the closed loop, gauge-invariant objects up to $O(\kappa^{6})$ that contribute to the operator signal automatically. This is more expensive than an explicit construction using gauge fields. However, the advantage of this method is that it may easily be extended to higher powers in $\kappa$. 

In our program, there are distinct differences between the noise subtraction part and the calculation of the perturbative VEV. The perturbative VEV constructs gauge invariant objects that contribute to the signal. To create these gauge invariant objects the propagator expands from the current lattice site and only constructs invariant, closed objects. The noise subtraction part, on the other hand, constructs all the quark propagators to a given order in kappa that contribute to the noise at each site. Hence, the difference between the perturbative VEV and the noise-subtraction method is that the VEV propagators are moving outward from a given lattice site to create objects that contribute to the signal, while the noise method uses all the contributions from every quark propagator ending at the same site. A diagram showing the distinction between these two processes is in Figure \ref{fig:vevsub}. 

{\setlength{\abovecaptionskip}{0mm}
\setlength{\belowcaptionskip}{0mm}
\begin{figure}[h!]
\centering
\includegraphics[scale=1, height=12cm, width=10cm]{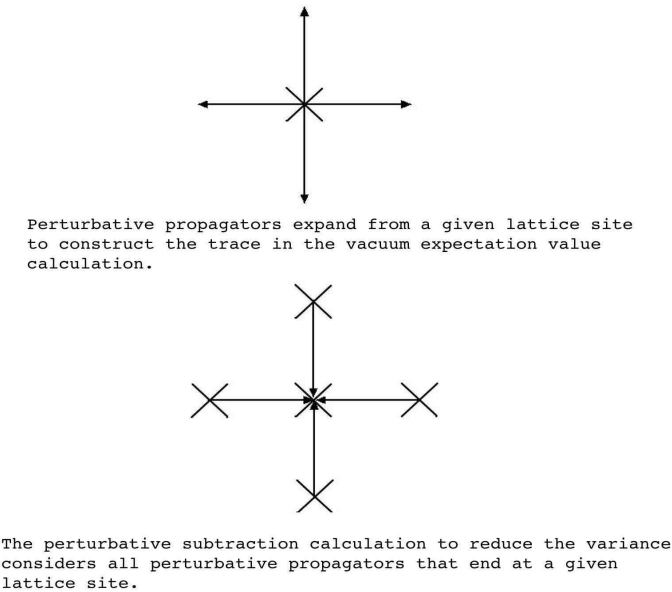}
\caption{Perturbative VEV and Subtraction Diagrams.}
\label{fig:vevsub}
\end{figure}}
\vspace{-12pt}
To create the perturbative VEV contributions mentioned above, a trace over Dirac and color indices at each lattice site is used for the perturbative VEV operator calculation. The trace is a result of the operator construction in (\ref{loopop}). In the noise-subtraction method, tracing would be incorrect because we wish to determine the contribution of the off-diagonal elements to the variance at each lattice site. Therefore, in the noise calculation we consider all the quark propagators ending at each lattice site as stated above. This information is then used to reduce the variance of the exact loop operator. \enlargethispage{\baselineskip}

Other perturbative methods have been proposed in references ~\cite{Micheal,SESAM}.

\subsection{Local Operators}
Since the disconnected calculation is delicate, it is always desirable to reduce the noise of the operators $O$. Each calculated operator has a real and imaginary part. However, for each operator in question, only the real or imaginary part is needed for the calculation. For the Wilson case, operator identities that determine whether the real or imaginary part contributes for each operator have been shown using the quark propagator identity $ S = \gamma_{5} S^{\dagger} \gamma_{5}$ in reference ~\cite{Wnoise}. The identities are (at each lattice site, x):
\begin{eqnarray}\label{disops}
Scalar &:& Re[\bar \psi(x) \psi(x)] \nonumber \\
Vector &:& Im[\bar \psi(x) \gamma_{\mu} \psi(x)] \nonumber \\
Axial  &:& Re[\bar\psi(x) \gamma_{5} \gamma_{\mu} \psi(x)] \\
Pseudoscalar &:& Re[\bar\psi(x) \gamma_{5}\psi(x)] \nonumber \\
Point-Split Vector &:& \kappa Im[\bar\psi (x+a_{\mu})(1+\gamma_{\mu})U^{\dagger}_{\mu}(x)\psi(x) - \bar\psi (x)(1-\gamma_{\mu})U_{\mu}(x)\psi (x+a_{\mu})] \nonumber \\
Tensor &:& Im[\bar\psi(x) \sigma_{\mu\nu}\psi(x) ]. \nonumber 
\end{eqnarray} 
In our calculation, the vector and scalar identities play an important role. Although these identities are only approximations, when a noise method is employed they allow the omitted part of each operator to be identified as noise and left out of the disconnected operator calculation. Ultimately, these identities reduce the variance in the form factor calculation. 
 
\section{Twisted Mass Disconnected Fermion Loops} 

Twisted mass fermions are becoming more popular in hadronic physics because they permit calculation of lower quark mass and $O(a)$ improvement is automatic in many quantities  ~\cite{Frezzotti:2003ni}. Ultimately, our goal is to use twisted fermions for the nucleon and disconnected operators to make realistic calculations of nucleon strange form factors. This section presents a method in which the perturbative subtraction method can be implemented to determine disconnected loop values using twisted fermions.

To make use of noise methods for twisted mass fermions we need to find an equivalent matrix to that in equation (\ref{Wilsonmatrix}). The twisted mass ferimonic matrix is 
\begin{eqnarray}\label{TMmatrix}
M_{TM} &=& I - 2i\mu \gamma_{5} \tau_{3} - \kappa P,
\end{eqnarray} 
where the matrix P is the same matrix used in the Wilson formulation. Let $\tan(\delta) = 2\kappa\mu$. The twisted fermion matrix can then be written as
\begin{eqnarray}
M_{TM} &=& I -i\gamma_{5}\tan(\delta) - \kappa P.
\end{eqnarray}
The inverse of this matrix can be shown to be
\begin{equation}\label{Mtpert}
M^{-1}_{TM} = e^{i\delta \gamma_{5}}\cos(\delta) (\frac{1}{1 - \kappa \cos(\delta) e^{i\delta \gamma_{5}}P}).
\end{equation}
In this form we can expand equation (\ref{Mtpert}) in a geometric series such that perturbative subtraction may be employed. The expansion of the twisted fermion matrix is 
\begin{eqnarray}\label{tmpert}
M^{-1}_{TM}(P) &=& \cos(\delta)e^{i\delta \gamma_{5}}(I + \cos(\delta) \kappa e^{i\delta \gamma_{5}} P + \cos^{2}(\delta ) \kappa^{2} e^{i\delta \gamma_{5}}Pe^{i\delta \gamma_{5}} \\
               &+& \cos^{3}(\delta ) \kappa^{3}e^{i\delta \gamma_{5}}Pe^{i\delta \gamma_{5}}Pe^{i\delta \gamma_{5}}P + ...) . \nonumber
\end{eqnarray}

The extension from the Wilson perturbative subtraction method to the twisted fermion method amounts to a chiral rotation of the original quark matrix. This expansion of the quark matrix is referred to as a left twisted mass expansion because the chiral twist projects onto the Wilson matrix from the left hand side. An equivalent expression can be formed by expanding the quark matrix such that the the rotation is applied from the right. These are numerically identical, but the right multiplication was found to be more expensive in terms of computational time. The extra time is a result of the rotation being done after each hypercube has been formed instead of a simple chiral rotation on the initial noise vector. 

The quark charge conjugation property in the full twisted mass formalism is $D_{u} = \gamma_{5} D^\dagger_{d} \gamma_{5}$ ~\cite{Frezzotti:2001ea}. Using this property we can show the local operators to have the same form as in equation (\ref{disops}) when averaged over ``tmU" and ``tmD" quarks. The calculation is different in that one has two flavors of quark propagators. In this formalism the charge conjugation property changes flavor as well as charge. For example, the scalar operator is only completely real when both flavors are considered in the trace, 
\begin{eqnarray}
O_{\bar{\psi}\psi} &=& \frac{1}{2}<\bar{\psi_{u}}\psi + \bar{\psi_{d}}\psi_{d}> \\
                   &=& \frac{1}{2}(M^{-1}_{u}  + M^{-1}_{d}).
\end{eqnarray}
It is important to realize that the $(u,d)$ subscripts on the quark propagators are not the physical up and down quark. Instead they are the unphysical twisted mass labels ~\cite{Abdel-Rehim:2006ve}. With this in mind, we now have two flavor doublets on the lattice. The doublets can be written as

\begin{equation}
\psi_{l} = (\frac{u}{d}),\space \psi_{h} = (\frac{s}{c}),
\end{equation}
where the subscripts $(l,h)$ are for light and heavy respectively. Each doublet is mass degenerate, thus the c quark is not the physical charm quark. Since this ``charmed quark" is not used in any of our quenched lattice calculations it is not necessary to include an explicit nondegeneracy in the doublet. This procedure is employed by the authors of ~\cite{Pena:2002wj}.

\section{Subtraction Results}

The first results of higher order subtraction in the twisted mass basis are presented in this section. In the Wilson case, for much heavier quark masses $(\kappa = .148)$, it has been shown that the effects of higher order subtraction can be dramatic for point-split vector, vector, and scalar operators ~\cite{Wnoise}. The estimated computer time used to do a perturbative subtraction calculation of the  disconnected operators is determined by the ratios of unsubtracted variance to the subtracted variance. \enlargethispage{\baselineskip}

% The currents all respond similarly to the subtraction methods. 
%     
% {\setlength{\abovecaptionskip}{0mm}
% \setlength{\belowcaptionskip}{0mm}
% \begin{figure}[h!]
% \centering
% \includegraphics[scale=1, height=8cm, width=13cm]{subratios.ps}
% \caption{Scalar Subtraciton to 10th order in $\kappa = .148$.}
% \label{fig:chargesub}
% \end{figure}}

The ratios of the variance were calculated for $\kappa^{4}$ and $\kappa^{6}$ with fifty twisted mass configurations to investigate the computational gains from the higher order subtraction. The hopping and twisted mass parameter for this calculation are $\mu=0.30$ and $\kappa=0.15679$, respectively. Each configuration uses the optimum number of noises in the determination of the perturbative quark matrix $M^{-1}_{TM}(P)$.  

The optimum number of noises to minimize the variance can be determined with the variances of the gauge configurations and noises, $V_{gauge}$ and $V_{noise}$, respectively ~\cite{Wnoise}. Given $N$-configurations and $M$-noises per configuration, the error bar on a given operator is 
\begin{equation}\label{sig}
\sigma = \sqrt{\frac{V_{noise}}{NM} + \frac{V_{gauge}}{N}}.
\end{equation} 
Clearly, equation \ref{sig} is minimized for $M=1$. This result can be modified to incorporate computational overhead. If it is assumed that there is an overhead associated with generating configuration and we assume a fixed amount of computer time for each configuration, then
\begin{equation}\label{nmin}
T = NM + G_{N}N,
\end{equation}
where $G_{N}$ is the time overhead for configuration generation. The minimization of equation (\ref{nmin}) gives
\begin{equation}
M = \frac{S_{noise}}{S_{gauge}}\sqrt{G_{N}},
\end{equation}
where $S_{noise}$ and $S_{gauge}$ are determined by their respective variances. In our calculation the ratio $S_{noise}/S_{gauge} \approx 1$ for the vectors, which results in an optimum number of noises of approximately 5. The number of noises was not optimized for the scalar but the vector operators.
%{\setlength{\abovecaptionskip}{0mm}
%\setlength{\belowcaptionskip}{0mm}
%\begin{figure}[h!]
%\centering
%\includegraphics[scale=1, height=8cm, width=13cm]{chargedensity2.ps}
%\caption{Charge Density Subtraciton to 6th order in $\kappa$.}
%\label{fig:chargesub}
%\end{figure}}
%By using higher order subtraction in the twisted mass basis one saves a factor of six in computer time. An equivalent statement is that the number of noises needed to produce a comparable result to unsubtracted noise method is reduced by a factor of six. These resutls for the other currents are given in the following tables. 
{\begin{table}[t!]
\caption{Sixth Order $\kappa$ Variance Ratios of Scalar Operators.}
\vspace{-12pt}
\begin{center}
\begin{footnotesize}
\begin{tabular}{l cc}
\hline \\

%Heading Line 1
     $Subtraction$      & $Scalar$    & $PseudoScalar$  \\
  
% \hline

%       $Unsbtracted$              &1.19298E-02 &1.51943E-02   \\
%       $Fourth Order (\kappa^{4})$&8.68650E-03 &1.04808E-02   \\
%       $Sixth  Order (\kappa^{6})$&8.40557E-03 &1.03756E-02   \\

\hline
%&&&&&&&&&&&&\\
% Variance Ratios
      $Fourth Order (\kappa^{4})$&1.9 &2.1   \\
      $Sixth  Order (\kappa^{6})$&2.0 &2.1   \\
\\
\hline
\vspace{-34pt}
\end{tabular}
\label{fig:ScalarOps}
\end{footnotesize}
\end{center}
\end{table}
}
{\begin{table}[t!]
\caption{Sixth Order $\kappa$ Variance Ratios of Vector Operators.}
\vspace{-12pt}
\begin{center}
\begin{footnotesize}
\begin{tabular}{l cccc}
\hline \\

%Heading Line 1
     $Subtraction$      & $Charge Density$ & $J1current$  & $J2current$     & $J3current$  \\
  
% \hline
%  
%       $Unsbtracted$              &5.26419E-03 &5.36814E-03 &5.24456E-03 &5.33668E-03  \\
%       $Fourth Order (\kappa^{4})$&2.35517E-03 &2.57696E-03 &2.68586E-03 &2.67192E-03  \\
%       $Sixth  Order (\kappa^{6})$&2.17420E-03 &2.40983E-03 &2.44170E-03 &2.52121E-03   \\

\hline
%&&&&&&&&&&&&\\
% Variance Ratios.
      $Fourth Order Ratio(\kappa^{4})$&5.0 &4.3 &3.8 &4.0  \\
      $Sixth  Order Ratio(\kappa^{6})$&5.9 &5.0 &4.6 &4.5   \\
\\
\hline
\vspace{-34pt}
\end{tabular}
\label{fig:VectoroPs}
\end{footnotesize}
\end{center}
\end{table}
}
% In tables \ref{fig:ScalarOps} and \ref{fig:VectoroPs} the first three rows represent the average error bar of a given operator for a specified subtraction level. The remaining rows are the ratio of the unsubtracted variance to the variance of the specified order in $\kappa$. 

In table \ref{fig:ScalarOps}, it is observed that the scalar operators do not respond as well to the perturbative subtraction as do the currents. This behavior is consistent with what is found in the Wilson case. The scalar operators gain a factor of 1.9 in computer time using fourth order subtraction. In comparison, when sixth order subtraction is used for the scalar operator a gain of approximately 2.0 is reported. This is an approximate 10 percent increase in algorithm speed.

According to table \ref{fig:VectoroPs}, using higher order subtraction one saves a factor of approximately five in computer time for the spatial currents. An equivalent statement is that the number of noises needed to produce a comparable result to unsubtracted noise method is reduced by a factor of five. The charge density responds better to the subtraction method than the spatial currents. The charge density operator saves a factor of approximately six in computer time. These results indicate an approximate 20 percent increase in algorithm speed from $O(\kappa^{4})$ to $O(\kappa^{6})$. 

{\setlength{\abovecaptionskip}{0mm}
\setlength{\belowcaptionskip}{0mm}
\begin{figure}[h!]
\centering
\includegraphics[scale=1, height=8cm, width=13cm]{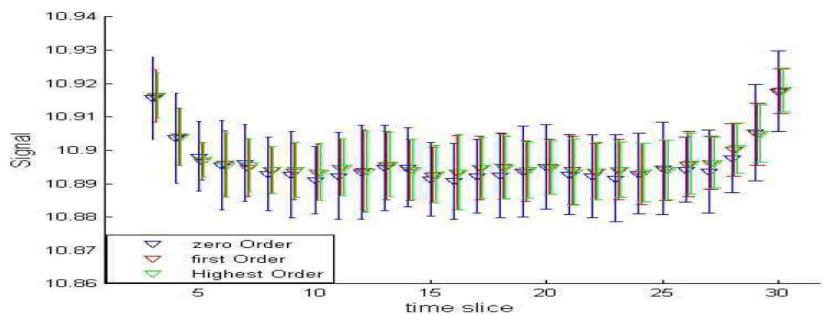}
\caption{Scalar Subtraction to 6th order in $\kappa$.}
\label{fig:scalarsub}
\end{figure}}

The scalar operator signal in figure \ref{fig:scalarsub} increases at time steps $1$ and $32$. This edge effect is a result of the non-periodic boundary condition in the time direction. Fortunately, these values are not used in the correlation function calculation and can be ignored. 

Similar subtraction diagrams for the psuedoscalar(Figure \ref{fig:pscalarsub}), charge density (Figure \ref{fig:chargesub}), and a spatial current (Figure \ref{fig:j1sub}) are below. These figures support the conclusion that the twisted vector operators respond better to subtraction methods than the scalar operators. \enlargethispage{\baselineskip}

{\setlength{\abovecaptionskip}{0mm}
\setlength{\belowcaptionskip}{0mm}
\begin{figure}[b!]
\centering
\includegraphics[scale=1, height=8cm, width=13cm]{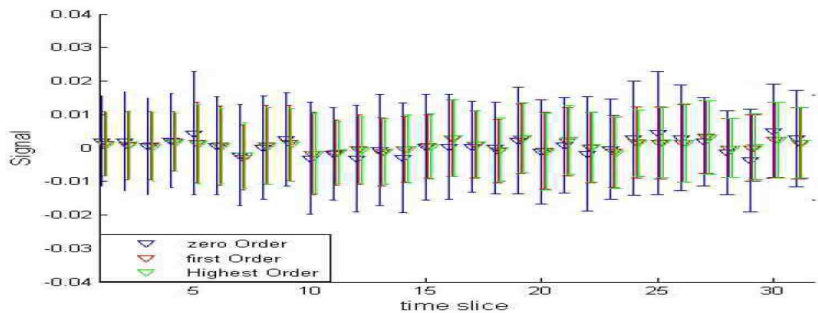}
\caption{PseudoScalar Subtraction to 6th order in $\kappa$.}
\label{fig:pscalarsub}
\end{figure}}

{\setlength{\abovecaptionskip}{0mm}
\setlength{\belowcaptionskip}{0mm}
\begin{figure}[t!]
\centering
\includegraphics[scale=1, height=8cm, width=13cm]{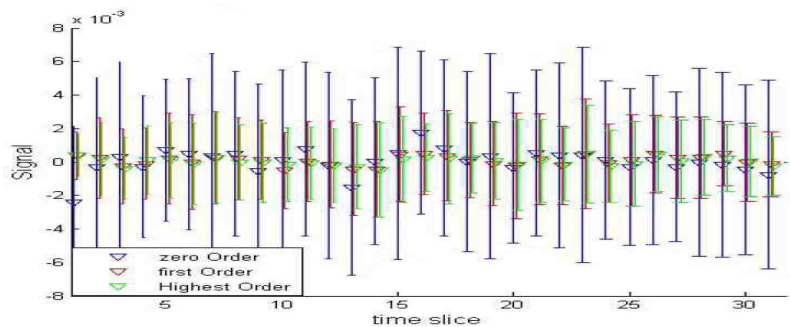}
\caption{Charge Density Subtraction to 6th order in $\kappa$.}
\label{fig:chargesub}
\end{figure}}

%{\setlength{\abovecaptionskip}{0mm}
%\setlength{\belowcaptionskip}{0mm}
%\begin{figure}[h!]
%\centering
%\includegraphics[scale=1, height=8cm, width=13cm]{j2current2.ps}
%\caption{Y-current Subtraciton.}
%\label{fig:j2sub}
%\end{figure}}
%
%{\setlength{\abovecaptionskip}{0mm}
%\setlength{\belowcaptionskip}{0mm}
%\begin{figure}[h!]
%\centering
%\includegraphics[scale=1, height=8cm, width=13cm]{j3currentfig2.ps}
%\caption{Z-current Subtraciton.}
%\label{fig:j3sub}
%\end{figure}}
In the tmQCD formalism it has been shown that the scalar-pseudoscalar and axial vector-vector operators mix ~\cite{frezzotti-2001-0108}. In this thesis the nucleons are calculated at maximal twist in which the physical constraint that there is no mixing between the charged psuedoscalar and vector is imposed to eliminate the axial-vector mixing ~\cite{abdel-rehim-2005-71}. However we have found that this does not eliminate the scalar-psuedoscalar operator mixing. Scalar mixing was observed by the authors of reference ~\cite{abdel-rehim-2005-71}. This mixing can be seen in equation \ref{tmpert}. The first term in the expansion is approximately 1 for a small rotation angle $\delta$. This guarantees that the scalar operator has a vacuum expectation value. So, independent of the maximal twist angle the scalar-pseudoscalar mixing will occur with this approach. Hopefully, other methods can be determined to remove the scalar-pseudoscalar mixing and promote a twisted disconnected noise method.\enlargethispage{\baselineskip}

{\setlength{\abovecaptionskip}{0mm}
\setlength{\belowcaptionskip}{0mm}
\begin{figure}[t!]
\centering
\includegraphics[scale=1, height=8cm, width=13cm]{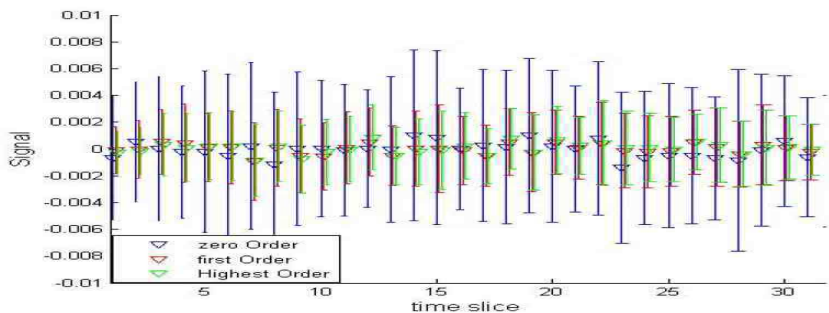}
\caption{$J_{1}$-current Subtraction.}
\label{fig:j1sub}
\end{figure}}

Using subtraction methods give a useful tool for exploring quark loops in the disconnected sector. In future disconnected nucleon calculations hopefully twisted perturbative subtraction can be used where scalar-pseudoscalar mixing has been removed to reach lighter quark masses.  
%OUTLINE :: (disconnected chapter)
%
%1) INTRO
%
%     a) Need to talk about operators being formed (Wilson) 
%     b) use of identity u = gamma5 d gamma5 ?? - This needs to be done above, and then shown for twisted operators here. 
%     c) results of the Wilson matrix subtraction....
%
%2) Formalism 
%
%     a) form the perturbative matrix
%     b) determine the twist angle. 
%
%3) resutls.  
%
%     a) show results for higher order subtraction, for all operators.
%     b) show comparalble results for twisted to Wilson.
% 

%
%\appendix
\chapter{Numerical Simulations and Results}
The gauge field configurations used in this study were generated from the unimproved Wilson gauge action at $\beta = 6.0$ on a $20^{3} \times 32$ lattice corresponding to a lattice spacing of
\begin{equation}
a = 0.1011(7) fm.
\end{equation}
as obtained from reference ~\cite{Gockeler:1997fn} from a physical string tension of $\sqrt{K}=$427MeV. This lattice spacing was used in the strangeness calculation in ~\cite{lewis-2003-67}. Our full ensemble of 200 configurations was produced from a thermalized Markov chain. Each ensemble configuration is generated with 2000 heatbath updates between saved configurations. 

The twisted mass lattice action at maximal twist was used to obtain four valence quark masses per configuration. Each mass has an associated hopping parameter $\kappa$ and twisted mass parameter $\mu$. These values are in Table (\ref{fig:twpairs}).

{\begin{table}[h!]
\caption{Maximally twisted mass pairs, ($\kappa$,$\mu$). }
\vspace{-12pt}
\begin{center}
\begin{footnotesize}
\begin{tabular}{l cc}
\hline \\
%Heading Line 1
 Mass Number     &     Hopping parameter, $\kappa$    & Twisted mass parameter, $\mu$  \\
 \hline 
% \hline

  1     & 0.15679 & 0.030  \\
  2     & 0.15708 & 0.015   \\
  3     & 0.15721 & 0.010   \\
  4     & 0.15728 & 0.005  \\
\\
\hline
\end{tabular}
\label{fig:twpairs}
\end{footnotesize}
\end{center}
\end{table}
}
\vspace{-34pt}
The valence quarks in our simulation are subject to Dirichlet time boundaries. The source is located at (1,1,1,4) which is four timesteps away from the boundary. 

Strangeness matrix elements are calculated using standard methods in which the three-point function is created by correlating a strange-quark loop with the nucleon propagators. The strange-quark loops are calculated with the perturbative subtraction techniques from chapter \ref{chapt:discon} with real $Z_{2}$ noise. The scalar loops in our calculation are determined to $O(\kappa^{6})$. The previous nucleon strangeness calculation employing this stochastic technique (Ref. ~\cite{lewis-2003-67}) was to lowest order subtraction, $O(\kappa^{4})$, from reference ~\cite{Thron-1998}. 

%The optimum number of noises to minimize the variance can be determined with the variances of the gauge configurations and noises, $V_{gauge}$ and $V_{noise}$, respectively ~\cite{Wnoise}. Given $N$ configurations and $M$ noises per configuration, the error bar on a given operator is 
%
%\begin{equation}\label{sig}
%\sigma = \sqrt{\frac{V_{noise}}{NM} + \frac{V_{gauge}}{N}}.
%\end{equation} 
%Clearly, equation \ref{sig} is minimized for $M=1$. This result can be modified to incorporate computational overhead. If it is assumed that there is an overhead associated with generating configuration and we assume a fixed amount of computer time for each configuration, then
%
%\begin{equation}\label{nmin}
%T = NM + G_{N}N,
%\end{equation}
%where $G_{N}$ is the time overhead for configuration generation. The minimization of equation \ref{nmin} gives
%\begin{equation}
%M = \frac{S_{noise}}{S_{gauge}}\sqrt{G_{N}},
%\end{equation}
%where $S_{noise}$ and $S_{gauge}$ are determined by their respective variances. In our calculation the ratio $S_{noise}/S_{gauge} \approx 1$, which results in an optimum number of noises of approximately 5. 

In our disconnected calculation we use loop values of $\kappa_{loop}=0.152$ and $\kappa_{loop}=0.154$. These kappa values correspond to vector meson masses of $912(8)$ MeV and $1066(4)$ MeV respectively ~\cite{Gockeler:1997fn,lewis-2003-67} which surrounds $1019$ MeV which ensures that our data will interpolate to a strange quark loop. In our lattice simulation the matrix elements are extracted from the ratio in equation (\ref{KTway}). A fixed loop background starting at the source and ending at time step 20 was used in these calculations. We consider the lowest five momentums given by,
\begin{equation}
a^{2}\vec{q}^{2} = n(\pi/10)^{2}, n = 0,1,2,3,4.
\end{equation}
The lowest three momentum are focused on in this chapter because the momentum associated with $n=3,4$ are still noisy and unpredictable due to a lack of configurations. 

% {\setlength{\abovecaptionskip}{10mm}
% \setlength{\belowcaptionskip}{0mm}
% \begin{figure}[h!]
% \centering
% \includegraphics[scale=1, height=8cm, width=12cm]{all_masses_152(4mom).ps}
% \caption{Fourth non-zero momentum scalar density diagram at $\kappa_{loop}$=.152.}
% \label{fig:4mom.152}
% \end{figure}}
Figures \ref{fig:mass1.152} - \ref{fig:mass4.154} plot the ratio of three to two-point functions for the lowest three momentums of all four masses in table (\ref{fig:twpairs}).

{\setlength{\abovecaptionskip}{0mm}
\setlength{\belowcaptionskip}{0mm}
\begin{figure}[t!]
\centering
\includegraphics[scale=1, height=8cm, width=12cm]{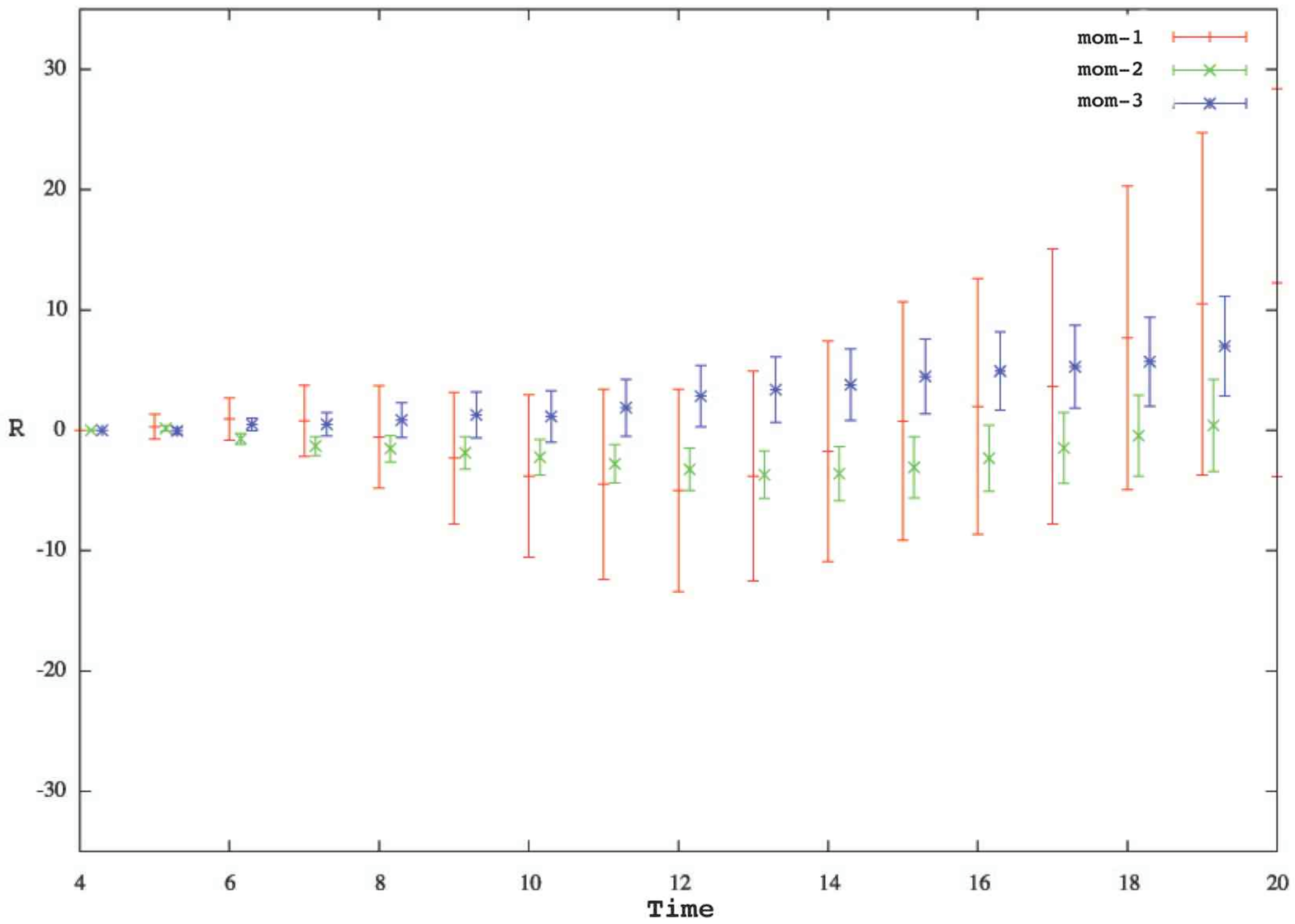}
\caption{Ratio in equation (\ref{KTway}) for the first three momenta for mass 1 scalar density diagram at $\kappa_{loop}$=.152.}
\label{fig:mass1.152}
\end{figure}}

{\setlength{\abovecaptionskip}{0mm}
\setlength{\belowcaptionskip}{0mm}
\begin{figure}[t!]
\centering
\includegraphics[scale=1, height=8cm, width=12cm]{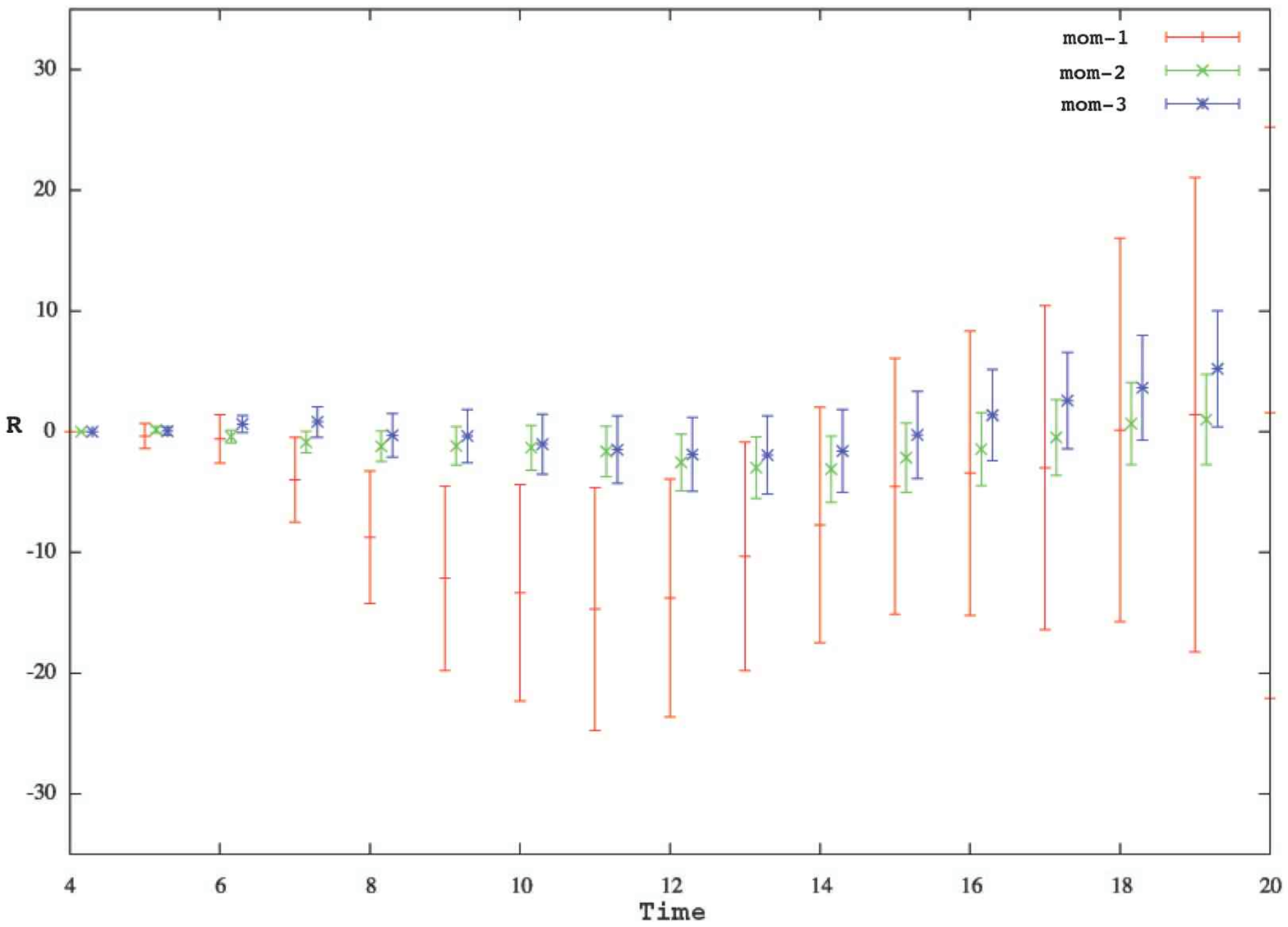}
\caption{Ratio in equation (\ref{KTway}) for the first three momenta for mass 1 scalar density diagram at $\kappa_{loop}$=.154.}
\label{fig:mass1.154}
\end{figure}}

{\setlength{\abovecaptionskip}{0mm}
\setlength{\belowcaptionskip}{0mm}
\begin{figure}[t!]
\centering
\includegraphics[scale=1, height=8cm, width=12cm]{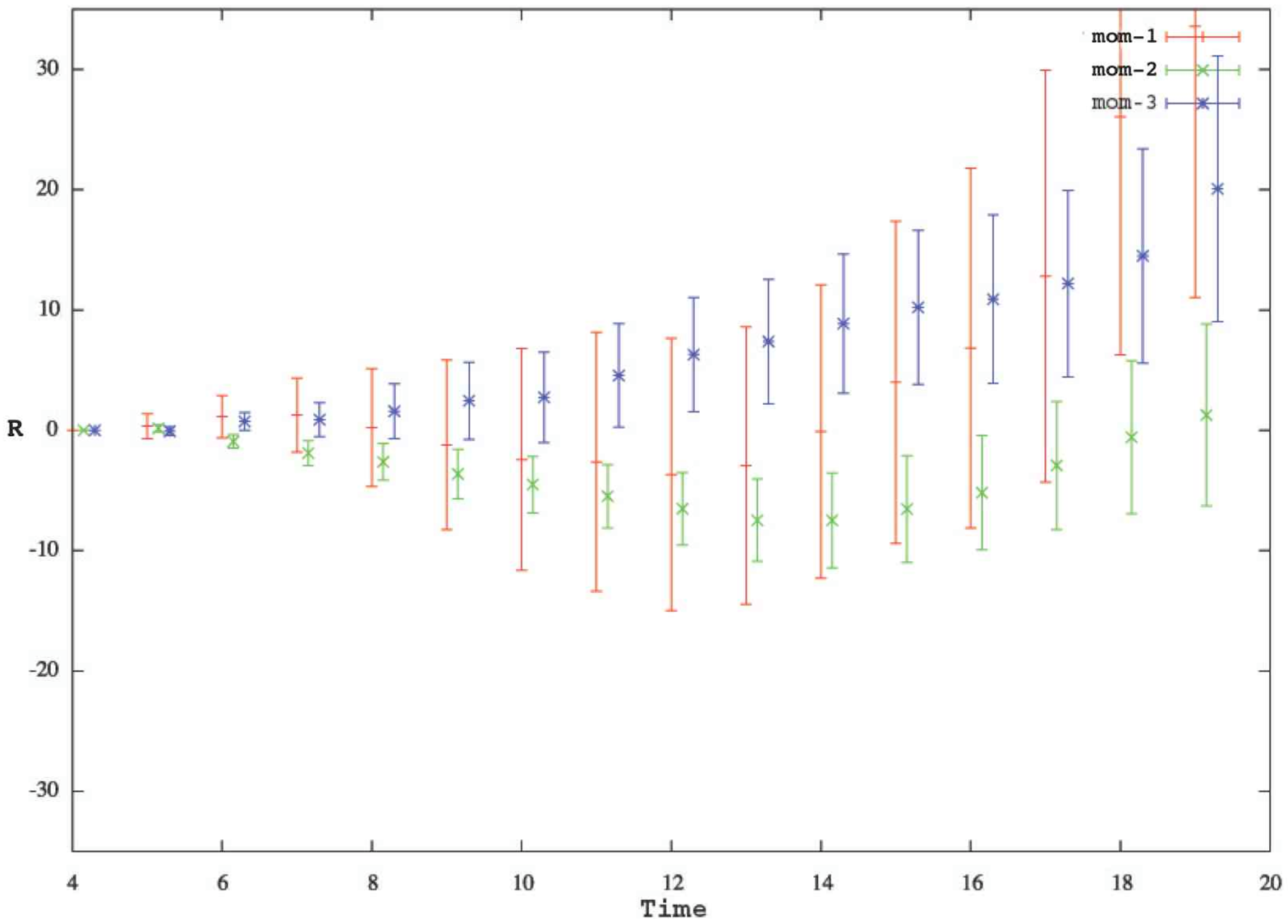}
\caption{Ratio in equation (\ref{KTway}) for the first three momenta for mass 2 scalar density diagram at $\kappa_{loop}$=.152.}
\label{fig:mass2.152}
\end{figure}}

{\setlength{\abovecaptionskip}{0mm}
\setlength{\belowcaptionskip}{0mm}
\begin{figure}[t!]
\centering
\includegraphics[scale=1, height=8cm, width=12cm]{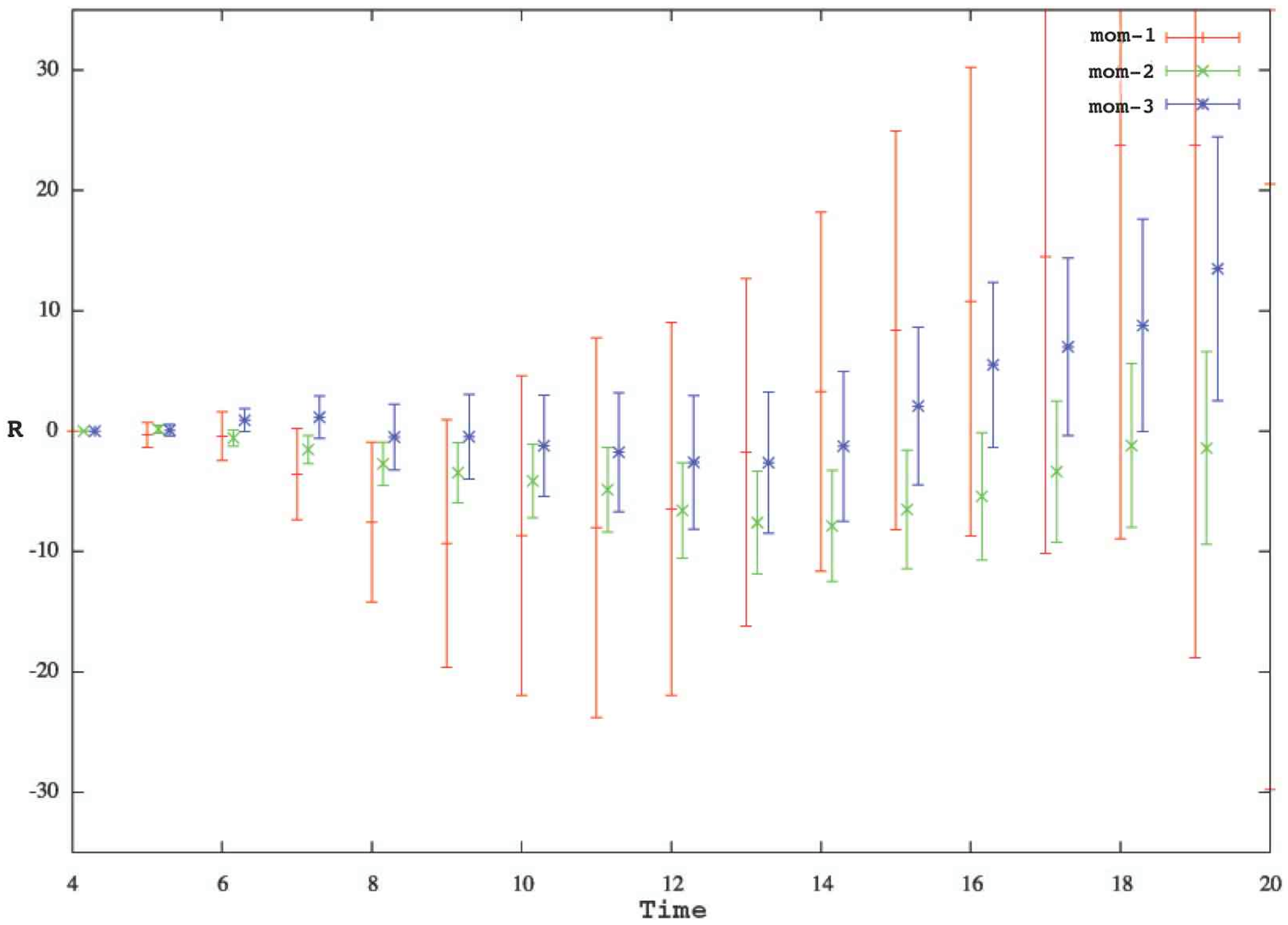}
\caption{Ratio in equation (\ref{KTway}) for the first three momenta for mass 2 scalar density diagram at $\kappa_{loop}$=.154.}
\label{fig:mass2.154}
\end{figure}}

{\setlength{\abovecaptionskip}{0mm}
\setlength{\belowcaptionskip}{0mm}
\begin{figure}[t!]
\centering
\includegraphics[scale=1, height=8cm, width=12cm]{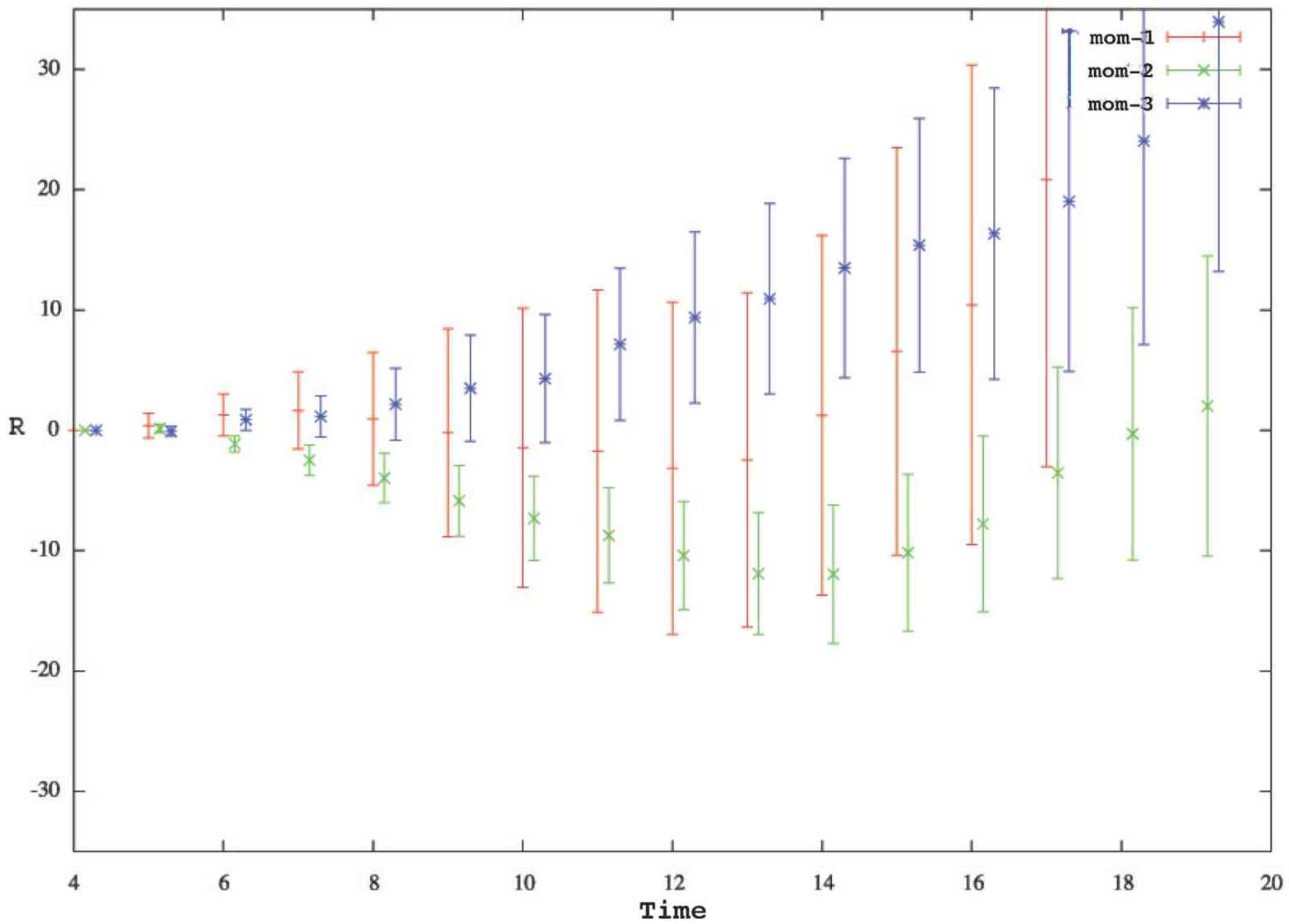}
\caption{Ratio in equation (\ref{KTway}) for the first three momenta for mass 3 scalar density diagram at $\kappa_{loop}$=.152.}
\label{fig:mass3.152}
\end{figure}}

{\setlength{\abovecaptionskip}{0mm}
\setlength{\belowcaptionskip}{0mm}
\begin{figure}[t!]
\centering
\includegraphics[scale=1, height=8cm, width=12cm]{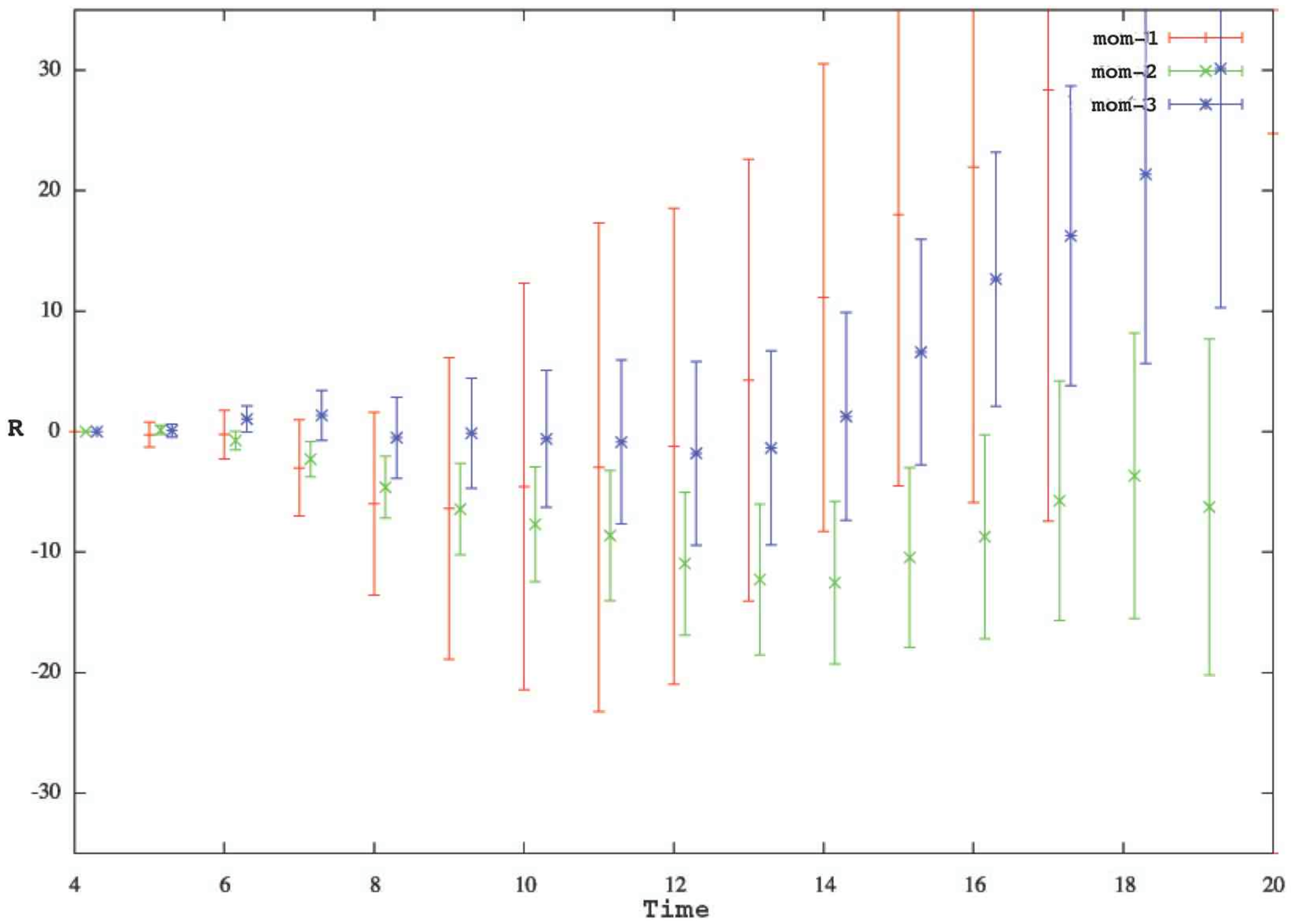}
\caption{Ratio in equation (\ref{KTway}) for the first three momenta for mass 3 scalar density diagram at $\kappa_{loop}$=.154.}
\label{fig:mass3.154}
\end{figure}}

{\setlength{\abovecaptionskip}{0mm}
\setlength{\belowcaptionskip}{0mm}
\begin{figure}[t!]
\centering
\includegraphics[scale=1, height=8cm, width=12cm]{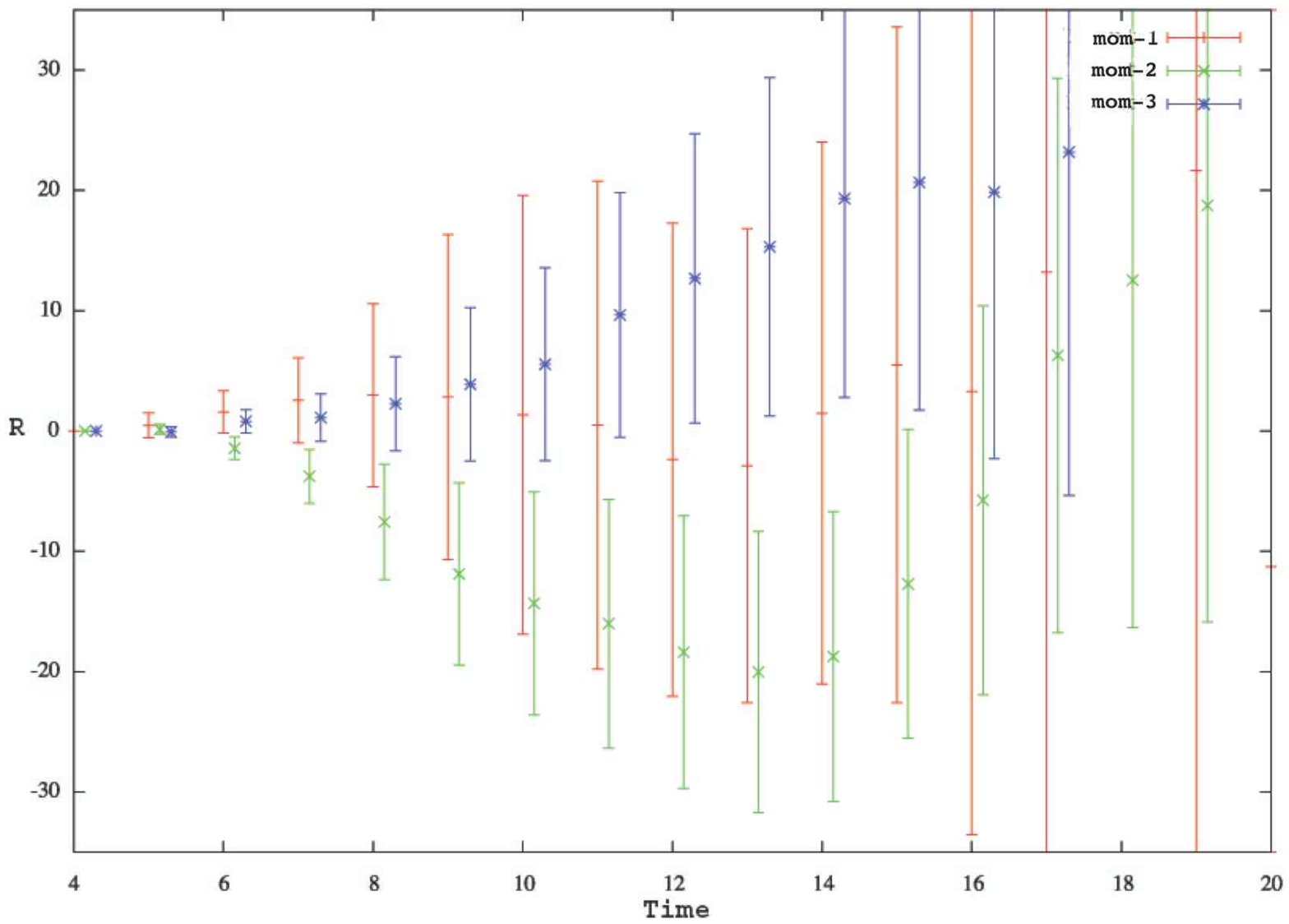}
\caption{Ratio in equation (\ref{KTway}) for the first three momenta for mass 4 scalar density diagram at $\kappa_{loop}$=.152.}
\label{fig:mass4.152}
\end{figure}}

{\setlength{\abovecaptionskip}{0mm}
\setlength{\belowcaptionskip}{0mm}
\begin{figure}[t!]
\centering
\includegraphics[scale=1, height=8cm, width=12cm]{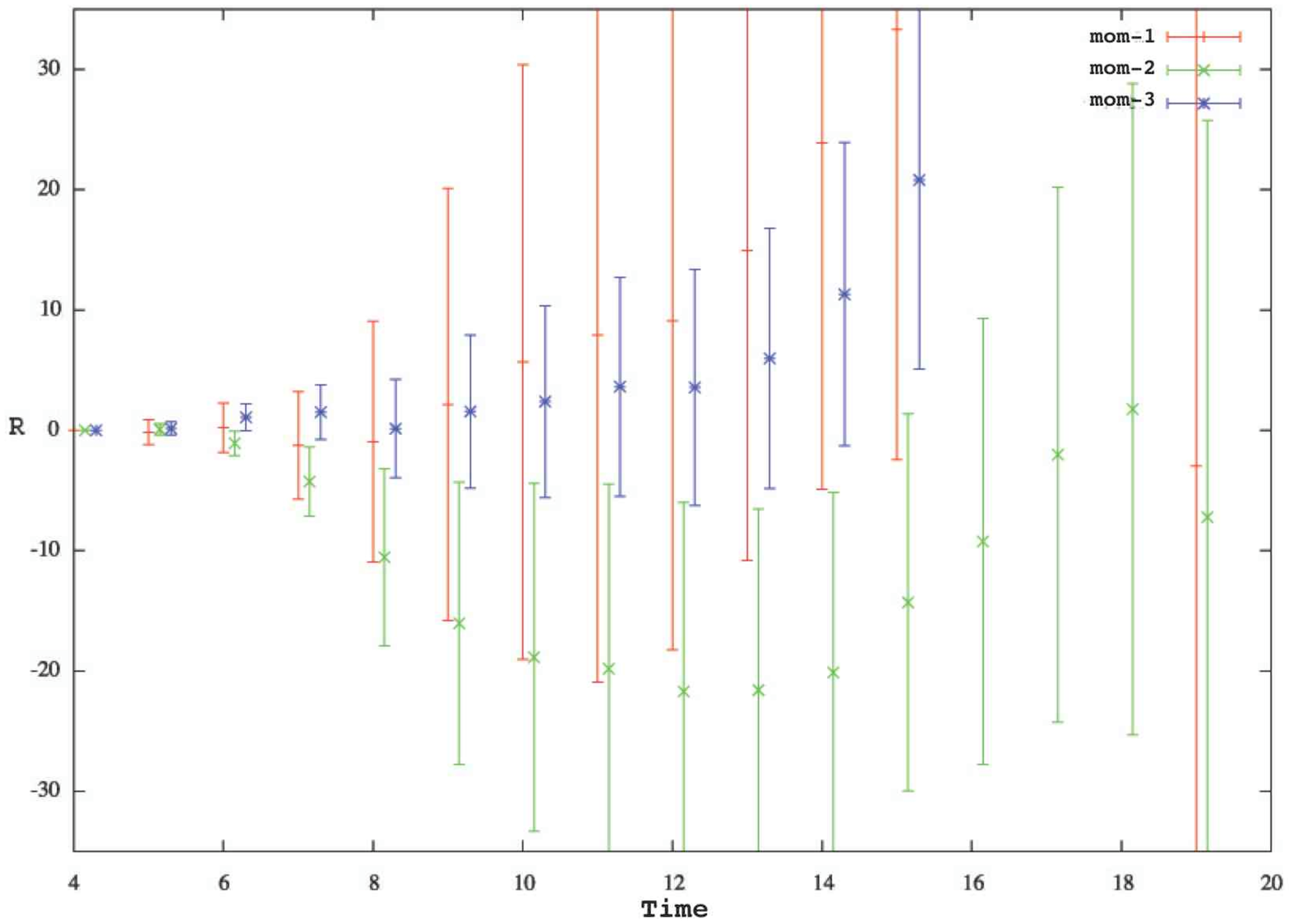}
\caption{Ratio in equation (\ref{KTway}) for the first three momenta for mass 4 scalar density diagram at $\kappa_{loop}$=.154.}
\label{fig:mass4.154}
\end{figure}}

\section{Jackknife Error Bars and Linear Fit}

The uncertainties for all the figures in this chapter are calculated with a jackknife error bar technique ~\cite{Efron,Young}. A jackknife error bar is calculated for the ratio in equation (\ref{KTway}) at every time step. The jackknife technique is referred to as a resampling method because it uses small changes from the original data set to determine the uncertainty in the data. The change in the data is a result of omitting each configuration from the ensemble average one-by-one and reproducing the ratio of the three and two-point functions at each time slice. The jackknife method is summarized below. 
\pagebreak
If we define the jackknife averages of the ratios from equation (\ref{KTway}) to be the configuration average of the ratios while omitting the $i^{th}$-ratio, then,
\begin{equation}\label{jackave}
R^{J}_{i}(\vec{q},t) \equiv \frac{1}{N-1}\sum_{i \not =j}R_{j}(\vec{q},t).
\end{equation}
 Also, we define the jackknife estimate of the ratio as the configuration average over all jackknife averages defined in equation (\ref{jackave}),
\begin{equation}
<R^{J}> \equiv \frac{1}{N}\sum^{N}_{i=1}R^{J}_{i}.
\end{equation}
The uncertainty in the ratio is then,
\begin{equation}
\sigma_{R^{J}} = \sqrt{\frac{1}{N-1}(<(R^{J})^{2}> - (<R^{J}>)^{2})}.
\end{equation}
The values of $\sigma_{R^{J}}$ are the uncertainties calculated for each time-slice and momentum for all masses in the figures above. 

Linear fits of the data in figures (\ref{fig:mass1.152}) - (\ref{fig:mass4.154}) are needed to extract the matrix elements. A least-squares fit to an arbitrary function is used to find the best fit over a specific range of time-steps ~\cite{Bevington}. Let $\chi=\chi(R,\sigma)$ be a fit parameter that is a function of the uncertainties generated with the jackknife technique and the ratio, $R$, from equation (\ref{KTway}). The method then searches the parameter space of $\chi$ to find its minimum value which corresponds to the best uncorrelated linear fit of the data. This method was adapted to consider correlated fits of the jackknife ratio data ~\cite{andersen-1997-255}. The fit parameter is multiplied by the covariance matrix, $C_{ij}$, defined in equation (\ref{covmat}). 
\begin{equation}\label{covmat}
C_{ij} = \frac{N-1}{N}\sum^{N}_{n=1}(R^{J}_{i}(n) - <{R}^{J}>)(R^{J}_{j}(n) - <{R}^{J}>),
\end{equation} 
where  N is the number of configurations, $\left\{i,j\right\}$ represent different time slices, $R^{J}_{i}(n)$ has jackknife ratio data, and $<{R}^{J}>$ is the configuration average that removes the bias. The covariance matrix considers correlations between ratio data at different time slices. These correlated fits are used in this thesis because it predicts the best linear fit over a specific time interval and the corresponding uncertainty in that fit. The fits and error bars presented in Table (\ref{tbl:152}) and Table (\ref{tbl:154}) are calculated with this correlated least-squares method. 

\section{Discussion}

The data in Tables (\ref{tbl:152}) and (\ref{tbl:154}) are for $\kappa_{loop}=.152$ and $\kappa_{loop}=.154$, respectively. In each table, the lowest three momentums are reported for all four twisted mass ($\kappa, \mu$) pairs. The range of the time interval for the linear least-squares fit and the associated error bar are given. The best linear fits for each mass and momentum are reported. Since this is a low statistics study giving preliminary results, the range of each fit is different for each mass and momentum.  
%\vspace{-24pt}
{\begin{table}[t!]
\caption{Fits for the matrix elements from equation (\ref{KTway}) for all 4 masses}
\centering{and $\kappa_{loop}=0.152$. The momentum squared is}
\centering{$a^{2}\vec{q}^{2}=n(\pi/10)^{2}$.}
\vspace{-12pt}
%\vspace{-5mm}
\begin{center}
\begin{footnotesize}
\begin{tabular}{l cccc}
\hline \\
%Heading Line 1
 ($\kappa_{v}$,$\mu$) & time steps& n   &    Scalar    \\

\hline

(0.15679, 0.030) &      &    &\\

                 &13-17 & 0  & 2.1 $\pm$ 1.5    \\
                 &15-17 & 1  & 0.77 $\pm$ 0.64    \\
                 &15-19 & 2  & 0.37 $\pm$ 0.30   \\
(0.15708, 0.015) &  & &\\

                 &13-16 & 0  & 2.3 $\pm$ 1.9     \\
                 &15-18 & 1  & 1.2 $\pm$ 0.89    \\
                 &12-16 & 2  & 0.88 $\pm$ 0.81 &   \\  
(0.15721, 0.010) &  & &\\

                 &15-19 & 0  & 2.5 $\pm$ 1.8    \\
                 &14-17 & 1  & 1.6 $\pm$ 1.2    \\
                 &9-13 & 2  & 0.90 $\pm$ 0.76  \\    
(0.15728, 0.005) &  & \\

                 &13-18 & 0  & 3.2 $\pm$ 3.0     \\
                 &14-18 & 1  & 1.3 $\pm$ 1.2   \\
                 &9-12  & 2  & 0.97 $\pm$ 0.79   \\
\\
\hline  
\vspace{-34pt}            
\end{tabular}
\label{tbl:152}
\end{footnotesize}
\end{center}
\end{table}
}
\nopagebreak
{\begin{table}[h!]
\caption{Fits for the matrix elements from equation (\ref{KTway})for all 4 masses} 
\centering{and $\kappa_{loop}=0.154$. The momentum squared is } 
\centering{$a^{2}\vec{q}^{2}=n(\pi/10)^{2}$.}
\vspace{-12pt}
%\vspace{-5mm}
\begin{center}
\begin{footnotesize}
\begin{tabular}{l cccc}
\hline \\
%Heading Line 1
 ($\kappa_{v}$,$\mu$) & time steps& n   &   Scalar    \\
 
\hline

(0.15679, 0.030) &      &    &\\

                 &13-16 & 0  & 2.1  $\pm$ 1.6    \\
                 &14-18 & 1  & 1.0  $\pm$ 0.75    \\
                 &15-19 & 2  & 0.49 $\pm$ 0.38   \\
(0.15708, 0.015) &  & &\\

                 &11-16 & 0  & 1.9 $\pm$ 1.5     \\
                 &14-18 & 1  & 1.4 $\pm$ 1.0    \\
                 &14-18 & 2  & 0.92 $\pm$ 0.74 &   \\  
(0.15721, 0.010) &  & &\\

                 &11-16 & 0  & 2.1 $\pm$ 1.7    \\
                 &14-18 & 1  & 1.8 $\pm$ 1.4    \\
                 &13-19 & 2  & 1.6 $\pm$ 1.3  \\    
(0.15728, 0.005) &  & &\\

                 &9-13 & 0  & 2.3 $\pm$ 2.0     \\
                 &15-19 & 1  & 1.3 $\pm$ 1.0     \\
                 &10-13 & 2  & 0.97 $\pm$ 0.89   \\  
\\
\hline  
\vspace{-24pt}     
\end{tabular}
\label{tbl:154}
\end{footnotesize}
\end{center}
\end{table}
}

The value of the scalar for the lightest quark mass at second momentum for $\kappa_{loop}=0.154$ is considered over the $15-19$ time slices in Table (\ref{tbl:154}). Figure (\ref{fig:mass4.154}) shows the plot of this data. The data point at the $19^{th}$ time slice is included because the fitting routine suggests that this point is highly correlated over this range and reasonable to fit with. The scalar value, without this point, in the range $15-18$ is 3.4 $\pm$ 3.3. This large change in the scalar value is a result on insufficient statistics and will be resolved with the addition of more configurations.

\begin{figure}[h!]
\centering
\includegraphics[scale=1, height=8cm, width=12cm]{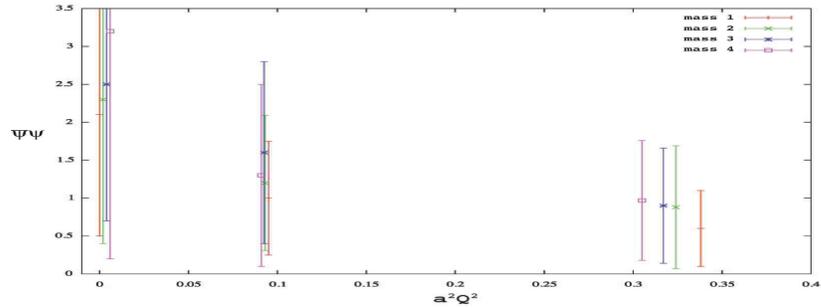}
\caption{The scalar density for the first three momentums and all momentums as a function of the dimensionless 4-momentum transfer $a^{2}Q^{2}$. These plots are for $\kappa_{loop}$=.152.}
\label{fig:152q2}
\end{figure}

\begin{figure}[h!]
\centering
\includegraphics[scale=1, height=8cm, width=12cm]{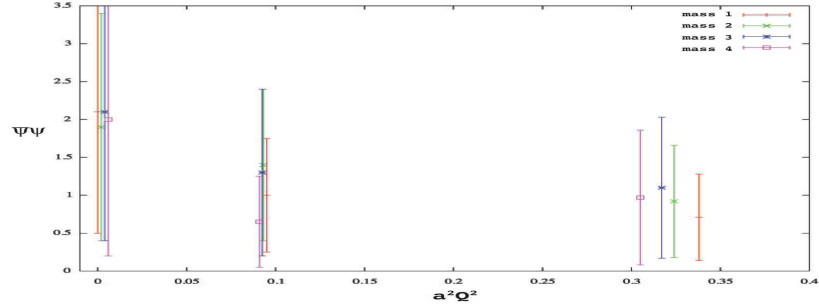}
\caption{The scalar density for the first three momentums and all momentums as a function of the dimensionless 4-momentum transfer $a^{2}Q^{2}$. These plots are for $\kappa_{loop}$=.154.}
\label{fig:154q2}
\end{figure}

Plots of the scalar $\bar{\psi}\psi$ as a function of the dimensionless 4-momentum transfer squared ($a^{2}Q^{2}$)  are plotted in Figures (\ref{fig:152q2}) and (\ref{fig:154q2}). The square of the 4-momentum transfer is 
\begin{equation}
Q^{2} = (q - q^{\prime})^{2},
\end{equation}
where $q=(E,\vec{q})$ and $q^{\prime}=(m_{N},0,0,0)$ is the final and initial momentum respectively, and $m_{N}$ is the nucleon mass from ~\cite{abdel-rehim-2005-71}. Then $Q^{2}$ can be written 
\begin{equation}
Q^{2} = 2m(E - m).
\end{equation}
The nucleon masses used in the 4-momentum transfer plots were calculated in reference ~\cite{abdel-rehim-2005-71}. One can see that the scalar density falls of smoothly and has similar behavior for both disconnected loop values. 

The zero momentum values in Figures (\ref{fig:152q2}) and (\ref{fig:154q2}) are offset slightly so that the data points can be clearly identified. 
  
\begin{figure}[h!]
\centering
\includegraphics[scale=1, height=8cm, width=12cm]{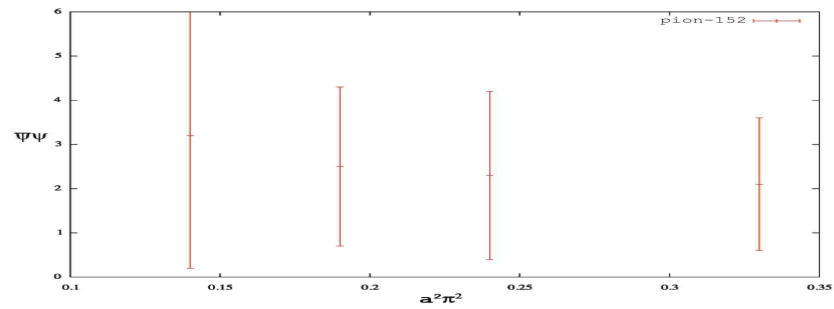}
\caption{The zero momentum scalar density as a function of pion mass squared. These plots are for $\kappa_{loop}$=.152.}
\label{fig:pion152}
\end{figure}

\begin{figure}[h!]
\centering
\includegraphics[scale=1, height=8cm, width=12cm]{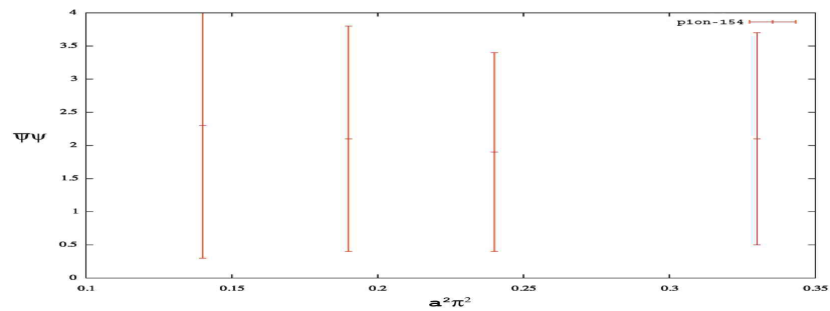}
\caption{The zero momentum scalar density as a function of pion mass squared. These plots are for $\kappa_{loop}$=.154.}
\label{fig:pion154}
\end{figure}

Another useful plot is the scalar density as a function of the pion mass. The pion mass squared is proportional to the quark mass. The pion mass is used in $\chi$PT to extrapolate to the physical quark masses. A plot of the scalar ($\bar{\psi}\psi$) as a function of the pion mass squared for each of the loop values is found in Figure (\ref{fig:pion152}) and Figure (\ref{fig:pion154}). The pion masses corresponding to our maximal twist masses used in these plots are reported in ~\cite{abdel-rehim-2005-71}.

A similar calculation of the strangeness scalar density matrix elements in the Wilson formalism is presented in reference ~\cite{lewis-2003-67}. Their calculation was performed on a $20^{3} \times 32$ lattice with $\beta=6.0$. The valence quark hopping parameters were $\kappa_{v}=\left\{0.152,0.153,0.154\right\}$ with loop values of $\kappa_{loop}=0.152$ and $\kappa_{loop}=0.154$. The results in this high statistics calculation are determined with 2000 configurations with statistical uncertainties obtained from 3000 bootstrap ensembles. The fit to the scalar density in this paper are found in Table (\ref{tbl:Wilsca}). The fits in this table begin 10 time steps from the source. \enlargethispage{\baselineskip}

{\begin{table}[t!]
\caption{Fits for the scalar matrix elements from reference ~\cite{lewis-2003-67}.}
\centering{These scalar values are calculated for}
\centering{ $\kappa_{loop}$=0.152 and $\kappa_{loop}$=0.154}
\vspace{-12pt}
% \vspace{-5mm}
\begin{center}
\begin{footnotesize}
\begin{tabular}{l cccc}
\hline \\
% Heading Line 1
$\kappa_{v}$ & n   & $\kappa_{loop}=0.152$  & $\kappa_{loop}=0.154$    \\

\hline

          0.152 &    &      &         \\

                & 0  & 2.6(4) & 3.7(13) \\
                & 1  & 1.7(2) & 2.1(6) \\
                & 2  & 1.2(2) & 1.1(6) \\

          0.153 &    &      &         \\

                & 0  & 2.7(5) & 4.0(14) \\
                & 1  & 1.8(3) & 2.2(7) \\
                & 2  & 1.3(2) & 1.3(11) \\

          0.154 &    &      &         \\

                & 0  & 2.9(5) & 4.2(5) \\
                & 1  & 1.8(3) & 2.3(8) \\
                & 2  & 1.3(3) & 1.3(8) \\
\\
\hline
\vspace{-34pt}
\end{tabular}
\label{tbl:Wilsca}
\end{footnotesize}
\end{center}
\end{table}}

% The results in ~\cite{lewis-2003-67} were found to be independent of the valence quark mass. 

The scalar results in ~\cite{lewis-2003-67} were found to decrease in amplitude as the momentum is increased. The scalar values in Tables (\ref{tbl:152}) and (\ref{tbl:154}) also decrease as the momentum increases. The scalar data reported in the high statistics study was found to be independent of the valence quark masses. Our data appears to increase slightly for smaller valence quark mass for both loop values. The nucleon quark masses in our study are the lightest valence masses used for the nucleon strangeness calculation to date and, therefore, our preliminary results are the first twisted mass calculation of the nucleon strangeness scalar density. 

The lightest valence quark mass in ~\cite{lewis-2003-67} is most comparable to the heaviest mass in our simulation. The matrix elements amplitudes of the lowest three momenta for the twisted and Wilson case are different. For example, in the Wilson case, the zero momentum scalar densities for $\kappa_{v}=0.154$ with $\kappa_{loop}=0.152$ and $\kappa_{loop}=0.154$ were reported to be 2.9 and 4.2 respectively. Our scalar values most comparable to $\kappa_{v}=0.154$ are 2.1 for both loop values. Our heaviest twisted quark mass pair is $(\kappa=0.15679, \mu=0.030)$. The values for the higher momentums compare similarly.    

%The lightest two quark masses (mass 3 and mass 4) in our simulation are nosier that the other two heavier masses. The increased noise in the lighter masses is expected. As one approaches lighter quark masses in the twisted formalism you approach the Wilson case. The Wilson case is known to suffer from large statistical errors at very light quark mass. It is for this reason that we expect the lighter quark masses to take more configuration ensembles to attain a reliable signal. 

% In the previous Wilson study, the lightest $\kappa_{v}=0.154$ value showed similar behavior. It was reported that if the uncertainties go like $1/\sqrt{N}$, then the ratio of uncertainties between the lightest and heaviest valence quark masses was 2.8.

Our preliminary results suggest that the raw data for the scalar elements are being calculated correctly. This calculation is aimed toward forming the renormalization group invariant quantity representing the fractional strange quark contribution to the nucleon mass in equation (\ref{snucmass}).

\begin{equation}\label{snucmass}
\frac{m_{s}<N|\bar{s}s|N>(0)}{m_{N}}.
\end{equation}
Once higher statistics are acquired the physical masses can be obtained in the continuum limit using $\chi$PT. The success of this calculation gives hope for future, high statistics, calculations using these methods of the electric and magnetic form factors to determine electric and magnetic properties of the nucleon in the presence of a strange quark loop.

\chapter{Conclusion}
%\section{Conclusion and Future Work}

A study of the strangeness contribution to the nucleon was conducted in this thesis. Our results show that the methods presented here are viable and will allow for a better study of the strangeness content of baryons. More specifically, we have shown preliminary results that indicate that the twisted mass formalism is a good approach to calculate the scalar form factor. To calculate the scalar many new and interesting techniques were developed to ``zero-in" on the form factor values using lighter valence quark masses so that we can make better contact with experimental results. Future work will include a high statistics calculation of the electric and magnetic form factors so that one may have a better understanding of the nucleon electromagnetic properties. 

We have shown many techniques to improve lattice calculations. Our results show that useful variations of the GMRES(m) algorithm can be employed to solve systems of linear equations that arise in Lattice QCD calculations efficiently. The saved matrix-vector products from these algorithms can reduce computational time dramatically over the life of a high statistics lattice calculation. Specifically, we have shown that GMRES-DRS(m,k) is a good technique to solve shifted systems of equations in the Wilson case by taking advantage of the properties of the Krylov subspace. As an extension to GMRES-DRS(m,k), we have developed another new technique to use a shifted GMRES(m)-Proj(k) method to solve subsequent right-hand simultaneously after the base system has used GMRES-DRS(m,k). 

The disconnected quark loop calculation used to form the disconnected three-point function was improved by expanding to higher orders in the perturbative expansion. This is an important result because going to higher order in kappa further reduces the variance of the loop operators and saves valuable computer time in the calculation. A twisted mass noise method was also presented in this thesis. This method responds well to the subtraction techniques in that the variance of the twisted loop operators is significantly reduced in our simulations. As noted in chapter \ref{chapt:discon}, these loops suffer from scalar-pseudoscalar mixing that causes both the scalar and the pseudoscalar to acquire a VEV. Future work to remove the mixing in the twisted perturbative subtraction method is necessary so that one may go to lower quark mass for the loops and produce more accurate strangeness calculations. 
%
%\chapter{Optical Unification:  Model FCREU1 Fields and Charges}
%\input{optuntab}
%
%\chapter{Broken Mirror Models:  States for Models 1 and 2}\label{chapt:mirrortables}
%\input{mirrortab}

%=========================================================================
%atbib
               
%\bigskip
%medskip
{
%\def\bibiteml#1#2{ }
%\begin{thebibliography}{99}
\bibliographystyle{phaip}
\nocite{*}
\bibliography{diss}
}
%\input{dissbib}
%\end{thebibliography}
%==============================================================================
\end{document}